\documentstyle[twocolumn,aps,prl,epsf]{revtex}

\newcommand{\di}{\mbox{d}}

\begin{document}
\title {%
Quantum-critical theory of the spin-fermion model
and its application to cuprates. Normal state analysis.}
\author{Ar. Abanov$^1$, Andrey V. Chubukov$^1$, 
and J. Schmalian$^2$}

\address{$^1$
Department of Physics, University of Wisconsin, Madison, WI 53706}
\address{$^2$ Department of Physics and Ames Laboratory, 
Iowa State University, Ames, IA 50011}  

\date{\today}
\maketitle

\begin{abstract}
We present the full analysis of the normal state of the spin-fermion 
model near
the antiferromagnetic instability in two dimensions. 
This model describes low-energy fermions interacting with their own collective
spin fluctuations, which soften at the antiferromagnetic transition. 
We argue that in 2D, the system has two typical energies - an effective
spin-fermion interaction $\bar g$ and an energy $\omega_{sf}$ below which the
system behaves as a Fermi liquid. The ratio of the two determines the
dimensionless coupling constant for spin-fermion interaction $\lambda^2 
\propto {\bar g}/\omega_{sf}$. We show that $\lambda$ 
scales with the spin correlation
length and diverges at criticality. This divergence implies that the 
conventional perturbative expansion breaks down. We 
developed a novel approach to the
problem - the expansion in either the inverse number of hot spots in 
the Brillouin zone, or the inverse number of fermionic flavors - 
which allowed us to explicitly account for
all terms which diverge as powers of $\lambda$, and treat the 
remaining, $O(\lambda)$ terms in the RG formalism.   
We applied this technique to study the properties of the spin-fermion model 
in various frequency and temperature regimes. We present the results for
the fermionic spectral function, 
spin susceptibility, optical conductivity and 
other observables.  We compare our 
results in detail with the normal state data 
for cuprates, and argue that the 
spin-fermion model is capable to explain the 
anomalous normal state properties of 
high $T_c$ materials. We also discuss the non
-applicability of the conventional 
$\phi^4$ theory of the quantum-critical behavior in 2D. 
\end{abstract}


\section{introduction}

The 15 years after the discovery of superconductivity in cuprate oxides~\cite{BM}
witnessed a large number of efforts to understand the mechanism of
superconductivity and numerous unusual normal state properties of 
cuprates. Parent compounds of cuprates (e.g., $YB_2C_3O_6$) are insulators and
also best examples of 2D Heisenberg antiferromagnets~\cite{AFM}. Upon hole doping, the
antiferromagnetism disappears, and the system instead becomes a superconductor.
Above optimal doping (which is defined such that superconducting 
$T_c$ is the largest), superconducting features, such as the gap, $\Delta$,
 in the spectral function, disappear at $T_c$~\cite{photo_over}. Below optimal doping, the 
gap survives above $T_c$, and disappears only at a higher temperature $T^*$~\cite{photo_under}. 
In addition, $T_c$ and $\Delta(T=0)$ 
become progressively  
uncorrelated with underdoping - $T_c$ decreases as doping decreases,  
while $\Delta (T=0)$ gets larger~\cite{tunn_under,girsh_science}.
This clearly contradicts BSC theory
 in which  $\Delta (T=0)$ scales with  $T_c$.
The region between $T_c$ and $T^{*}$ is called the pseudogap region. 

\begin{figure}[tbp]
\begin{center}
\epsfxsize=\columnwidth 
\epsffile{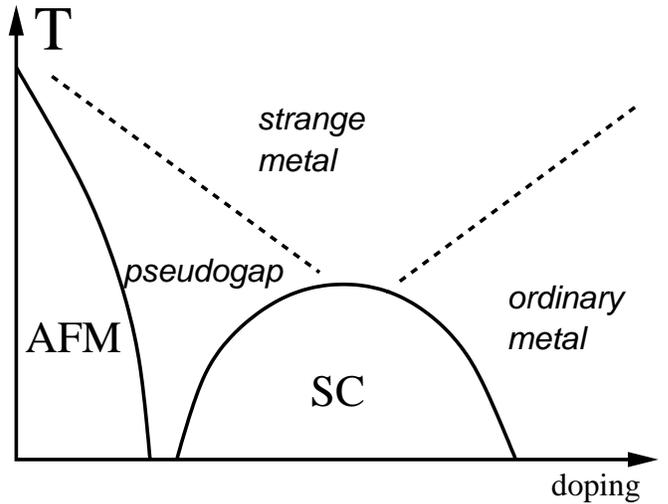}
\end{center}
\caption{ 
A schematic phase diagram of cuprates.
}
\label{phase-diagram}
\end{figure}

A schematic phase diagram of cuprates is presented 
in Fig.\ref{phase-diagram}.
There is general agreement among the researchers that the superconducting 
state predominantly has $d_{x^2-y^2}$ symmetry~\cite{tsue_RMP}. 
In this pairing state, the
superconducting gap has nodes for diagonal directions along the Fermi surface, 
i.e., for $k_x = \pm k_y$. Much less, however, researches 
agree at the present time about  
what is the origin of the pseudogap and also the 
anomalous normal state properties of cuprates. 
These two features (the pseudogap and the anomalous normal state behavior)
are challenging for researchers and keep high $T_c$ problem afloat 
over such large period of time. 

This paper is devoted to the normal state properties of cuprates,
 so to keep our discussion focused we discuss only  the experimental 
evidence for anomalous normal state behavior.
The key evidence collaborated by   
very detailed photoemission~\cite{valla,kaminski}, optical~\cite{timusk},
 transport~\cite{transport}
 and other studies is that
 near optimal doping, the inverse quasiparticle 
lifetime is nearly linear in both temperature and frequency over a  range
 of $T$ between $T_c$ and $1000K$, and  at frequencies roughly between few tens and few hundred meV. 
The Fermi liquid theory on the other hand predicts that the inverse quasiparticle 
lifetime is quadratic in $\omega$ and $T$ at the lowest energies. 
In overdoped cuprates,
 the low-frequency behavior of the quasiparticle lifetime is more
consistent with the Fermi liquid theory~\cite{extra_over}.

While there is no experimental proof at the moment that the 
measured non-Fermi liquid behavior at optimal doping 
extends down to $\omega, T =0$ (i.e., that there is no
crossover to the Fermi liquid behavior at the lowest energies),
 the very fact that
the system behavior is markedly different from that in a Fermi liquid 
over a wide range of energies implies that theoretical description of 
cuprates should
necessary involve strong coupling effect. 

From theoretical perspective, there exist two very distinct proposals about
what may cause non-Fermi-liquid behavior. First proposal is that this 
behavior is generic to doped Mott insulators in two dimensions
 and is 
due to the fact that 
strong Coulomb interaction  not only makes  
the picture of weakly interacting electrons invalid, but electron even
 ceases to exist as a quasiparticle (its residue is identically zero),
 and should instead be viewed as a convolution of two other objects 
one of which (spinon) carry spin but no charge, while the other (holon) 
carry charge but no spin~\cite{anderson}. The theories based on the idea that spin and 
charge excitations in
 cuprates are separated have been worked out in great detail~\cite{spin_charge_sep} and 
substantially  deepened our understanding of the physics of very 
strongly coupled electrons.

A second proposal is based on the assumption that the anomalous 
normal state behavior is caused by the closeness to a quantum 
phase transition of one type or another~
\cite{varma1,grilli,subir21,david_review,chub2}.
  Near phase transitions, 
quantum fluctuations are generically enhanced, and this should give rise to 
deviations from Fermi-liquid behavior~\cite{NFL_at_crit_general}.
 Moreover,
 in low $D$ systems,
the region of Fermi liquid behavior progressively shrinks to lower and 
lower energies as the system approaches quantum criticality.
  In distinction to the first approach, the idea 
that normal state behavior is due to a near quantum criticality 
necessary implies that at least on one side of a transition, the 
system behaves as a Fermi liquid at the lowest frequencies. 

The idea of quantum criticality emerged in the early days of high $T_c$ era
as a way to justify the successful 
 phenomenology~\cite{varma1}
 which used the linear energy dependence of
 the quasiparticle scattering rate as an input (the Marginal Fermi 
Liquid theory (MFL)). Another early idea of quantum criticality was focused on 
magnetic properties of cuprates, 
  and was based on the assumption that the behavior of weakly doped 
antiferromagnet is close to that of undoped systems near the quantum 
transition between magnetically ordered and disordered states~\cite{subir}. 

At present, there are three suggestions as to what kind of
 quantum criticality 
may cause the observed non Fermi-liquid behavior. The first is a 
phenomenological idea that there exists a quantum critical point 
somewhere around optimal doping, 
 which separates Fermi-liquid
and pseudogap phases~\cite{varma}.
 In this phenomenology, superconducting region 
emerges as a dome on top of this point. The second, also phenomenological, 
suggestion
 is that
quantum criticality and pseudogap behavior may be caused by a competition
 between superconductivity and some other yet undiscovered 
ordering (specific suggestion involves $d-$density wave~\cite{ddw}).
 The third suggestion
is that non Fermi liquid physics is due to the closeness to a magnetic 
quantum critical point which separates paramagnetic and antiferromagnetic 
phases~\cite{david_review,chub2}.

In this paper we explore the third possibility and discuss what kind of 
critical behavior one can expect near antiferromagnetic transition. 
The magnetic scenario emerged from the early days of high $T_c$ era~\cite{david_review,scal_pr} 
and was primarily based on the experimental observations            
 that although antiferromagnetism observed in parent compounds 
quickly disappears upon doping, 
 short-range antiferromagnetic correlations persist
up to much larger doping concentrations as is evident from 
 NMR~\cite{nmr} and neutron scattering~\cite{aeppli,keimer1} 
data, and these concentrations possibly cover the whole doping range of
a superconducting behavior.    
Another favorable feature of a  magnetic scenario, known even before 
high $T_c$, is that near half-filling, the  spin fluctuation exchange  
 gives rise to
an attraction  in a $d_{x^2 -y^2}-$ pairing channel~\cite{scal_pr} and hence can 
cause $d-$ wave superconductivity~\cite{mont_pines}.
 A simple way to understand this is to 
assume that  spin-fluctuations in high-$T_c$ materials
play the
same role as phonons in ordinary superconductors and 
create an analog of a deformation potential 
in the form 
\begin{equation}
{\cal H}_{sf} = g \int d{\vec k} d {\vec q}
 ~c^{\dagger}_{k,\alpha} {\vec \sigma}_{\alpha,\beta}
c_{k+q,\beta} {\vec S}_{-q},
\label{1}
\end{equation}
where $\sigma^i_{\alpha,\beta}$
 are Pauli matrices and $g$ is the coupling constant.
Using this Hamiltonian, one can obtain an effective  
pairing interaction between fermions 
 which amplitude is proportional to the propagator of the
bosonic field $S_q$, i.e., to the dynamical spin
susceptibility. In a weak coupling BSC theory, 
spin dynamics is irrelevant, and the gap
equation in a spin singlet channel has the form~\cite{landau} 
\begin{equation}
\Delta_k = - \frac{3}{2} g^2 \int \chi ({\bf k}-{\bf p}) \Delta_p 
\frac{\tanh(\epsilon_p/2T)}{2
\epsilon_p} d{\vec p}
\label{gap}
\end{equation} 
In distinction to phonons,
 the r.h.s. of this
equation has an extra minus sign due to a summation over spin components~
\cite{mont_pines} 
(a projection of 
$\vec{\sigma_{\alpha \beta}} \vec{\sigma_{\gamma \delta}}$ onto a
singlet spin channel yields $-(3/2) (\delta_{\alpha
\beta} \delta_{\gamma \delta} - \delta_{\alpha \delta} \delta_{\beta \gamma})$
whereas a conventional spin-independent interaction yields no 
overall minus sign). 
Because of a sign change, an $s-$wave solution $\Delta_k = const$
is impossible. 
However, since $\chi ({\bf q})$ is peaked near an 
antiferromagnetic ${\bf Q} =(\pi,\pi)$,
 pairing interaction relates
the gap at momenta ${\bf k}$ and ${\bf k}+{\bf Q}$.  In this situation, 
one can eliminate the overall
minus sign by using an ansatz $\Delta_k = - \Delta_{k+Q}$.
For tetragonal lattice, 
this ansatz implies $d_{x^2 - y^2}$ symmetry of the pairing gap~\cite{agr}, i.e., 
 spin fluctuation exchange gives rise 
  to a $d-$wave superconductivity.

\begin{figure}[tbp]
\begin{center}
\epsfxsize=\columnwidth 
\epsffile{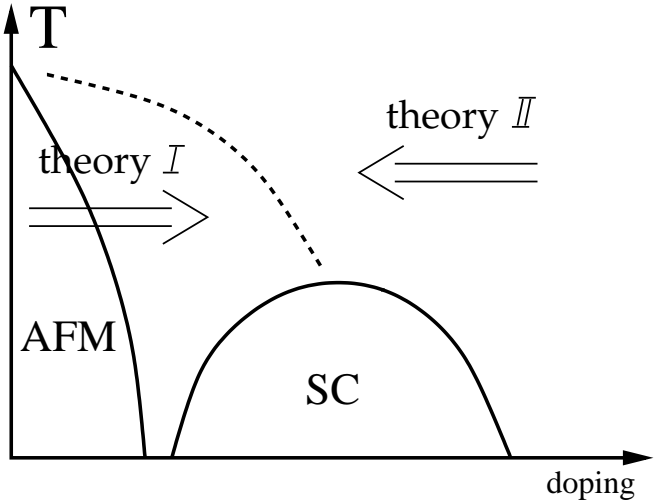}
\end{center}
\caption{ 
Two different theoretical paths to study the physics of underdoped and optimally doped cuprates. In the first path, labeled as ``theory I'', the point of departure is the antiferromagnetic state at half-filling. In the second path, labeled as ``theory II'' the 
point of departure is the conventional metallic state in the overdoped regime.
}
\label{phase-diagram-theory}
\end{figure}

The studies of the normal state properties of cuprates within 
the magnetic scenario could also be
divided into two subclasses (see Fig. \ref{phase-diagram-theory}). 
In the first class, which we labeled as ``theory I'',
 the  point of departure is
Heisenberg antiferromagnetism at half-filling, and the issue addressed is 
how the system evolves with increasing doping~\cite{subir21,scs,subir,shraiman_siggia,swz,dagotto,castro_neto,hanke}. These studies mostly consider
small concentrations of electrons at which either long-range or 
quasi-long range magnetic order keeps majority of electrons localized, 
and the Fermi surface consists of separated patches (hole pockets). 
In the second class, labeled in Fig~\ref{phase-diagram-theory} as ``theory II'', the point of
departure is the Fermi liquid  metallic state at large dopings, and 
the issue addressed is how this Fermi liquid behavior modifies and 
eventually disappears as the system approaches the magnetic transition. 
In this second approach, the
physics is governed by the interaction between electrons and their collective 
 spin bosonic excitations  which become soft modes near the transition. This
 model is often called the spin-fermion model.  
Indeed, both approaches describe the same strongly interacting 
electronic system near magnetic instability
 and should yield the same results
for the same regions of doping and temperatures, unless 
there are instabilities 
at large energies 
(comparable to e.g., fermionic bandwidth), which  
cannot be detected in the low-energy theory and have to be taken as inputs. 
From this perspective, the choice of the model is somewhat subjective 
and is mostly determined  by 
which starting point is thought to be closer to real situation at a given 
doping. Very near half-filling, the system clearly displays Heisenberg 
antiferromagnetism, and
the Fermi surface should consist of hole pockets. In this limit, 
one has to work rather hard within the spin-fermion model to reproduce 
the features which are readily obtained if one departs from 
antiferromagnetic state at half-filling.  On the other hand, we believe that 
the spin-fermion model is a better starting point 
if one wants to describe the physics 
at and somewhat below optimal doping mostly because 
at present the majority of photoemission experiments even in heavily 
underdoped cuprates do not show the emergence of the shadow Fermi 
surface (the extra piece of the Fermi surface 
shifted from the original one by antiferromagnetic ${\bf Q}$),
 which is the necessary ingredient for the formation of 
hole pockets~\cite{shadow_FS}. 

The present paper summarizes our analysis of the spin-fermion model 
in the normal state, or, more precisely, when the pairing correlations can be 
neglected.  We argue that in $2D$, a dimensionless coupling
constant for the spin-fermion interaction $\lambda$
 scales as inverse magnetic correlation length $\xi$ and diverges
as the system approaches a quantum-critical point (QCP) of a 
transition to a magnetically ordered state. We show
 that due to a divergence of $\lambda$, there exists a wide
region in the $T,\lambda$ plane where the system is in the 
quantum-critical regime near the magnetic transition.
We will show that in this regime, the system behavior is 
qualitatively different from that in a Fermi liquid. 

\begin{figure}[tbp]
\begin{center}
\epsfxsize=\columnwidth 
\epsffile{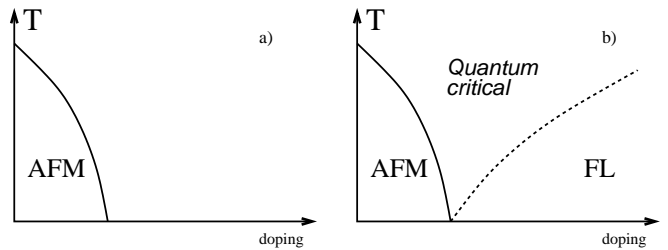}
\end{center}
\caption{ 
A cartoon of the system behavior near antiferromagnetic 
quantum-critical point
}
\label{phase-diagram-normal}
\end{figure}

Our point of departure is the ``zero-order'' phase diagram 
with antiferromagnetism included as an input 
( Fig.\ref{phase-diagram-normal}a). The deviations 
from the magnetic transition are measured by
 the doping dependent 
spin correlation length $\xi (x)$. 
We assume that all other features on the phase 
diagram, such as  superconductivity, 
pseudogap regime, and the crossover from a Fermi-liquid to not Fermi 
liquid behavior in the normal state are {\it low-energy phenomena} 
which should be 
{\it obtained} within the model. Alternatively speaking, we assume 
that antiferromagnetism is produced by fermions with energies comparable 
to the bandwidth, while other phenomena are produced by fermions near 
the Fermi surface. In cuprates, the separation of scales is not very strong 
($T^*$ for strongly underdoped materials is a fraction of an exchange integral $J$), i.e., to some extent 
high-energy fermions do contribute to pseudogap phenomenon and may 
also contribute to anomalous normal state behavior. 
This contribution is inevitably lost in our approach. However, we believe 
that the fundamental physics can be understood by separating the scales 
perhaps more strongly than in reality, and the inclusion of fermions 
with energies comparable to the bandwidth will change the results 
quantitatively but not qualitatively.

The phase diagram which emerges from our studies 
is presented in Fig. \ref{phase-diagram-normal}b.
We argue that the region near the QCP is divided into a
  Fermi liquid regime and a quantum-critical regime where the system behavior
is the same as at the critical point. The upper boundary of the 
quantum-critical behavior is 
roughly located at frequencies comparable to spin-fermion coupling constant 
${\bar g}$. The crossover from a Fermi liquid to a 
quantum-critical behavior on the other hand occurs at energies of order
 $\omega_{sf} \sim {\bar g}\lambda^2$  where $\lambda$, which 
we already introduced above, is the 
dimensionless coupling constant in the problem  
 ($\lambda \propto {\bar g}/v_F \xi^{-1}$, where $v_F$
 is the Fermi velocity). 
At weak coupling, $\omega_{sf}$ exceeds ${\bar g}$, and the 
quantum-critical behavior is not realized. However, at strong 
coupling, $\omega_{sf} \ll 
{\bar g}$, and the system displays quantum-critical behavior 
in a wide range of frequencies. This strong coupling behavior 
necessary occurs very near the transition as $\lambda \propto \xi$, 
but it can also be reached at intermediate $\xi$ when the spin-fermion 
interaction ${\bar g}$ increases.

 In the Fermi liquid regime, fermionic self-energy behaves as 
$\Sigma^{\prime \prime} \propto \omega^2/\omega_{sf}$. 
In the opposite limit, when $\omega_{sf} =0$, we found that it scales,
up to logarithmical corrections, as
 $\Sigma^{\prime \prime} \propto \sqrt{{\bar g E}}$, 
where $E= \mbox{max}(\omega, T)$.
 At a finite $\omega_{sf}$, we find 
that there also exists a wide intermediate region, roughly between 
$0.5 \omega_{sf}$ and $6-8 \omega_{sf}$, where 
$\Sigma^{\prime \prime} \propto E$, i.e., it is linear 
in both frequency and temperature. This behavior of the self-energy causes a
cascade of crossovers in the fermionic spectral function and conductivity measured as functions of temperature and frequency. 

These are the key results of the paper. We derive them in 
Sec.~\ref{n_s} -\ref{finiteT}  below in a formal $1/N$ expansion 
where $N$ can be regarded  either as 
the number of hot spots in the Brillouin zone 
(crossing points between the Fermi surface and the magnetic zone boundary), 
or the number of electron flavors.

\begin{figure}[tbp]
\begin{center}
\epsfxsize=\columnwidth 
\epsffile{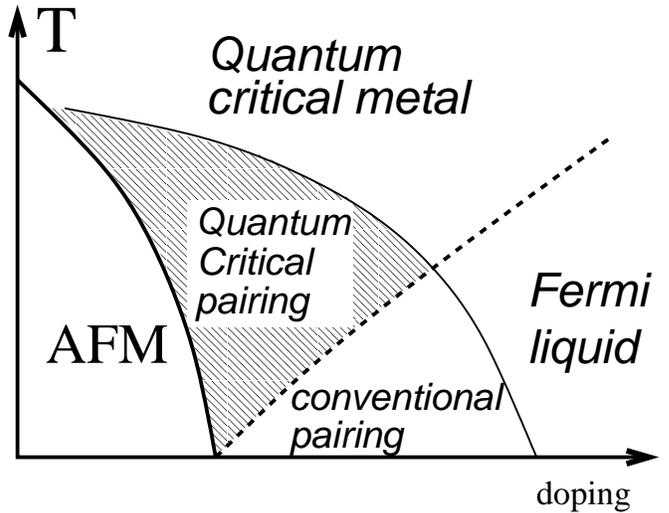}
\end{center}
\caption{ 
A schematic phase diagram of cuprates obtained in
\protect\cite{acs} 
in the
 Eliashberg-type formalism}
\label{phase-diagram-normalSC}
\end{figure}

As we said, in this paper we focus on the normal state properties 
of the spin-fermion model.
For completeness, we also present in Fig. \ref{phase-diagram-normalSC}  
the phase diagram 
which we obtained~\cite{acs} in the Eliashberg-type formalism 
(extended compared to that for phonons~\cite{eli_ph} 
 to include the feedback from pairing on magnetically mediated interaction)  
by using the normal state results as a base and adding the interaction 
in the pairing channel. 
Out key point here is that the two scales $\omega_{sf}$ and ${\bar g}$ which 
we find in the normal state are also present in the pairing problem. Namely,
the pairing instability temperature $T_{ins}$ tends to a finite value 
$\sim {\bar g}$ (more accurately, ${\bar g}/N$) at the quantum-critical 
point. In a conventional situation, this would imply that superconducting 
properties gradually evolve below $T_{ins}$. In our situation, this, 
however, is not the case as 
at strong coupling, when $T_{ins} \propto {\bar g} 
\gg \omega_{sf}$, the pairing comes from frequencies 
which well exceed $\omega_{sf}$,
 and thus involves incoherent fermions~\cite{acf}. This pairing is
 qualitatively different from that in the BCS theory. 
We found that in this situation, immediately 
below $T_{ins}$ electrons form singlet pairs, but the
 pairs still behave incoherently and do not propagate. Only when the 
temperature is reduced well below $T_{ins}$, the feedback from 
superconductivity
eventually ``clears up'' fermionic excitations, and the system  
behaves as a conventional superconductor. 
We found that the conventional superconducting behavior with e.g., a
sharp quasiparticle peak in the spectral function) emerges
 only when 
the temperature becomes smaller 
than  the fraction of $(\omega_{sf} {\bar g})^{1/2} \sim {\bar g}/\lambda 
\ll {\bar g}$. Physically, this second scale (modified $\omega_{sf}$) is 
 the gap in the spectrum 
collective spin excitations which become propagating, magnon-like 
in a $d-$wave superconductor~\cite{sw_dsc}.

The cuprates are not the only objects which motivate our study. 
Over the last few years, there is a growing interest in the behavior of
various heavy-fermion materials near the magnetic instability which can 
be achieved either by chemical variation (i.e., doping) or by applying 
external perturbation such as pressure~\cite{hf_exp,hf_dome,frac_power}.
 The magnetic instability can be either
 ferromagnetic or antiferromagnetic, depending on material.
 There is a convincing 
experimental evidence that at least in some of heavy fermion
 materials, there exists a dome on top of a quantum critical point 
where the system is in a superconducting phase~\cite{hf_dome}. 
From this perspective 
the behavior of heavy-fermion materials  may be not very different from 
that of cuprates. There is also a number of experimental data which show 
that the normal state behavior near magnetic instability likely deviates 
from that in a Fermi liquid down to very low energies~\cite{nfl_hf,hf_th}. 
These deviations 
have been observed in both fermionic and spin
 properties~\cite{frac_power}.
 We caution, however, that heavy-fermion materials are 
mostly three-dimensional~\cite{hf_th} while our analysis is 
valid in  two dimensions. Still, we believe that the computational 
scheme we developed for $2D$ can be applied also to $3D$ materials 
at strong coupling.

Our approach to the spin-fermion model 
is an extension of the earlier detailed studies 
 by D. Pines and co-workers~\cite{david_review,mont_pines,st_pines}. 
The only qualitative difference is that we do 
not assume a'priori the form of the dynamical susceptibility but rather derive it.  From this perspective, our 
work should be regarded as an attempt to put the studies
 of the spin-fermion model on a solid theoretical basis and  to understand 
whether one can attack the problem analytically, 
in a controllable way. Previous studies of the spin-fermion model 
were mostly  performed numerically.

Also, our approach and computational technique are
 in many respects similar to the studies  by 
B. Altshuler, A. Millis and L. Ioffe~\cite{ami,im} of fermions interacting with low-energy
 collective modes.  Our studies also agree, in most part, with
 numerical studies of the spin-fermion model by Benemman and co-authors~\cite{karl}. Our analysis of the effective bosonic theory of the model at criticality agrees in most parts with the studies of Lercher and Wheatley~\cite{wheatley}.

Finally, our analysis bears some parallels 
with large $D$ studies of fermionic systems~\cite{infD}. 
Just as in $D \rightarrow \infty$ 
theories, our self-energy predominantly depends on frequency 
( at the lowest energies 
the renormalization of $\epsilon_k$ 
 in the fermionic propagator is by  $\log \lambda/\lambda$, 
smaller that the renormalization of $\omega$). The difference, as we 
understand it, is that large $D$ theories were chiefly applied to 
explain the effects, such as metal-insulator transition, 
for which fermions with energies comparable to the bandwidth play 
an important role. We, on the contrary, focus on 
 universal features of the system behavior at the lowest frequencies,
when the fermionic density of states can be approximated by a constant, i.e., 
 the bandwidth can be set to infinity. 
The paper is 
organized as follows. In the next section, we discuss the spin-fermion model
and its relation to lattice models of cuprates. In Sec~\ref{n_s} we present the
calculations at $T=0$. We first discuss ordinary perturbation theory and 
then $1/N$ expansion in the strong coupling ($\lambda \gg 1$) limit. 
We then proceed in Sec.~\ref{sec_ninf} 
with the calculations at $N \rightarrow \infty$. In the 
next Sec~\ref{per_th} we discuss $1/N$ expansion at $T=0$. In this Section
 we also compare 
our results at the QCP with the conventional $\phi^4$ 
theory of such transition~\cite{hertz,millis}.
 We argue that spin-fermion model at criticality 
is {\it not} described by a standard $\phi^4$ theory because of complex 
momentum and frequency dependences of the effective four-boson vertex 
made out of fermions. In the next Sec~\ref{finiteT} we present the 
results of our calculations at finite temperature. Here we  obtain 
the full form of the fermionic spectral function which can be directly 
compared to the photoemission data. 
In Sec.~\ref{opt_cond} we discus in detail 
f the behavior of optical conductivity 
$\sigma = \sigma_1 + i \sigma_2$, effective 
scattering rate $1/\tau$ and the effective quasiparticle mass. These 
expressions can  be directly compared with optical and transport data.
In Sec~\ref{other_observables} we present the results for other 
observables: fermionic density of states, dynamical spin susceptibility, 
 and Raman intensity.  
Finally, in Sec~\ref{pairing_vertex} we discuss the behavior of the 
pairing vertex and show that 
although pairing fluctuations in the normal state generally affect fermionic 
propagator only to order $1/N$, 
 the  ladder corrections to the pairing vertex are not small in $1/N$. 
 We argue that the ladder series 
 give rise to a $d-$wave pairing instability.
The results of our analysis are summarized in Sec~\ref{summary}. 
In this last section 
 we also compare our results for the fermionic spectral function 
and conductivity 
with the photoemission and optical data for near optimally doped cuprates.

Some of the results presented in the paper have already been discussed 
in short communications~\cite{chub2,acs,acf,chub1,cs,ch_m,cmm,ac_qcp,ach}.
 
\section{spin-fermion model}
\label{sf_m}

In this section, we present the  microscopic justification for
 the spin-fermion model. We
use renormalization group (RG) arguments and general considerations to obtain 
the effective Hamiltonian which describes the interaction between low-energy
fermions and their collective spin degrees of freedom.
We then develop a new method to compute 
the forms of the fermionic propagator and dynamical 
spin susceptibility for strong spin-fermion interaction.

The first-principle approach to
cuprates  demands that one  
chooses a microscopic model which describes fermions on a lattice with a
short-range, Hubbard-type interaction. If the choice is correct and if
one manages to solve this model and understand the result,  
there is no need to introduce any low-energy phenomenology.
Unfortunately, this microscopic model is unknown and presumably is
rather involved.  The Hubbard model with on-site interaction 
is a undoubtly  a reasonable first choice~\cite{hubbard,dagotto}, 
but one has have in mind that
it indeed gives 
only an approximate description of the actual physics in
 cuprates. Furthermore, even
for the single band Hubbard model no 
exact  solution has
 been yet obtained in two dimensions,
and one has to either study the model
numerically~\cite{hubbard,dagotto}, or use
 an approximate theoretical scheme such as small U
expansion~\cite{heinz}, or an expansion in the artificially extended 
number of electronic orbitals per site~\cite{exp_orb}.

In the low-energy approach which we advocate, we
 are not interested in the behavior at high
energies (comparable to  the
 fermionic bandwidth $W$), and want to get only a limited
information about the system, namely, how it behaves at
 energies smaller than some cutoff energy $\Lambda \leq W$.
In the RG sense, this effective theory
can be regarded as
obtained by integrating out fermionic degrees of freedom
with energies between $W$ and $\Lambda$.
\begin{figure}[tbp]
\begin{center}
\epsfxsize=\columnwidth 
\epsffile{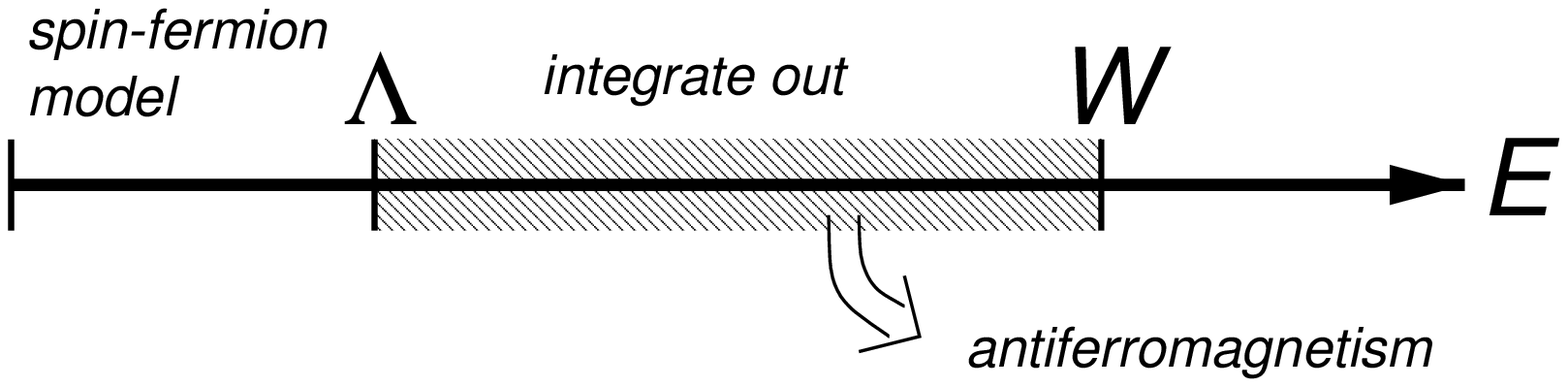}
\end{center}
\label{integratingOut}
\end{figure}
It is generally believed
that this procedure 
 does not take the system away from the basin of
attraction of the Fermi-liquid fixed point. 
We are not aware of any solid calculations which would indicate the opposite. 
In other words, we assume that 
the fermionic interactions at high energies 
do not destroy
the Fermi-liquid behavior. 
This, however,  does not imply the throughout applicability of
the conventional Landau Fermi liquid theory because of possible infrared 
singularities. In fact we show below that in some range of parameters 
the fermionic  self-energy corrections are indeed singular, and give rise to a 
non-Fermi liquid behavior.

Strictly speaking, the RG procedure is justified only if the 
 spin-fermion coupling 
does not exceed fermionic bandwidth which we assume to hold. If this
 condition is not satisfied, the separation between low-energy and high-energy
 excitations becomes problematic. We believe that our results are valid, with
 some minor modifications, also in the limit of very large couplings provided, indeed, that there is no destruction of the Fermi liquid due by fluctuations at lattice scales. However, universal results which we report here are only valid if 
 the coupling are smaller than the bandwidth, and typical fermionic 
momenta are located near the Fermi surface, when the fermionic dispersion can be linearized.    

In general,  the low-energy behavior of fermionic systems is governed by 
degrees of freedom which
have low-energy excitations.   
One such degree of freedom is obviously given by the fermion itself since
 it possesses an arbitrary low energy near the Fermi surface. 
 Potential
candidates  for other low energy excitations are the bosonic collective
modes of fermions. In a general situation, these
collective modes are gaped already at energies comparable to 
the bandwidth. However,
if the fermionic system is close to an instability of some kind, then the
corresponding bosonic 
collective mode has a much smaller gap which vanishes at the
instability point. The closeness to antiferromagnetism naturally makes
spin excitations relevant candidates for such low-energy degrees of freedom. 

Throughout this paper we assume that  
\begin{enumerate}
 \item  spin fluctuations 
can be treated as low-energy excitations in the wide range of dopings,  
\item one can neglect the effects of all
 other low-energy collective degrees of freedom, independent
of spin excitations
\end{enumerate}

These are two basic assumptions of our approach. Their 
applicability is solely 
determined by  comparisons with the experimental data. 
We believe that the
NMR and neutron scattering data in cuprates 
 indicate that for all doping ranges
(including overdoped materials), the spin susceptibility possesses a
substantial momentum and frequency dependence near
the antiferromagnetic momentum $(\pi,\pi)$, and the typical energies
associated
with these dependencies are at least by order of magnitude smaller than the 
fermionic bandwidth.  We also believe that there is  
no clear experimental indication of the presence of the low-energy
excitations in other interaction channels. In principle, this does not 
preclude moderately strong effects from other channels which can account for
 moderate values of the Landau parameters~\cite{ioffe_larkin}.
 These effects may change numbers but 
 certainly do not  change the basics physics. Note also that our approach does not rule out the effects in other channels
 which are secondary in the sense that they emerge
 within the spin-fermion model.
For example, charge excitations may
acquire a low-energy dynamics due to a coupling with 
spin fluctuations. This coupling 
 may eventually lead to phase separation~\cite{dombre,cm2} and, potentially,
 (when lattice effects are included) to 
the  formation of stripes~\cite{stripes}. 

The treatment of spin fluctuations as  separate bosonic degrees of freedom
 requires some extra clarifications.  
The point is that in  distinction to  
heavy-fermion materials,  
there is only one  spice of fermions in cuprates, 
and spin fluctuations just result from a multiple
interactions between particles and holes, i.e., are
 collective modes of the fermions. From this perspective,
the introduction of spin excitations as an extra low-energy degree of freedom 
 is just a convenient way to separate the energy scales: 
we assume that there exists a single dominant channel for fermion-fermion
interaction at energies smaller than $\Lambda$, and introduce a (spin) 
collective mode which mediates this interaction. At the same time, the
static propagator of this collective mode (and, in particular, 
 the value of the spin correlation length) is obviously determined by high 
energy
fermions about which we have little information. We only expect, as we already said,  that the integration over high fermionic energies
does not give rise to any singularity in the bare 
spin susceptibility. The latter then should have 
 a  regular Ornstein-Zernike form
\begin{equation}
\chi_0 ({\bf q, \omega}) = \frac{\chi_0}{\xi^{-2} + 
({\bf q}-{\bf Q})^2 - (\omega/v_s)^2}\, .                      \label{chi0}
\end{equation}

The  input parameters in Eq. (\ref{chi0}) are
 the spin correlation length, $\xi$
and the spin velocity $v_s$ which is obviously  of order $v_F$
 as spins are made out of fermions. The overall factor  $\chi_0$    
can  be absorbed into the renormalization of the
coupling constant (see below) and should not be counted as an extra variable.

Note that the bare $\chi_0 ({\bf q},\omega)$ is real. 
This is an essential point
in our consideration. 
We argue that the 
imaginary part of the susceptibility is determined by fermions with energies
smaller than $\Lambda$ and has to be computed within a low-energy theory 
rather than taken as an input.
Indeed,  since the spins have
no   source of damping  other  than to decay into a
particle hole pair, the  inverse lifetime of a spin fluctuation coincides with
 the imaginary
part of the fully renormalized particle-hole bubble. 
Due to the necessity to conserve energy and momentum in a decay process,
 it involves only low-energy fermions 
with frequencies {\it smaller} than the external 
spin frequency.  

In earlier phenomenological studies of the spin-fermion model, the full spin 
susceptibility, including its imaginary
part,  was considered as an input~\cite{david_review}. 
The spin dynamics was assumed to be a simple
 relaxational one with
 $\chi^{-1} ({\bf Q},\omega) \propto 1 -i \omega/\omega_{sf}$.
 The value of $\omega_{sf}$ was taken from experiments and 
assumed to vary with doping 
independently of other input parameters. The phenomenological
 assumption about the form of the full susceptibility is roughly
 consistent with our findings. At the same time, we will see that 
$\omega_{sf}$ is expressed 
in terms of other input parameters and cannot be varied independently of the
 spin-fermion coupling.     

Note that the damping term in the spin susceptibility only appears if 
 the Fermi surface 
contains 
 hot spots (points separated by ${\bf Q}$)~\cite{hr}. 
For  a Fermi surface without hot spots 
the spin
 decay is forbidden at low
 frequencies due to energy constraint.
In this situation,
 the full spin  susceptibility is real at small frequencies and differs from
(\ref{chi0}) only due to  effects of quantum criticality.
 On the other hand, for a Fermi surface with hot
spots, the spin damping is permittable down to the lowest energies and
overshadows the quadratic in $\omega$ term in the bare susceptibility
which therefore becomes irrelevant.

The topology of the Fermi surface in cuprates 
is non-universal and in principle can vary from
one material to the other. ARPES data indicate that at least in the best
studied Bi-based cuprates, the Fermi surface is centered at $(\pi,\pi)$, i.e.,
it is an open electron Fermi surface~\cite{photo_over,ARPES_FS}. 
Applying  Luttinger theorem which states that
the area of electron states, measured in units of the Brillouin zone area,
 equals to the density of electrons~\cite{lutt}, one can immediately make sure  that such  Fermi surface necessarily crosses the 
magnetic Brillouin zone boundary and therefore contains hot spots.
As we just said, in this situation, 
the $\omega^2$ term in the bare spin susceptibility is 
irrelevant, i.e., $\chi_0 ({\bf q}, \omega)$ can be approximated by 
its static part. This in turn implies 
 implies that the spin-fermion model can be described by the Hamiltonian. 
This Hamiltonian involves
 fermions which live near the Fermi surface, spin fluctuations, and 
the interaction between the two degrees of freedom, and 
can generally be written as
\begin{eqnarray}
{\cal H} &=&  \sum_{{\bf k},\alpha} 
 {\bf v}_k ({\bf k}-{\bf k}_F)  
 c^{\dagger}_{{\bf k},\alpha} c_{{\bf k},\alpha}
+ \sum_q \chi_0^{-1} ({\bf q}) {\bf S}_q {\bf S}_{-q}          \nonumber \\
&&+  g \sum_{{\bf q,k},\alpha,\beta}~
c^{\dagger}_{{\bf k+ q}, \alpha}\,
{\bf \sigma}_{\alpha,\beta}\, c_{{\bf k},\beta} 
\cdot {\bf S}_{\bf -q}\, .                                     \label{intham}
\end{eqnarray}
Here,  $c^{\dagger}_{{\bf k}, \alpha} $ is the fermionic creation operator
for an electron with crystal momentum ${\bf k}$ and spin projection $\alpha$,
$\sigma_i$ are the Pauli matrices,  and 
 $g$  is the coupling constant which measures the strength of the 
interaction between fermionic spins and the collective spin
degrees of freedom described by  bosonic variables $S_q$. 

Alternatively to the Hamiltonian in Eq.~\ref{intham},
 we can also write down the effective action for the model
\begin{eqnarray}
S &=&-\int_0^\beta d\tau \int_0^\beta d\tau' \sum_{{\bf k},\sigma}
c^\dagger_{{\bf k}\sigma} (\tau)
G^{-1}_0 ( {\bf k}, \tau-\tau') c_{{\bf k}\sigma}(\tau')       \nonumber \\
&& +  \frac{1}{2} \int_0^\beta d\tau \int_0^\beta d\tau'  
\sum_{{\bf q}} \chi_0^{-1} ({\bf q}) \, {\bf S}_{\bf q}(\tau) 
\cdot {\bf S}_{-{\bf q}} (\tau')\,                               \nonumber \\
&&  + g \int_0^\beta d\tau \sum_{{\bf q}}   \, 
{\bf s}_{\bf q}(\tau) \cdot {\bf S}_{-{\bf q}} (\tau)\, .        \label{sfm}
\end{eqnarray}
where $G^{-1}_0 ( {\bf k}, \tau) = \partial_{\tau} - 
 {\bf v}_k ({\bf k}-{\bf k}_F)$
 is the bare Fermionic propagator. 
In general, this representation is more 
advantageous as it 
allows a time dependence of the bare 
spin propagator $\chi_0 ({\bf q}, \tau)$. In our case, as we said, this time 
dependence is irrelevant, and one can therefore use a conventional 
Hamiltonian description. 

\begin{figure}[tbp]
\begin{center}
\epsfxsize=\columnwidth 
\epsffile{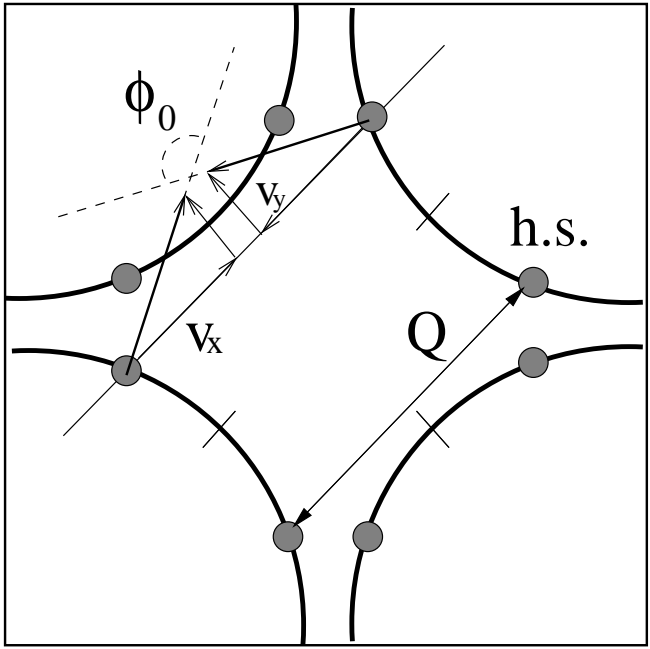}
\end{center}
\caption{ A schematic picture of the Fermi surface. 
The notation h.s. stands for  ``hot spot'' --  the point on the 
Fermi surface separated by another point on the Fermi surface
by the antiferromagnetic vector ${\bf Q}$. The arrows denote 
the velocities at hot spots.  $\phi_0$ is the angle between 
the velocities at hot spots separated by ${\bf Q}$.}
\label{fermi-surface}
\end{figure}

The input parameters in (\ref{intham}) are $\xi$, the
fermionic velocity ${\bf v}_k$ and the spin-fermion coupling $g$.
We demonstrate below that the physics associated with the closeness to
an antiferromagnetic instability mostly involves fermions near hot spots. 
For ${\bf Q}=(\pi,\pi)$, there are $8$ hot
spots in the Brillouin zone with the same $v_F\equiv |{\bf v}_k|$.
The direction of  ${\bf v}_k$ at two hot spots separated by ${\bf Q}$ 
differs by an angle $\phi_0$. This angle, however, should not be counted as
a relevant input parameter as we will see that the 
physics in insensitive
to the actual value of $\phi_0$ as long as $\phi_0 \neq \pi$, i.e., 
 there is no nesting at hot spots. The results  for the 
nested Fermi surface are 
different, and we will not discuss this special 
case in the paper. Experimentally, in cuprates the hot spots are located not 
far away from the corners of the Brillouin zone (i.e., from $(0,\pi)$ 
and symmetry
related points). In this situation, $\phi_0$ is obviously close to $\pi/2$.  
 For practical purposes, it is more convenient to introduce, 
 instead of $v_F$ and $\phi_0$, the two velocities $v_x$ and $v_y$ via
$\epsilon_k = v_x k_x + v_y k_y$ and $\epsilon_{k+Q} = - v_x k_x + v_y k_y$,
where $k$ measure a deviation from a hot spot. Obviously, $v^2_F = v^2_x + 
v^2_y$ and $\phi_0 = 2 {\rm tan}^{-1} v_x/v_y$. The limit 
$\phi_0 = \pi$ corresponds to
$v_y =0$. (see Fig. \ref{fermi-surface})

The  spin-fermion
 coupling constant $g$  is
a fully renormalized irreducible vertex for the particle-hole channel.
In the Born approximation,  $g$  equals to the Hubbard $U$. 
Beyond Born approximation $g$ 
acquires a strong doping dependence and
rapidly decreases with increasing doping. 
For example, an RPA renormalization of $g$ in the particle-particle channel  
yields
$g = U/(1 + (U/t) F(x))$ where $F(x)$, subject to $F(0) =0$,
 is an increasing function of the doping, $x$. 
This form shows that $g \sim U$ only at small doping
while at larger $x$ it progressively decreases down to $g \sim t$.
This form of $g$ is indeed only an approximate one as 
there is no justification to restrict with one particular
renormalization channel at high frequencies. 
We can only quite generally assume
that $g$ is some  doping dependent coupling
constant which increases as the system approaches half-filling.

To summarize, the relevant input parameters for our model are $\xi$, 
$v_F$, and the effective coupling 
${\bar g} = g^2 \chi_0$, which is a 
 combination in which $g$ and $\chi_0$ appear in perturbation series. 
The first two parameters can  be merged into an 
energy scale $v_F \xi^{-1}$.
Out of the two energies, ${\bar g}$ and $v_F \xi^{-1}$, 
 one can  construct an overall scale and a
 single 
dimensionless ratio 
\[
\lambda = 3{\bar g}/(4 \pi v_F \xi^{-1})
\]
which, as we will see in the next section,
 determines the
strength  of both fermionic and bosonic self-energy corrections (the
numerical factor in $\lambda$  is introduced for further convenience).
Physically, $\lambda$ measures the ratio of the 
effective coupling constant and a fermionic energy  
at a typical fermionic $|k-k_F|\sim \xi^{-1}$,
 which sets the momentum range for 
 spin-fermion coupling.
When $\lambda$ is small, fermions are nearly decoupled from spin 
fluctuations and behave
as an almost ideal Fermi gas. 
On the contrary, when $\lambda$ is large,  
the bare fermionic dispersion  is almost completely
overshadowed by the interaction, and this may give rise to a non-Fermi liquid
behavior. 

The  fact that a large number of experimental data for 
cuprates already near optimal doping differ
from the predictions of the Fermi liquid theory 
 indicates that if spin-fermion coupling a relevant mechanism for 
deviations from the Fermi-liquid theory,   $\lambda$ should be large
 already for optimally doped 
samples, and  increase with decreasing doping. 
Our estimates of $\lambda$ using the photoemission\cite{kaminski,johnson1,bogdanov1},
 neutron~\cite{neutron_extract}, and NMR\cite{nmr_extract} data yields
$\lambda \sim 2$ near optimal doping (see Sec~\ref{summary}).
This observation is a challenge to a theory as
real systems turns out to be outside the basin of applicability of a
conventional, weak coupling perturbation theory and one should instead try
unconventional procedures in a search for a strong-coupling solution. 

We now proceed with the evaluations of the fermionic and bosonic self-energies.

\section{Perturbation theory for the spin-fermion model at $T=0$}
\label{n_s}

In this Section, we analyze
the spin-fermion model at $T=0$. 
Our strategy is the following: we first present
the results of a formal perturbation theory for both fermionic and bosonic
propagators. We show that perturbation expansion for the fermionic self-energy
 and the spin-fermion vertex 
holds in powers of $\lambda$ and obviously breaks down for 
 $\lambda >1$. 
We next show that at $\lambda \geq 1$ the self-energy corrections to the spin
propagator also become relevant and change the dynamics of spin fluctuations 
at frequencies which are mostly relevant for fermionic self-energy and 
vertex renormalization. We then demonstrate that even for large $\lambda$, 
one can construct a variant of perturbation theory 
in which dominant, $O(\lambda)$ 
fluctuation contributions are included into new zero-order fermionic and
 bosonic propagators for which we obtain explicit answers. 
 The remaining fluctuation corrections are 
logarithmical in $\lambda$, and we  analyze them in the one-loop RG formalism. 

\subsection{A formal perturbation theory}

We begin with the direct  zero-temperature
perturbation expansion in powers of $\lambda$. Direct
perturbative expansion means that all diagrams are evaluated using  the bare
values of the spin susceptibility and the fermionic propagator.

\subsubsection{fermionic self-energy}

We start with the fermionic self-energy, $\Sigma (k, \Omega)$ related to
the fermionic propagator by $G^{-1} (k, \omega) = \Omega + \Sigma (k, \Omega) 
- \epsilon_k$. Let's first compute $\Sigma (k, \Omega)$ near a hot spot expanding  
to first order in frequency and in the quasiparticle energy. We will demonstrate that this supposedly straightforward 
computation has to be performed with more care than one might expect.  

\begin{figure}[tbp]
\begin{center}
\epsfxsize=\columnwidth
\epsffile{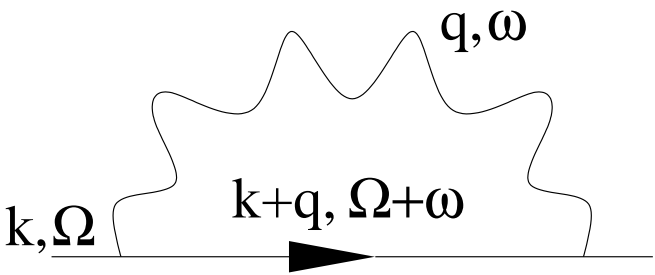}
\end{center}
\caption{ 
The lowest-order diagram for the fermionic self-energy.
}
\label{self-energy}
\end{figure}

The lowest-order self-energy correction 
involves a single spin-fluctuation
exchange and is presented
in Fig. \ref{self-energy}. In the  analytical form we have in Matsubara
frequencies
\begin{eqnarray}
&&\Sigma ({\bf k}, \Omega_m) = -3 g^2 \int \frac {d^2 q 
d\omega_m}{(2\pi)^3}\nonumber \\
&\times & G_0 ({\bf {k+q}},\omega_m + \Omega_m) 
\chi_0 ({\bf q},\omega_m)\, ,                                   \label{pert}
\end{eqnarray} 
where 
\begin{equation}
G_0^{-1} ({\bf k}, \omega_m) = i \omega_m   
-{\bf v}_F ({\bf k} - {\bf k}_{hs}).
\end{equation}

Subtracting from (\ref{pert}) the
 self-energy at $\Omega_m= 0$ and ${\bf k} = {\bf k}_{hs}$ which 
 can be absorbed into the renormalization of the chemical potential, 
and expanding to first order
in  ${\bf k} -{\bf k}_{hs}$ and in frequency, 
we obtain after simple manipulations
\begin{equation}
\Sigma ({\bf k}, \Omega_m) = 
(i \Omega_m - \epsilon_{k+Q})~I({\bf k}, \Omega_m)              \label{s1-I}
\end{equation}
where  
\begin{eqnarray}
I(k, \Omega_m) &=&  \frac{3 ~{\bar g}}{(2\pi)^3} 
\int d^2 {\tilde q} d \omega_m
~\frac{1}{\xi^{-2} + {\tilde {\bf q}}^2 + (\omega/v_s)^2}    \nonumber \\ 
&\times&\frac{1}{i \Omega_m 
 - \epsilon_{k+Q} + i \omega_m  - v_F {\tilde q}_x}~ 
~\frac{1}{i \omega_m - v_F {\tilde q}_x}~                    \label{pertnew0} 
\end{eqnarray} 
Here ${\tilde {\bf q}} = {\bf q}-{\bf Q}$, 
and  $x$-axis is chosen along   
 ${\bf v}_F$ at ${\bf k}+{\bf Q}$.

 In the spirit on a conventional expansion in $\Omega_m$ and $\epsilon_{k+Q}$, 
 it is tempting to evaluate 
$I({\bf k},\Omega_m)$ right at ${\bf k} = {\bf k}_{hs}$ and $\Omega_m =0$.
The corresponding computations are performed in Appendix A, and the result is
\begin{equation}
I({\bf k}_{hs}, 0) = - \lambda~\frac{v_{s}}{v_F + v_{s}}     \label{pert33}
\end{equation}
Substituting this into (\ref{s1-I}) one would obtain that the self-energy only accounts for the renormalization of the quasiparticle residue, and the magnitude of the renormalization depends on the ratio $v_s/v_F$. Apparently this is the whole story. However, it turns out that Eq. (\ref{pert33}) is  not the full result, and 
\begin{equation}
\lim_{\Omega_m \rightarrow 0, k \rightarrow k_{hs}}
 I ({\bf k}, \Omega_m) \neq I ({\bf k}_{hs}, 0)
\end{equation} 
To show this we evaluate $I({\bf k}, \Omega_m)$ 
 keeping both  $\Omega_m$ and ${\bf v}_F {\tilde{\bf k}}$ 
in the integrand in (\ref{s1-I}). Analyzing the integrand, we observe 
that there exists a tiny region of frequencies where the poles in
the two fermionic propagators are close to each other, but still are  
located in different halfs of a complex ${\tilde q}_x$ plane. 
For $\Omega_m >0$, this region is sandwiched between 
 $\omega_m > -\Omega_m$ and $\omega_m <0$. 
The integration
over ${\tilde q}_x$ in  (\ref{s1-I}) in this range of frequencies, and
a subsequent integration over $\omega_m$ 
yields an extra contribution to 
$I(k, \Omega_m)$ in the form  
\begin{equation}
I_{an}( {\bf k}, \Omega_m) = \lambda~
\frac{i\Omega_m}{i\Omega_m - \epsilon_{k+Q}}                   \label{extra}
\end{equation}

We see  that $I_{an}( {\bf k}, \Omega_m)$ 
comes from the integration over internal
frequencies $\omega_m$  
which are {\it smaller} than the external $\Omega_m$.
Clearly, this anomalous piece 
 could not be obtained in a conventional perturbation
expansion over $\Omega_m$ as in the latter one assumes that 
the internal energies are much larger
than the external one. 
This assumption is generally justified by a simple phase space argument, i.e., 
 the contribution from  $|\omega_m|\leq \Omega_m$ is normally small due to the
smallness of the integration range. 
Here, however, 
 the  smallness of the phase space  is compensated by
the smallness of the denominator in (\ref{s1-I}) as
without $\Omega_m$ and $\epsilon_{k+Q}$, 
the product of two Green's functions contains a double pole.
At finite $\Omega_m$ and $\epsilon_{k+Q}$, the double pole  splits into 
two single poles, but
the energy difference between them 
is still only $i\Omega_m - \epsilon_{k+Q}$, i.e., is of the same order as the
integration range over frequency.
 The same reasoning is used to extract the  effects due to 
  chiral anomaly in quantum chromodynamics~\cite{chiral}.

Substituting the total expression for $I({\bf k},\Omega_m)$ into the 
 self-energy, we obtain 
 \begin{equation}
\Sigma ({\bf k}, \Omega) = \lambda 
\left[i\Omega_m - (i\Omega_m - \epsilon_{k+Q})
\frac{v_{s}}{v_s + v_F} \right].                             \label{pert22}
\end{equation}
 
We see that both terms in $\Sigma$ are of order $\lambda$, i.e., there
is no strong distinction between ``anomalous'' and ``normal'' contributions
to the fermionic self-energy. This could be anticipated as, e.g., 
for  $v_s = \infty$,  $\chi_0 ({\bf q}, \omega)$ is purely static, and 
$\Sigma ({\bf k},\Omega)$ should 
only depend on $\epsilon_{k+Q}$ as 
the  dependence on external $\Omega_m$  is eliminated by 
  shifting the internal frequency in the fermionic propagator 
in (\ref{pert}). We indeed reproduce this result:  
for $v_s = \infty$ Eq. (\ref{pert22}) yields 
$\Sigma ({\bf k}, \Omega_m) = \lambda \epsilon_{k+Q}$.

\subsubsection{vertex renormalization}

We next compute the lowest-order correction to the
 spin-fermion vertex for fermions at hot spots. 
\begin{figure}[tbp]
\begin{center}
\epsfxsize=2.5in 
\epsfysize=1.4in
\epsffile{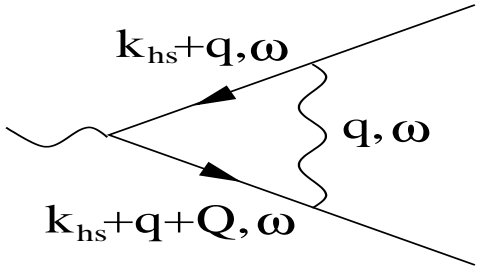}
\end{center}
\caption{ 
The lowest-order diagram for the correction to the spin-fermion vertex.
}
\label{vertex-correction}
\end{figure}
The lowest-order vertex correction diagram is
 presented in Fig. \ref{vertex-correction}. The vertex correction 
is obviously the largest for a 
fermion at a hot spot. For ${\bf k} = {\bf k}_{hs}$ and $\omega =0$ we obtain
$g^R = g + \Delta g$ where 
\begin{eqnarray}
\frac{\Delta g}{g} &=& i{\bar g} 
\int \frac{d \omega_m d^{2}q}{(2 \pi)^3}
\chi ({\bf q}, \omega_m) \nonumber \\
&\times&G(\omega_m,{\bf k}_{hs} +{\bf q})
~G(\omega_m, {\bf k}_{hs}+q +Q)                        \label{vert1}
\end{eqnarray}
Observe that the overall factor (the result of
 the summation over spin components) is different from that in the
fermionic self-energy ($-1$ instead of of $3$). 
Expanding, as above, near hot spots as 
$\epsilon_{k_{hs} +q} = v_x q_x + v_y q_y$, 
$\epsilon_{k_{hs} +q+Q} = -v_x q_x + v_y q_y$ where 
$v^2_x + v^2_y = v^2_F$, and performing 
the computations described in Appendix A,  we obtain
\begin{equation}
 \frac{\Delta g}{g} = \frac{\lambda}{3}
\frac{v_{F}}{v_{s}}\frac{1}{\beta \sqrt{1-\alpha ^{2}}}
\ln \frac{\alpha \left(\sqrt{1-\alpha ^{2}}-\beta  \right)}
{\sqrt{\left(\beta ^{2}+\alpha ^{2} \right)
\left(1-\alpha ^{2}\right)}-\beta}                                 \label{dg}
\end{equation}
where $\alpha = v_y/v_s$ and $\beta = v_x/v_s$. 
For purely  static bare susceptibility, i.e., $v_F/v_s \rightarrow 0$,
$\Delta g/g$  is a smooth function of 
$v_x$ and $ v_y$.
\begin{equation}
 \frac{\Delta g}{g} = 
\frac{\lambda}{3}
\frac{v_{F}}{v_{x}}\sinh^{-1}\frac{v_{x}}{v_{y}}                  \label{dg1}
\end{equation}
In the opposite limit $v_s \ll v_F$, we obtain from (\ref{dg})
\begin{equation}
 \frac{\Delta g}{g}  = 
\frac{\lambda}{3}~\frac{v_s v_F}{v_{x}v_{y}}
~ \tan^{-1}\frac{v_x}{v_y}                                       \label{dg2}
\end{equation}
Observe that the correction to the spin-fermion vertex 
vanishes when the velocity of spin 
excitations  
tends to zero. 
 
We also verified that the correction to the spin-fermion vertex does not
contain a singularity, similar to what we found for a self-energy correction.
In other words, the limiting
 value of $\Delta g$ at vanishing frequency  coincides with 
 $\Delta g$ evaluated right at zero frequency. 
The absence of a singular correction to a  vertex  is 
 associated with the fact that the two internal fermions in the 
vertex correction diagram have different directions of the velocities, 
and hence there is no double pole which might give an extra piece after
regularization.

\subsubsection{higher-order diagrams and the structure of the perturbation theory}

We see that the perturbative expansion for the fermionic
self-energy and the spin-fermion vertex
 holds in the dimensionless parameter $\lambda$. 
Higher-order diagrams can be easily estimated. They all scale as
higher powers of $\lambda$.  
To this end, diagrammatic, perturbative approach clearly
fails if $\lambda$ is large.

There is, however, a  hint already at this stage that there are some
peculiarities in the perturbation theory at strong coupling. Namely,
we expect (and we show below) that at strong coupling, spin fluctuations
are completely overdamped, i.e., the frequency dependence of the spin
susceptibility is dominated by $i\omega$ term instead of $\omega^2$ term. 
Crudely, this effect can be modeled by reducing $v_s$ 
in (\ref{pert22}) and (\ref{dg}).  We see, that when $v_F/v_s$ 
becomes very large, 
the regular part of the fermionic self-energy 
vanishes leaving only the anomalous piece 
in (\ref{pert22}).
Simultaneously, $\Delta g/g$, i.e., 
vertex correction also vanishes. 
One can easily make sure that this tendency persists to all orders
of the perturbation theory. 

Alternatively speaking,  at $v_s \rightarrow 0$,
the vertex correction and the regular piece of the  fermionic self-energy both
vanish leaving the anomalous piece in the fermionic self-energy as the only 
term one has to evaluate.

\subsubsection{spin polarization operator}

To verify that the spin-fermion interaction reduces $v_s$ and 
 makes spin fluctuations softer,   
 we now compute the bosonic self-energy.
It comes  in our model from a  virtual 
processes in which a spin fluctuation decays 
into a particle-hole bubble, and hence coincides, up to
an overall factor, with the
fully renormalized spin polarization operator  $ \Pi({\bf q},\omega)$. 
We define 
$\Pi ({\bf q}, \omega)$ via 
$\chi^{-1} ({\bf q}, \omega) = \chi^{-1}_0 ({\bf q},\omega) - 
\Pi ({\bf q},\omega)/(\chi_0 \xi^2)$, or  
\begin{equation}
\chi ({\bf q},\omega) = \frac{\chi_0 \xi^{2}}{ 1 + 
\xi^{2} ({\bf q}-{\bf Q})^2 - \Pi ({\bf q},\omega)}.          \label{chi}
\end{equation}
The contributions to  the particle-hole bubble
from high-energy fermions  
are already absorbed into the bare susceptibility, and in our consideration 
we have to consider only fermions with energies smaller than $\Lambda$.
Also, as static $\chi_{0} ({\bf q})$ is
assumed to be peaked at ${\bf q}={\bf Q}$, it is sufficient to compute 
the polarization operator only at ${\bf q} ={\bf Q}$.

\begin{figure}[tbp]
\begin{center}
\epsfxsize=\columnwidth
\epsffile{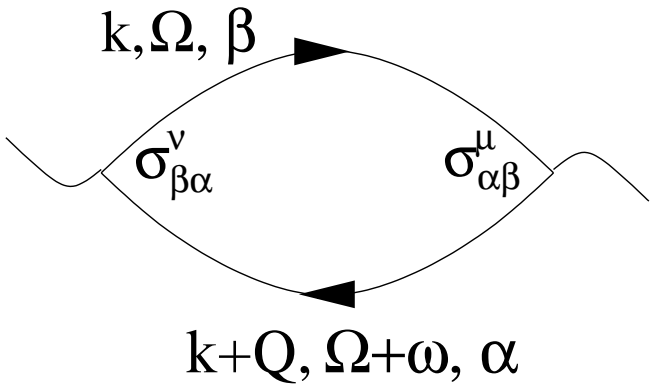}
\end{center}
\caption{ 
The lowest-order diagram for the spin polarization operator.
}
\label{polarization}
\end{figure}

Let's again start with the lowest order perturbation theory 
in the 
spin-fermion coupling. The lowest-oder diagram for $\Pi ({\bf Q}, \omega)$ 
is presented in Fig. \ref{polarization}. In analytical form, we have 
in real frequencies 
\begin{equation}
\Pi ({\bf Q},\omega) = 2~{\bar g} \xi^2
 \int \frac {d^2{\bf  k} d \Omega}{(2\pi)^3}
~G_0 ({\bf {k+Q}},\Omega + \omega)~G_0 ({\bf {k}},\Omega)        \label{Pii}
\end{equation}
(a factor of 2 comes from a summation over spin components).
 Expanding both 
fermionic energies at ${\bf k}$ and ${\bf k}+{\bf Q}$ to 
linear order in the deviations from hot spots, integrating
over momentum and frequency in (\ref{Pii})
and multiplying the result by the number of hot spots $N=8$,
we obtain,  
\begin{equation}
\Pi ({\bf Q},\omega) = i~\omega/\omega_{\rm sf}                 \label{Pi1}
\end{equation}
where
\begin{equation}
\omega_{\rm sf} = \frac{4\pi}{N}~ \frac{v_x v_y \xi^{-2}}{\bar g} 
 \equiv \frac{3}{N}~
\frac{v_x v_y}{v^2_F}~\frac{v_F \xi^{-1}}{\lambda}              \label{omsf}
\end{equation} 
 Instead of explicitly inserting $N=8$,  we 
will keep $N$ as a variable - later we perform  a 
formal expansion in $1/N$.

The independence of ${\rm Im} \Pi ({\bf Q},\omega)$ of the 
fermionic cutoff $\Lambda$ is indeed the consequence of the 
energy conservation requirement which confines fermions to hot spots.
Observe, however, that 
Eq. (\ref{Pi1}) is the full expression for 
$\Pi ({\bf Q},\omega)$, not only its imaginary part. In other words,
as long as one restricts with the linearized fermionic dispersion near 
the Fermi surface, $Re \Pi ({\bf Q},\omega) =0$.  This last result 
implies that 
there is no universal correction to $\xi$ from low energy fermions.
 The corrections to $\xi$ only 
appear when we expand quasiparticle energies  further in 
deviations from the Fermi surface. It is easy to check that these 
corrections scale as $\Lambda/W$ and are therefore just minor leftovers
from the contributions from high fermionic energies.  

We see that the lowest-order spin polarization operator has a simple
relaxational, linear in $\omega$ form. This implies that at low frequencies,
the renormalized spin susceptibility more strongly depends on frequency 
than the bare susceptibility. This confirms
 our early conjecture that 
the effect of the bosonic self-energy  
can be crudely modeled by increasing $v_s$. 

We next estimate what
bosonic momenta and
frequencies mostly contribute to the fermionic self-energy in (\ref{pert})
and to the vertex correction in (\ref{vert1}).
Simple power counting  shows that 
the integrals in (\ref{pert}) and (\ref{vert1}) are dominated by
$|{\bf q}-{\bf Q}| \sim \xi^{-1}$ and $\omega \sim v_F \xi^{-1}$.
Comparing  the full and bare susceptibilities at 
these momenta and frequencies, 
we immediately see that for $\lambda \geq 1$ 
$\Pi ({\bf Q},\omega) \sim \lambda$ dominates over bare
 $\chi_0 ({\bf q},\omega)$.  
This implies that for large $\lambda$, 
spin fluctuations  which mostly contribute to the fermionic self-energy, 
are completely overdamped due to a strong decay into a
particle-hole pair.
Clearly, in this situation, one cannot
 neglect  the transmutation of the spin dynamics in evaluating the 
fermionic
self-energy and the correction to the spin-fermion vertex.

\subsection{renormalized perturbation theory}
 
Our next strategy is the following. We assume for a moment that Eq.
(\ref{Pi1}) for $\Pi ({\bf Q},\omega)$ with $\omega_{sf}$ given by 
(\ref{omsf}) is valid for all couplings, and   
re-evaluate
the lowest-order fermionic self-energy and vertex correction  
using the renormalized form of the spin susceptibility.
We then check the self-consistency of this procedure by re-evaluating 
$\Pi ({\bf Q}, \omega)$ with renormalized fermionic propagators and vertices.

\subsubsection{fermionic self-energy}

Consider first the fermionic self-energy. 
Substituting the renormalized  
$\chi ({\bf q},\omega)$ instead of $\chi_0 ({\bf q}, \omega)$ 
into  Eq.\ref{pert},
 subtracting, as before,  the self-energy at $\omega= \epsilon_{k+Q}=0$, 
and expanding to linear order in 
${\bf k} - {\bf k}_{hs}$, we obtain
\begin{equation}
\Sigma ({\bf k}, \Omega) = 
(i\Omega - \epsilon_{k+Q})~I({\bf k}, \omega)               \label{i}
\end{equation}
but now
\begin{eqnarray}
I({\bf k}, \Omega) &=&  3 {\bar g} \xi^2 
\int \frac {d^2 {\tilde q} d \omega}{(2\pi)^3}
~\frac{1}{1 + ({\tilde {\bf q}} \xi)^2 + 
|\omega|/\omega_{\rm sf} +\frac{\omega^2}{v^2_s \xi^{-2}}}        \nonumber \\ 
& \times &\frac{1}{i\Omega  - \epsilon_{k+Q} + 
i\omega  - v_F {\tilde q}_x}~ 
~\frac{1}{i\omega - v_F {\tilde q}_x}                           \label{pertnew}
\end{eqnarray} 

Consider, as before, the limit $\Omega, \epsilon_{k+Q} \rightarrow 0$
and compute first the regular piece $I({\bf k}_{hs},0)$. 
Without $\epsilon_{k+Q}$  and $\Omega$, 
the last two terms in (\ref{pertnew}) are equal and 
produce a double pole at 
$v_F {\tilde q}_x = i\omega$. 
Closing the integration contour over ${\tilde q}_x$ by a semi-circle 
in a half-plane with no double pole,   
we find that we only  have to consider the pole in
 the spin
susceptibility. This pole is located at a much larger 
${\tilde q}_x \propto
\sqrt{\omega}$ than the double pole. In this situation,
 one can neglect $\omega$ in 
the Green's functions in comparison with ${\tilde q}_x \propto \sqrt{\omega}$.
 The remaining computation of $I(k_{hs}, 0)$ is
 straightforward. The details are
presented in Appendix A, and for $\lambda \gg 1$, 
the result is, with logarithmic accuracy
\begin{equation}
I_{reg} ({\bf k}_{hs}, 0) =  
-\frac{12 v_x v_y}{\pi N v^2_F} ~\log \lambda                    \label{Ireg}
\end{equation}
We see that the regular piece in $I({\bf k}, \Omega)$
 only logarithmically depends on $\lambda$, 
in distinction to a direct perturbation theory 
where it was of order $\lambda$. 
This result is indeed consistent with our 
earlier observation that $I_{reg} ({\bf k}_{hs}, 0)$ in (\ref{pert33})
 vanishes in the limit $v_s \rightarrow 0$.
 The still presence of the $\log \lambda$ factor in  (\ref{Ireg})
is a consequence of 
$\Pi ({\bf Q}, \omega) \propto \omega$, and 
  could not, indeed, be anticipated in a perturbation theory with a 
bare spin susceptibility.  

For completeness, we also present the expression for 
$I_{reg} ({\bf k}_{hs}, 0)$ at arbitrary $\lambda$. 
The calculations are presented in Appendix A. The result is particularly
simple for $v_s \rightarrow \infty$, i.e., for purely static bare 
susceptibility. We obtained
\begin{equation}
I_{reg} ({\bf k}_{hs}, 0) = - \lambda~ \frac{1 + \frac{a}{\pi} 
\log{\frac{a}{2}}}{1 + \frac{a^2}{4}}                           \label{Iint}
\end{equation}
where $a = (v_F \xi^{-1}/\omega_{sf}) = 
(N\lambda/3)~(v^2_F/v_x v_y)$.
For $a \ll 1$, we recover the result of a direct perturbation theory, 
$I_{reg} ({\bf k}_{hs}, 0) = - \lambda$. For $a \gg 1$, Eq. (\ref{Iint})
reduces to (\ref{Ireg}).

We next compute an anomalous term in $I({\bf k}, \Omega)$ which, we
remind, is a contribution from a regularized 
double pole in  $I({\bf k}, \Omega)$.
Integrating, as before, in Eq. (\ref{pertnew}) over a tiny region
 $-\Omega < \omega <0$ where the  poles in the two fermionic propagators
are located in different half-planes, we obtain 
\begin{equation}
I_{an}( {\bf k}, \Omega) = \lambda~
\frac{i\Omega}{i\Omega - \epsilon_{k+Q}}                      \label{extra'}
\end{equation}

The fact that $I_{an}( {\bf k}, \Omega)$ remains the same as in 
 a direct perturbation theory can be easily explained. Indeed, 
 the integration over $\omega$ in  $I_{an}( {\bf k}, \Omega)$
 is confined to vanishingly
small frequencies for which the dynamical term in the spin 
susceptibility is small no mater whether 
the  dynamics is  ballistic or
relaxational. 

Substituting the full form for $I( k, \Omega)$ into (\ref{i}) 
we obtain $\Sigma ({\bf k}, \Omega) = \Sigma_{reg} ({\bf k}, \Omega) + 
\Sigma_{an} ({\bf k}, \Omega)$, where for $\lambda \gg 1$
\begin{eqnarray}
\Sigma_{an} ({\bf k}, \Omega) &=& i \lambda \Omega \nonumber \\
 \Sigma_{reg} ({\bf k}, \Omega) &=&   
 (i\Omega - \epsilon_{k+Q})~\frac{12 v_x v_y}{\pi N v^2_F} ~\log \lambda 
\label{se_new1}
\end{eqnarray}

We see therefore that the fermionic self-energy  consists of two
parts -- a regular part which accounts for the renormalization of the quasiparticle residue, and a singular part which 
depends only on $\omega$. The regular part is strongly reduced compared to the 
direct
perturbation theory, and scales as 
$\log \lambda$ rather than $\lambda$. 
This reduction is due to the fact that at  frequencies which 
determine the fermionic self-energy, the renormalized 
spin excitations obey relaxational dynamics.
 On the other hand, 
the singular part of the self-energy comes from very 
low internal frequencies, comparable to the external one, 
and  is totally insensitive to the change of the spin dynamics.
At strong coupling, 
the singular, nonperturbative piece in
 $\Sigma$ is much larger than the regular piece and 
obviously should play a central role in all our analysis.

In more general terms, the appearance of $I_{an} ({\bf k}, \Omega)$
 is the reflection of the singularity 
in the particle-hole bubble at small momentum
and frequency transfer. This follows from the observation that
 for small
$v_F {\tilde q}_x$ and $\omega^\prime$ 
the  spin susceptibility, integrated over ${\tilde q}_y$,
 can be approximated by a constant, and the anomalous contribution
to $I({\bf k}, \omega)$ can be reexpressed as
\begin{equation}
I_{an} ({\bf k}, \Omega) =  
\lambda~ {\tilde \Pi} ({\bf k}, \Omega)\, ,                 \label{sigmaanom}
\end{equation}
where 
\begin{equation}
{\tilde \Pi} ({\bf k}, \Omega) = -i~
\int \frac{d{\tilde q}_x~ d\omega}{(2\pi)^2}~
G_0 ({\tilde q}_x, \omega)~G_0 ({\bf k} + 
{\tilde q}_x, \omega + \Omega)                               \label{polar}
\end{equation}
is the particle-hole polarization bubble with a small momentum/frequency 
transfer (it  differs by an overall 
 constant
from $\Pi ({\bf k},\omega)$ introduced in (\ref{chi})).
At vanishing ${\bf k}$ and $\Omega$,  
this polarization bubble 
is formally ultraviolet  divergent, and its value hence 
depends on how the regularization is performed~\cite{landau}, i.e., 
on the order of the integration over frequency 
and over momentum. Doing frequency integration first, one obtains that 
${\tilde \Pi} ({\bf k}, \Omega)$ vanishes, while 
doing momentum integration   first, one obtains that 
\begin{equation}
{\tilde \Pi} ({\bf k}, \Omega) = 
\frac{\Omega}{\Omega - {\bf v}_F {\bf k}}\, .                \label{polar2}
\end{equation}  
In our case, 
the regularization of the self-energy is imposed by the 
the fact that   $I({\bf k}, \omega)$ is actually a convolution
 of the polarization bubble and the spin susceptibility. In the above 
calculations we neglected the dependence of 
$\chi ({\tilde {\bf }}, \omega)$ on 
${\tilde q}_x$ and $\omega$. In reality, however, 
$\chi ({\tilde {\bf q}}, \omega)$
indeed vanishes  when  either 
${\tilde q}_x$ or $\omega$ diverge.
The general rule of the ultraviolet 
regularization procedure is that it has to be
performed such that to avoid generating  extra poles at 
energies smaller than the energies of the
 poles in  the particle-hole bubble, Eq. 
(\ref{polar}).  The pole in  (\ref{polar})
is located at 
 $\omega \propto | {\tilde {\bf  q}}|$. The pole in the spin susceptibility 
$\chi ({\bf q},\omega)$ is located at
$\omega \propto {\tilde {\bf q}}^2$, 
i.e., at a much {\it smaller} frequency, and 
at a much {\it larger} ${\tilde q}_x$ 
 than those for the 
pole in (\ref{polar}).
Clearly, then, the correct way to regularized (\ref{polar}) is to
 first integrate over momentum in (\ref{polar}), and only then integrate over 
frequency. 
This sequence of integrations yields the anomalous piece in 
${\tilde \Pi} ({\bf k}, \Omega)$, Eq. (\ref{polar2}),
which in turn gives rise to 
the anomalous piece in the fermionic self-energy (\ref{sigmaanom}).

Note by passing that in a direct  perturbation theory (which operates 
with the  bare susceptibility), 
 a typical frequency scales with a typical momentum.
In this situation, the result of regularization strongly depends on
the ratio of  $v_F$ and $v_s$
(see  Eq. (\ref{pert22})). In particular, the absence
of the frequency dependence of the self-energy at 
 $v_s = \infty$ 
is caused the fact that in this situation 
 $\chi_0 ({\bf q}, \omega)$ is purely static, and the regularization 
obviously has to  be  performed by doing  frequency
integration first.

\subsubsection{vertex correction}

We now re-evaluate  the vertex correction with the relaxational
form of the  
spin susceptibility.  
The  analytical expression for $\Delta g/g$
is the same as in (\ref{vert1}), but 
with $\chi ({\bf q},\omega)$ instead of $\chi_0 ({\bf q})$.
Linearizing, as before, the fermionic dispersion near the Fermi surface, and
performing momentum and frequency integration as 
 described in Appendix A, we obtain for $\lambda \gg 1$, with logarithmical
 accuracy,
\begin{equation}
\frac{\Delta g}{g} =  \frac{Q(v)}{N}~ \log \lambda
\label{vert2}
\end{equation}
where 
\begin{equation}
Q(v) =\frac{4}{\pi}~ {\rm tan}^{-1} \frac{v_{x}}{v_{y}}
\label{vert3}
\end{equation}
is a smooth function of the ratio of velocities interpolating between
$Q=1$ for  $v_x = v_y$, and $Q=2$ for $v_y \rightarrow 0$. 
The last limit corresponds to 
almost nested Fermi surface at hot spots.
A similar logarithmical form of the vertex correction has been obtained by 
Altshuler,Ioffe and Millis~\cite{ami}. 

We see that the change of the spin dynamics 
 has a strong effect on the strength of the 
vertex correction: in a direct perturbation 
theory, it was of order of $\lambda$, now
 it scales only as $\log \lambda$. This result is fully consistent with our
earlier observation from a direct perturbation theory  (Eq. (\ref{dg2})) that
 $\Delta g/g$  vanishes when
$v_s \rightarrow 0$. As with the $k-$dependent piece in the self-energy, 
the subleading  
 logarithmical dependence of $\Delta g/g$ 
  could not, indeed, be anticipated in a direct perturbation theory with the
bare spin susceptibility.  

\subsubsection{higher-order diagrams}

We next need  to understand what happens when we 
go to higher-orders in the self-consistent perturbation theory for 
fermionic self-energy and the spin-fermion vertex.
  The issue is indeed whether
higher order diagrams scale as higher powers of $\lambda$ or $\log \lambda$.  
We argue that higher-order terms contribute higher powers of $\log \lambda$ but
 not higher powers of $\lambda$. Our reasoning is based on the fact
 that in the second-order theory, the  vertex correction and the 
regular part of the fermionic self-energy are $O(\log\lambda)$, 
and $O(\lambda)$ term in the second-order 
fermionic self-energy  emerges only as a result 
of a proper regularization of the double pole in the integrand 
for $\partial \Sigma/\partial \Omega$. 
Higher-order self-energy and vertex correction 
diagrams contain more fermionic propagators with different 
momenta, and  one can easily make
 sure that there is no  phase space  
for a double pole. Accordingly,  higher-order vertex correction diagrams just
 contribute higher powers of $\log \lambda$, while higher-order 
diagrams for the fermionic self-energy 
 either simply contribute higher powers of $\log \lambda$, or 
account for $\log \lambda$ corrections 
to $\Sigma_{an} ({\bf k}, \Omega)$. We verified this argument 
by explicitly computing next order self-energy and vertex corrections.      

\subsubsection{spin polarization operator}
 
A closely related issue is how the fermionic self-energy and 
vertex renormalization affects the form of the spin polarization operator. 
We recall that  above we evaluated the fermionic self-energy 
and $\Delta g/g$ assuming 
 that that the spin polarization operator
has the same form as in the direct perturbation theory. We now need  to
 verify to which extent this assumption is correct. 
We show that the renormalization of  $\Pi ({\bf Q}, \omega)$ from its 
free-fermion form, Eq. (\ref{Pi1}) is only due to 
vertex corrections and to the regular part of $\Sigma ({\bf k}, \Omega)$, 
while a much larger
$\Sigma_{an} = \lambda \Omega$ does not  affect 
the spin polarization operator. 
To  demonstrate this we re-evaluate  $\Pi ({\bf Q}, \omega)$, 
Eq. (\ref{Pii}), with 
$G({\bf k}, \Omega) = \Omega + \Sigma_{an} (\Omega) - \epsilon_k$.
The computation is rather straightforward as the inclusion of 
$\Sigma_{an}$ into fermionic propagator 
 modifies  the quasiparticle residue $Z$ 
 and the quasiparticle mass (defined as 
$v_F = p_F/m$) as $Z = 1/(1 + \lambda)$, $m^* = m (1 + \lambda)$.
Substituting the renormalized form of $G(k, \Omega)$ into  (\ref{Pii})
 and performing elementary manipulations, we find that $Z$ and $m^*/m$ 
appear in $\Pi ({\bf Q}, \omega)$ in the
 combination $Z^2 (m^*/m)^2$ which is {\it independent} of $\lambda$. 
In other words, as long as we restrict with only 
$\Sigma_{an} ({\bf k}, \Omega)$,
$\Pi ({\bf Q}, \omega)$ remains the same as for free fermions. 

The inclusion of $\Sigma_{ref} ({\bf k}, \Omega)$ 
into the fermionic propagator 
and the corrections to the spin-fermion vertex 
does affect the form of $\Pi ({\bf Q}, \omega)$, but the corrections 
obviously hold in powers of $\log \lambda$. 

\subsection{summary of Sec~\protect\ref{n_s}}

We  now summarize what we obtained in this section.

\begin{enumerate}
\renewcommand{\labelenumi}{\roman{enumi}}
\item
We found that the direct perturbation theory for fermionic 
self-energy and spin-fermion vertex 
holds in powers of $\lambda \propto {\bar g}/(v_F \xi^{-1})$ and does not 
converge at strong coupling.
\item
We found that for typical bosonic energies which mostly contribute to
fermionic $\Sigma ({\bf k}, \Omega)$ and to vertex renormalization, 
the bosonic self-energy 
also becomes relevant for $\lambda \geq 1$ and accounts for the 
change in the bosonic dynamics from a 
ballistic one to a  relaxational one.
\item
We performed the self-consistent perturbative expansion in which we 
used the relaxational form of the spin susceptibility as an input, and found
that it holds in powers of $\log \lambda$, plus there exists the 
anomalous, nonperturbative  piece
in the fermionic self-energy which still scales as $\lambda$. 
At strong coupling, 
 this piece  much larger than the regular self-energy  and 
obviously should play a central role.
\item

To logarithmical accuracy, the anomalous piece in the fermionic self-energy 
is given by the second-order diagram, higher-order terms just add 
extra powers of $\log \lambda$.
\item
We found  that as long as we restrict with only $O(\lambda)$ 
piece in the self-energy and neglect vertex corrections and
the renormalization of the spin-fermion vertex,  the spin-polarization 
operator remain the same as for free fermions, i.e., the 
self-consistent perturbation theory is justified.
\end{enumerate}

\section{ $N \rightarrow \infty$ limit, $T=0$}
\label{sec_ninf}

These results of the previous section imply that     
 neglecting 
logarithms, one can construct a fully self-consistent new ``zero-order'' 
theory. In this theory, spin excitations are purely relaxational and  vertex 
corrections are absent. At the same time, the (anomalous) fermionic self-energy is strong and progressively, as $\lambda$ increases, destroys fermionic coherence. 

This new ``zero-order'' theory 
 indeed makes sense only if one can specify the limit when it becomes ``exact'' and also construct a 
 controllable perturbative expansion around it
 We notice in this regard that all $\log \lambda$ terms contain 
a factor $1/N$ where, we remind,  $N (=8)$ is the number of hot spots in the 
Brillouin zone. The presence of $1/N$ is a direct consequence of the fact that the $\log \lambda$  
terms appear due to a change from a ballistic to a 
relaxational spin dynamics. The prefactor for logarithm 
should then be inversely proportional to 
the spin damping rate as when this rate is zero, the self-energy and vertex 
corrections are much larger and scale as  $O(\lambda)$.
 A spin fluctuation with momentum ${\bf Q}$ has $N/2$ channels to decay into 
particle-hole excitation near hot spots, hence the damping rate scales as $N$, and the prefactor for the
 logarithm scales as $1/N$.  Below we formally treat 
$N$ as an arbitrary number. 
Then one can define the $N \rightarrow \infty$ limit, when logarithmical corrections can be totally neglected. In this limit, our new ``zero-order'' theory
 becomes exact. 
At finite but small $1/N$, 
 the perturbation theory around our new vacuum is the expansion in $1/N \log \lambda$ . 

Instead of introducing an artificially large  number of hot spots, 
one can also extend the spin-fermion model to 
 a large number of electron flavors $M$ ($=1$ in the physical case)  and expand in $1/(8M)$. This later expansion is more appealing from physics perspective 
 as the extension to large $N$ requires a rearrangement of the 
 Fermi surface to a large number of ``hot'' segments which 
is difficult to visualize.  In both cases, however, the main 
idea is to enhance the effect of the bosonic damping and therefore reduce the strength of regular self-energy and vertex corrections compared to the anomalous term in the fermionic self-energy. The computations are identical in both cases, and below we will just label the expansion as ``large N'' theory.
We discuss some formal aspects of the $1/N$ expansion in Appendix~
\ref{a_formal}.

The limit $N = \infty$ bears some similarity
 to the Migdal-Eliashberg limit for electron-phonon problem~\cite{eliashberg,migdal,mahan} although in our case bosonic modes are not independent degrees of freedom. Nevertheless, the smallness in $1/N$ has the same consequence as the smallness of $m/M$ where $m$ is the electronic mass, and $M$ is the ionic mass: one can neglect vertex corrections and also the momentum dependent piece in the fermionic self-energy. Below we will occasionally refer to the $N=\infty$ limit as the Eliashberg-type theory.

We now proceed with the more detailed discussion of the $N = \infty$ limit.

\subsection{derivation of the Eliashberg-type equations} 

So far our analysis of the fermionic self-energy 
was restricted to the limit of vanishing  
 fermionic frequency $\Omega$. 
 In this limit,  we have at $N \rightarrow \infty$, 
$\Sigma ({\bf k}, \Omega) = \lambda \Omega$. We have also 
checked that for this $\Sigma ({\bf k}, \Omega)$, the spin 
polarization operator is the same as for free fermions.

We now derive  the full form of the fermionic self-energy at $N = \infty$,
 valid for arbitrary frequencies.  
The input for the derivation is the knowledge that  
 vertex corrections and the corrections to $\epsilon_k$  both 
scale as $1/N$ and vanish at $N = \infty$.  
Without them, we have, quite generally, 
\begin{equation}
G({\bf k}, \Omega) = \frac{1}{{\tilde \Sigma} (k_{\parallel} \Omega) 
- \epsilon_k},                                                   \label{arbG}
\end{equation}
where ${\tilde \Sigma} (\Omega ) = \Omega + \Sigma (k_{\parallel} \Omega)$,
 and $k_{\parallel}$ is the momentum component along the Fermi surface.
The dependence of the self-energy on $k_{\parallel}$ is a subleading 
effect which we will be discussing below. This effect is relevant at 
the lowest frequencies but becomes progressively weaker as $\Omega$ 
increases. We explicitly verified  a'posteriori that at 
$N \rightarrow \infty$, typical frequencies are such that the 
dependence of $G({\bf k}, \Omega)$ of
 intermediate fermions on $k_{\parallel}$ can be totally neglected. We
therefore just ignore it in the derivation below. 

\begin{figure}[tbp]
\begin{center}
\epsfxsize=\columnwidth
\epsffile{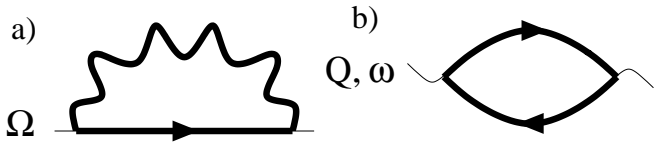}
\end{center}
\caption{ 
The diagrams for the fermionic self-energy and spin polarization operator 
at $N = \infty$.}
\label{Ninfinity}
\end{figure}

The full diagram for the self-energy at $N = \infty$ is presented in 
Fig \ref{Ninfinity}a.
Another diagram in Fig \ref{Ninfinity}b is 
for the full spin polarization operator. 
We cannot a`priori use the free fermion form for $\Pi ({\bf Q}, \omega)$ 
as so far we
have only demonstrated that it is preserved if we use the small $\Omega$ form
 of the fermionic self-energy. What we have to do is to derive and 
solve a set of two coupled equations for $\Sigma (k_{\parallel} \Omega)$ and $\Pi ({\bf Q}, \omega)$. 

The evaluation of the diagrams is straightforward. Let's first consider fermions near hot spots when $\Sigma (k_\parallel \Omega) \approx \Sigma (\Omega)$.
Substituting Eqs. (\ref{arbG}) and (\ref{chi}) into the diagram for the self-energy and integrating over momenta along  the
Fermi surface  at ${\bf k}_{hs} +{\bf Q}$,  we obtain
\begin{eqnarray}
\Sigma (\Omega) &=& \frac{3 i {\bar g}}{8 \pi^2}~
\int d \omega d {\tilde q}_x \frac{1}{[{\tilde q}^2_x + \xi^2 (1 - \Pi 
({\bf Q}, \omega))]^{1/2}}                       \nonumber\\
&\times & \frac{\Sigma (\omega) - \Sigma (\omega + \Omega)}
{(v_F {\tilde q}_x - \Sigma (\omega))
(v_F {\tilde q}_x - \Sigma (\omega +\Omega))}                    \label{sinew}
\end{eqnarray}
where as before ${\tilde q} = {\bf q}- {\bf Q}$.

We already know that the ``regular'' part of the self-energy 
vanishes at $N \rightarrow \infty$, so let's focus on the 
``anomalous'' part which should reduce 
 to $\Sigma_{an} (\Omega)$
in the $\Omega \rightarrow 0$ limit. This piece comes
 from  the range of frequencies where the two
fermionic poles are in different half-planes of ${\tilde q}_x$. 
 Since ${\rm sign} \Sigma (\omega) = {\rm sign} \omega$,
 this $\omega$ range is sandwiched between 
$- \Omega < \omega <0$ (for definiteness 
we set $\Omega >0$). The next step is to integrate over ${\tilde q}_x$.
 At small $\Omega$, we neglected ${\tilde q}_x$ as its inclusion would 
yield higher powers of $\Omega$. For arbitrary $\Omega$, we cannot use this argument. Still, it turns out that even for high frequencies,
 keeping ${\tilde q}_x$ in the spin susceptibility give rise to corrections 
which are at least small by $1/N^2$. We explicitly show this below in a separate subsection. At this moment, we just neglect ${\tilde q}_x$ 
 in the spin propagator and explicitly 
integrate over ${\tilde q}_x$ in (\ref{sinew}). 
 Evaluating the integral, we find that the 
fermionic self-energy is
{\it canceled out}, and $\Sigma (\Omega)$ takes the form
\begin{equation}
\Sigma (\Omega) = 2 \lambda \int_0^\Omega 
\frac{d \omega}{\sqrt{1 - \Pi ({\bf Q}, \omega)}}                \label{set1}
\end{equation}
This cancellation of the self-energy for intermediate fermions implies 
that self-consistent, FLEX-type
calculations in the normal state 
 are not necessary as the dominant piece in $\Sigma (\Omega)$ is
captured already by using free-fermion form of the propagator of intermediate fermions.

We next evaluate the spin polarization operator for 
arbitrary $\Sigma (\Omega)$. 
 Substituting the spin propagator (\ref{arbG}) into
(\ref{Pii}), replacing, as before the momentum integration by the 
integration over $d \epsilon_k d \epsilon_{k+Q}$, and integrating over
energies,  we obtain
\begin{equation}
\Pi ({\bf Q},\omega')\!\!=\!
\frac{i}{\omega_{\rm sf}}\!\!\! \int\limits_{-\infty}^{\infty}\!\! d \omega
\left( 1 \!-\! \frac{\Sigma (\omega)\Sigma(\omega + 
\omega')}{\sqrt{\Sigma^{2} (\omega)}
\sqrt{\Sigma^{2} (\omega + \omega')}}\right).                     \label{Pinew}
\end{equation}

Eqs. (\ref{set1}) and (\ref{Pinew}) form the set of two 
 coupled integral equations for the fermionic self-energy and the 
spin polarization operator. 

\subsection{solution of the Eliashberg-type equations}

A more careful look at Eqs. (\ref{set1}) and (\ref{Pinew}) shows that 
 the solution of the set is straightforward. 
Indeed, we clearly see that 
the magnitude and the functional 
dependence of $\Sigma (\omega)$ 
is totally irrelevant for $\Pi ({\bf Q}, \omega)$, 
all that matters is the fact that ${\rm sign}~\Sigma (\omega) = 
{\rm sign} \omega$. The frequency integration in (\ref{Pinew}) is then 
straightforward, and performing it  
we find that 
\begin{equation}
\Pi ({\bf Q}, \omega) = i\frac{\omega}{\omega _{\rm  sf}}
\label{setpi11}
\end{equation} 
i.e., for arbitrary $\Sigma (\Omega)$ it preserves 
exactly the same form  as for free fermions. 
Substituting this result into (\ref{set1}) we immediately find that 
\begin{equation}
\Sigma (\Omega) = \frac{2 \lambda \Omega}{1
 + \sqrt{1 - i \frac{|\Omega|}{\omega_{sf}}}}.                \label{setot}
\end{equation}
This expression can also be re-written in the scaling form
\begin{equation}
\Sigma ( \Omega) = \lambda \omega_{sf} 
g \left(\frac{\omega}{\omega_{sf}}\right),
~~g(x) = \frac{2x}{ 1 + \sqrt{1-i|x|}}
                                    \label{setot1}
\end{equation}
 At small
 $\Omega$  we indeed recover the previous result 
$\Sigma (\Omega) = \lambda \Omega$. 

The  general explanation
why the $k-$ independent 
fermionic self-energy $\Sigma (\Omega)$ 
does not affect the polarization bubble at
a finite  momentum was given by
Kadanoff \cite{kadanoff}.  He pointed out that the expansion of
the fermionic energy to first order in momentum
deviation from $k_F$ is equivalent to imposing  an approximate Migdal sum
rule on the spectral function $A ({\bf k}, \omega) = (1/\pi)~ 
Im {\tilde G}_0 ({\bf k}, \omega)$
\begin{equation}
\int d \epsilon_{\bf k} A ({\bf k}, \omega) =1
\label{A}
\end{equation}
Expressing ${\tilde G}_0 ({\bf k}, \omega)$ in terms of 
$A({\bf k}, \omega)$ via
Kramers-Kronig relation and making use of (\ref{A}), one finds
\begin{equation}
\Pi ({\bf Q},\omega') = 
- \frac{i}{\omega_{\rm sf}} \int_{-\infty}^{\infty} d \omega 
\frac{ d f(\omega)}{d \omega} \omega' + {\cal O}(\omega'^2)
\end{equation}
where $f(\omega)$ is the Fermi function.
Eq. (\ref{Pi1}) then follows from the fact that $f(\omega)$ is $1$ at $\omega =
-\infty$ and $0$ at $\omega = +\infty$.

Eqs. (\ref{setpi11}) and (\ref{setot}) are the central results of $N = \infty$ analysis. We see that the scale for nonlinear effects in frequency in 
$\Sigma (\Omega)$ is set by the typical spin relaxation frequency 
$\omega_{sf} = 4\pi v_x v_y \xi^{-2}/(N\bar g)$ (the $1/N$ factor in $\omega_{sf}$ can be eliminated by rescaling 
$v \rightarrow v N$ and ${\bar g} \rightarrow {\bar g} N$ as we discuss in Appendix\ref{a_formal}). 
This $\omega_{sf}$ is obviously a measure of the deviation from the 
 QCP. 
At scales smaller than $\omega_{sf}$, spin susceptibility is nearly static, and $\Sigma (\Omega)$ is expandable in $\Omega$. In this regime, we should generally expect a Fermi liquid behavior with 
$\Sigma^{\prime \prime} (\Omega) \propto \omega^2$.
However, above $\omega_{sf}$, the system should cross over into a 
quantum-critical regime, where the system behavior is 
determined by the fixed point at $\xi = \infty$. In this regime, 
 spin excitations become massless diffusons, and fermionic self-energy should
 modify such that to eliminate the dependence on $\xi$.

We indeed find this behavior in our $\Sigma(\Omega)$. We see from 
(\ref{setot}) that for $|\Omega| < \omega_{sf}$, 
the self-energy behaves as 
 \begin{equation}
\Sigma (\Omega) = 
\lambda \Omega \left( 1 + i \frac{|\Omega|}{4 \omega_{sf}}\right),
\end{equation}
For  $|\Omega| > \omega_{sf}$, however, 
it crosses over to a different behavior 
 \begin{equation}
\Sigma (\Omega) = 
(i |\Omega| {\bar \omega)^{1/2}~\mbox{sign} \Omega},
\end{equation}    
We see therefore that above $\omega_{sf}$, 
 both $\Sigma^{\prime} (\Omega)$ and 
$\Sigma^{\prime \prime} (\Omega)$ scale as $\sqrt{|\Omega|}$.
The normalization energy 
\begin{equation}
{\bar \omega} = 4 \lambda^2 \omega_{sf} = \frac{9 {\bar g}}{2 \pi N}
~\frac{2 v_x v_y}{v^2_F}
\end{equation}
does not depend on $\xi$ as it should be in the quantum-critical regime.
This ${\bar \omega}$ 
sets the upper cutoff for the quantum-critical 
behavior (its dependence  on $N$ can again be eliminated by rescaling $g \rightarrow N {\bar g}$. 

The $\sqrt{\Omega}$ form of the self-energy at the antiferromagnetic transition was first obtained by Millis~\cite{andy92}. The $\sqrt{\omega}$ form also emerges in the quantum-critical models with disorder~\cite{georges}. 
 
 Substituting the quantum-critical form of the self-energy 
into the fermionic propagator, we obtain 
that at ${\bf k} \approx {\bf k}_{hs}$ and 
$|\Omega| < {\bar \omega}$  
\begin{equation}
G(k, \Omega) = \frac{1}{\bar \omega}~
\frac{\sqrt{i |\Omega| {\bar \omega}}
 {\rm sign} \Omega + \epsilon_k}
{i |\Omega| -\epsilon^2_k/\bar \omega}.                       \label{ginc}
\end{equation}
We see that in the quantum-critical regime, the system behavior is 
qualitatively different from that in a Fermi liquid: there is no 
pole in the fermionic propagator
at real frequencies. Instead, $G(k, \Omega)$  has a pole along
an imaginary frequency axis, at $|\Omega| = \epsilon^{eff}_k = 
\epsilon^2_k/{\bar \omega}$. 
This non-Fermi-liquid pole gives rise to a 
broad maximum at $\Omega = \epsilon^{eff}_k$. 
 We also verified that in the crossover region 
$\Omega \sim \omega_{sf}$, 
the pole in the fermionic propagator 
gradually moves, with increasing $\Omega$, 
 from  real to imaginary frequency axis   

At frequencies larger than $\bar \omega$, the bare $\omega$ term
in the fermionic
 propagator wins over the self-energy, 
and the spectral function recovers the  peak
 at $\Omega = \epsilon_k$. Still, this behavior is not a Fermi-liquid one 
as the width of the peak scales as $\sqrt{\Omega}$.   

\begin{figure}[tbp]
\begin{center}
\epsfxsize= \columnwidth
\epsffile{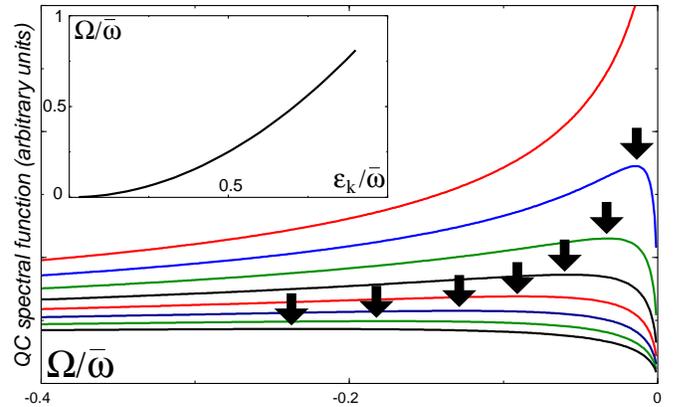}
\end{center}
\caption{ 
The behavior of $Im G({\bf k}, \Omega)$ at $\omega_{sf} =0$ (\ref{ginc}).
The inset shows the ``quasiparticle'' dispersion inferred from the position
of the maximum in the spectral function (indicated by arrows).}
\label{qcspectr}
\end{figure}

The behavior of $Im G({\bf k}, \Omega)$ at $\omega_{sf} =0$ 
is shown in Fig. \ref{qcspectr} (see Eq. (\ref{ginc})).
The inset shows the ``quasiparticle'' dispersion inferred from the position
of the maximum in the spectral function. 
We see that the destruction of the Fermi liquid behavior accounts for the
effective flattening of the ``quasiparticle'' dispersion at frequencies
below $\bar \omega$.

\begin{figure}[tbp]
\begin{center}
\epsfxsize=\columnwidth
\epsffile{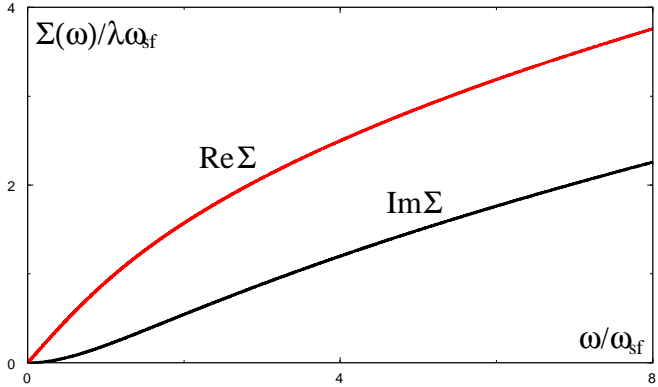}
\end{center}
\caption{ 
Real and imaginary parts of the fermionic 
self-energy, Eq.(\protect\ref{setot1}), vs frequency.
}
\label{sigmaT0}
\end{figure}

In Fig. \ref{sigmaT0} we present both real and 
imaginary parts of the fermionic self-energy vs frequency
 at a finite $\omega_{sf}$. We clearly see
a Fermi-liquid behavior at the smallest frequencies, and  the $\sqrt{\omega}$
behavior well above $\omega_{sf}$. By numerical reasons, the crossover 
region in between these two limiting regimes is rather wide - it stretches
between $0.5 \omega_{sf}$ and $6-8 \omega_{sf}$. In the crossover regime,
$\Sigma^{\prime \prime} ({\bf k}, \omega)$ is, 
to a surprisingly good accuracy, 
a linear function of frequency. This linearity, however, does not follow
from any theory considerations, and is just a hidden numerical 
property of the self-energy, Eq. (\ref{setot}).
This linearity, however, is relevant for the 
explanation of the photoemission and optics data in cuprates.

\begin{figure}[tbp]
\begin{center}
\epsfxsize=\columnwidth
\epsffile{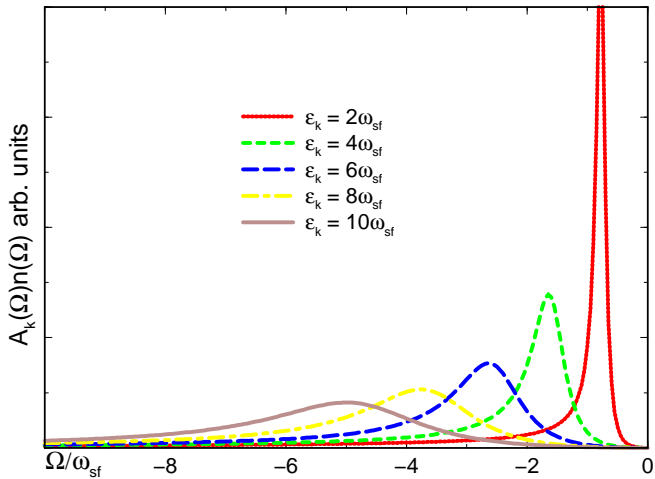}
\end{center}
\caption{ 
The behavior of $Im G({\bf k}, \Omega)$ at a finite $\omega_{sf}$.
For definiteness we used $\lambda =1.7$ in which case ${\bar \omega} =11.56 \omega_{sf}$.
The high-frequency behavior of $Im G({\bf k}, \Omega)$ is the same as in 
 Fig.(\protect\ref{qcspectr}).
}
\label{spectr_T0}
\end{figure}
In Fig~\ref{spectr_T0} we present the forms of the spectral function  
$Im G({\bf k}, \Omega)$ for different values of $\epsilon_k$ and finite $\omega_{sf}$.
Comparing this result with Fig~\ref{qcspectr} we see that the quasiparticle peak sharpens up at
frequencies comparable to $\omega_{sf}$, as it indeed should as the system crosses over to the 
Fermi-liquid behavior.  Notice, however,
 that even at $\omega \sim \omega_{sf}$, the width of the
 peak is comparable to its amplitude.

We next discuss to  
 which extent the
full anomalous self-energy  depends on the momentum along the Fermi surface. 
Repeating the same calculations which lead to (\ref{setot})
but for a finite ${\tilde {\bf k}} = 
{\bf k} - {\bf k}_{hs}$, we find  that 
$\Sigma ({\bf k}, \Omega) \equiv \Sigma (k_\parallel, \Omega)$ 
can be rewritten as
\begin{equation}
\Sigma (k, \Omega) = \lambda (k)  
\frac{2 \Omega}{1 + \sqrt{1 - 
i \frac{|\Omega|}{\omega_{sf} (k)}}}                          \label{setot2}
\end{equation}
i.e.,  it still has the same form as at a hot spot, 
but with $k$ dependent 
\begin{equation}
\lambda (k) = \lambda/(1 + ({\tilde k} \xi)^2)^{1/2}, \,\,\,\,
\omega_{sf} (k) = \omega_{sf} (1 +  ({\tilde k} \xi)^2)        \label{kdep}
\end{equation}
We see  that away from the hot spots, 
the effective coupling gets smaller, 
and the crossover frequency $\omega_{sf} (k)$ increases. Still,
however, at frequencies which exceed 
$\omega_{sf} (k)$, the system 
displays the same non-Fermi liquid behavior as at hot spots, i.e., 
$\Sigma^{\prime \prime}(k, \Omega)$  first scales 
linearly with $\Omega$, 
and crosses over to a $\sqrt{\Omega}$ form at larger frequencies.

In the Fermi liquid regime, the slope  of $\Sigma^{\prime \prime}$
  depends on $k$ as $(1 + ({\tilde k} \xi)^2)^{-3/2}$. In the non-Fermi liquid regime, this dependence gets weaker and completely vanishes in the 
$\sqrt{\Omega}$ regime
where the overall factor in $\Sigma (k, \Omega)$ 
 becomes $\lambda (k) \sqrt{\omega_{sf} (k)} = 
\lambda \sqrt{\omega_{sf}}$. This can be seen 
 already from (\ref{setot}). This implies the
momentum range where $\Sigma (k, \Omega)$ weakly depends  momentum 
(i.e., the ``size'' of a hot spot) depends on frequency. 
For  $\Omega < \omega_{sf}$, 
this range is obviously constrained by 
$|{\tilde k}| \xi \leq 1 $, but for larger
frequencies it increases and eventually extends to 
$|{\tilde k}| \xi < (|\Omega|/\omega_{sf})^{1/2}$.
In particular, deep in the quantum-critical regime, i.e., for 
 frequencies comparable ${\bar \omega} = 
4 \lambda^2 \omega_{sf}$,
the weak momentum dependence of the fermionic self-energy extends to
$|{\bf k} - {\bf k}_{hs}| \leq {\bar g}/v_F$, i.e, the ``size'' of a hot spot 
remains {\it finite} even at 
 $\xi \rightarrow \infty $. This result implies that in the quantum-critical 
regime the momentum variation of the self-energy does not
play any significant role and can be safely neglected. 

\subsection{the accuracy of the Eliashberg-type theory} 

We now have to go back and verify the accuracy  
of the Eliashberg-type Eqs. (\protect\ref{set1}) and (\protect\ref{Pinew}).
We begin with the fermionic self-energy. 
Recall that in deriving  (\protect\ref{set1}) we neglected the 
dependence on ${\tilde q}_x$ in the spin propagator.
Near the singular pole, ${\tilde q}^2_x \sim (\Sigma (\omega)/v_F)^2$. 
To estimate the relative strength of this term, let's evaluate it with
$\Sigma (\omega)$ from (\ref{setot}) 
obtained by neglecting this contribution. 
At $\omega < \omega_{sf}$, $\Sigma (\omega) \approx \lambda \omega$, 
hence typical ${\tilde q}^2_x \sim \lambda^2 \omega^2/v^2_F = 
(\omega \gamma) (\omega/(4 N^2 \omega_{sf}))$ are negligible compared to 
$ \omega \gamma$. Clearly, 
${\tilde q}^2_x$ term can be omitted. In the quantum-critical regime,
$\omega \gg \omega_{sf}$, we have 
${\tilde q}^2_x \sim \omega {\bar \omega}/v^2_F = (\omega \gamma)  
(6 v_x v_y/N v^2_F)^2$.
%
%
We see that the
term which we omitted in (\ref{set1}) and, hence, (\ref{setot}) 
has the same frequency dependence as the 
damping term in the spin susceptibility, 
but has an extra factor $(1/N)^2$  and therefore can be neglected in the $N \rightarrow \infty$ limit.  

To understand how $1/N$ expansion works in terms of numbers, we evaluated the
overall factor for 
$\Sigma (\omega)$ in the quantum-critical regime by solving Eq. (\ref{sinew})
self-consistently for a physical $N=8$. Substituting $\Sigma (\Omega) = 
A (i |\Omega| {\bar \omega})^{1/2} {\rm sign} \omega$ as an input 
into (\ref{sinew}), we
indeed obtained the same form in the output. Solving for $A$, we obtained
$A= 0.94$ which is very close to $A=1$ -- the result without $1/N$
corrections.  

We also compared the $N = \infty$ result with the full second-order
 expression for 
$\Sigma (\Omega)$  at arbitrary $\Omega$. 
The calculations are presented in the Appendix A. The result is
particularly simple  for $v_{s} = \infty $ in which case we have
from (\ref{sigma-full})
\begin{equation}
\Sigma (\Omega ) = 2\lambda \omega _{sf} a~
\ln \frac{i\sqrt{K_{\Omega }-1}+\sqrt{K_{\Omega } +1}}
{i\sqrt{K_{\Omega }\!\!-\!1\!\!+\!A_{\Omega }} + 
\sqrt{K_{\Omega}\! +1 \!\! -\!A_{\Omega }\!}}\label{fullhs}
\end{equation}
where
\begin{equation}
K_{\Omega }^{2} = 1+\frac{4}{a^2}\left(1-i\frac{\Omega }{\omega _{sf}}\right);
\,\,\,\,A_{\Omega } = \frac{2i}{a^2} \frac{\Omega}{\omega _{sf}}
\end{equation}
and, we remind, $a = (v_{F}\xi ^{-1})/\omega_{sf} = 
(\lambda N/3)(v^2_F/v_{x}v_{y})$.
Our $N = \infty$  result, 
Eq. (\ref{setot}), is
the limiting from of (\ref{fullhs}) at $a \rightarrow \infty$. 
The expansion to a linear order in $1/a$ yields a 
correction to (\ref{setot}) 
which at low frequencies is logarithmical in $\lambda$
and coincides with  Eqs. 
(\ref{i}) and (\ref{Ireg}).
   
\begin{figure}[tbp]
\begin{center}
\epsfxsize=\columnwidth 
\epsffile{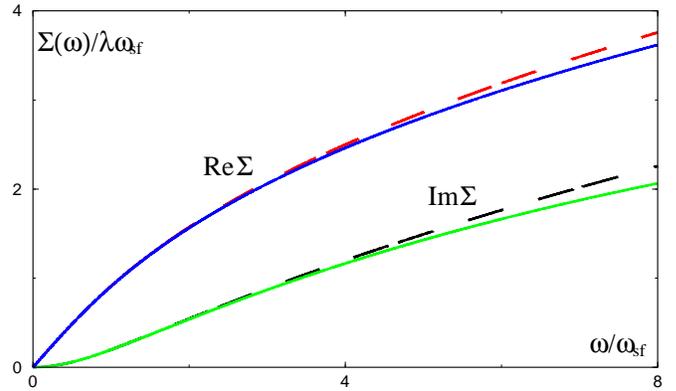}
\end{center}
\caption{ 
Real and imaginary parts of the full fermionic 
self-energy at finite $N$ as a function of frequency. Dashed lines represent
the self-energy without the regular part (see Fig. \protect\ref{sigmaT0}).
}
\label{sigmaT0N}
\end{figure}

In Fig. \ref{sigmaT0N} we plot the full second-order result for the 
self-energy at $N=8$ and our $N = \infty$ result which is its  anomalous
part. For definiteness, we set $v_x = v_y$ and $\lambda =2$
in which case $a = 32/3$.
We see that the two
curves are very close to each other up at least to 
$\Omega = {\bar \omega}$, which, we remind, 
is the upper cutoff frequency for the quantum-critical behavior.    
We consider this agreement as another
 evidence that the restriction with only the anomalous contribution
to the self-energy works well for the physical $N=8$.  

A minor comment is in order here.
In the quantum-critical regime, the second-order expression for
$\Sigma (\omega)$ contains a leading   $\Omega^{1/2}$ term and a 
subleading  $\Omega \log \Omega$ term. 
On the other hand, performing computations with the
full fermionic propagators, we found a subleading logarithmical term in
$\Sigma ({\bf k}_{hs}, \Omega)$ but no $\Omega \log \Omega$ term.
 A careful look into this discrepancy 
 shows that the reason why
 the subleading (regular) piece in the second-order self-energy has 
$\Omega \log \Omega$ form 
is because for free fermions,
$\partial G_0^{-1}({\bf k}, \omega)/\partial \omega = - 
 \partial G_0^{-1}({\bf k}, \omega)/\partial \epsilon_k = 1$.
In the self-consistent calculations, 
we still have $\partial G^{-1}({\bf k}, \omega)/\partial \epsilon_k = -1$, but
now $\partial G^{-1}({\bf k}, \omega)/\partial \omega = \partial 
\Sigma (\omega)/\partial \omega$. In the quantum-critical regime, 
$\partial \Sigma (\omega)/\partial \omega \propto \omega^{-1/2}$. The 
 appearance of the
extra $\omega^{1/2}$ in the denominator, 
transforms $\Omega \log \Omega$ term
 into $\sqrt{\Omega}/N^2$, exactly as we found.

Finally, we also verify that the renormalization of $\epsilon_k$ and
of the spin-fermion vertex scale as $\log \lambda /N$ even when we use 
quantum-critical forms of the self-energy for intermediate fermions.
Lets, first analyze $\Sigma ({\bf k}, 0)$.
Since the fermionic self-energy in (\ref{arbG}) does not affect $\epsilon_k$,
we can straightforwardly expand $\Sigma ({\bf k}, 0)$
 in $\epsilon_{k+Q}$. Performing then a 
 simple power counting, we find that in the quantum-critical regime,
 the inclusion of the 
self-energy  of intermediate fermions 
only gives rise to $O(1/N^2)$ corrections to the second-order result. 

The same result holds for vertex renormalization. 
Substituting (\ref{arbG}) into (\ref{vert1}) we obtain
\begin{eqnarray}
\frac{\Delta g}{g} &=& \frac{i {\bar g}}{(2\pi)^3}~\int d^2{\tilde q} d \omega
~\frac{1}{\xi^{-2} - i |\omega| \gamma + {\tilde {\bf q}}^2}      \nonumber \\
&\times&\frac{1}{\Sigma (\omega) - ( v_x {\tilde q}_x + v_y {\tilde q}_y)}
~\frac{1}{\Sigma (\omega) - ( -v_x {\tilde q}_x + v_y {\tilde q}_y)}
\label{newvert}
\end{eqnarray}
We immediately see that if
 we neglect the momentum dependence of the spin susceptibility, then
the fermionic self-energy is totally eliminated 
after the integration over $d^2 {\tilde q}$, and the 
vertex correction remains exactly the
 same as if we use free fermions (see Eq. (\ref{vert2})). 
If we keep the momentum
dependence in $\chi ({\bf q}, \omega)$, the functional form of $\Delta g/g$
does not change, but
the coefficient changes by $1 + O(1/N^2)$.

\subsection{vertices at small momentum transfer}
\label{vert_zero}

Before we proceed with the $1/N$ expansion, let's consider in more length the 
issue of vertex renormalization for different bosonic momenta $q$.
Earlier we found that the particle-hole 
vertex with momentum
transfer ${\bf Q}$ is only weakly (as $\log \lambda /N$) 
renormalized by spin-fermion 
interaction. Mathematically, the smallness of the vertex correction 
is due to the fact that the two internal fermions in the vertex 
correction diagram had different 
directions of Fermi velocities, and hence one can independently integrate 
over $\epsilon_k$ and $\epsilon_{k+Q}$. The vertex correction  
then can roughly be written as 
\begin{equation}
\frac{\delta g}{g} \sim \frac{g^2}{v^2_F}
 \int d \omega N^2 (\omega) \chi ({\bf Q}, \omega)
\label{oo1}
\end{equation}
 where   $N(\omega) = \int d \epsilon_k G ({\bf k}, \omega) = 
-i\pi \mbox{sign} (\omega)$.
The frequency dependence then comes only from $\chi({\bf Q}, \omega) \propto  
v^2_{F}/(N g^2 \omega)$, and frequency integration  yields 
$\delta g/g \sim (L/N)$ where $L$ is the logarithm of the lower cutoff. 

This reasoning is valid for all external $q$, not necessary $Q = (\pi,\pi)$
(for ${\bf q} \neq {\bf Q}$, the logarithm is cut also by $|{\bf q}-{\bf Q}|$).
 However, it breaks up when one considers a particle-hole
 vertex with zero momentum transfer. For such vertex, the 
    Ward identity implies~\cite{mahan} that at vanishing bosonic frequency
$\delta g/g = 1 + \partial \Sigma ({\bf k}, \omega)/\partial \omega$ for a 
scalar vertex (the one which does not depend on fermionic momentum and 
conserves spin projection) and 
$\delta g/g = 1 + \partial \Sigma ({\bf k}, \omega)/\partial \epsilon_k$ for a 
current vertex (which has an extra $\partial \epsilon_k/\partial {\bf k})$. 
In our situation, $\partial \Sigma ({\bf k}, \omega)/\partial \epsilon_k$ is
 small in $1/N$, but $ \partial \Sigma ({\bf k}, \omega)/\partial \omega$ 
is not. 
Moreover, as $\Sigma (\omega) \gg \omega$ at strong coupling and 
$\omega < {\bar \omega}$, the renormalized scalar vertex well 
exceeds the bare one. 
        
That the renormalization of the scalar vertex with zero momentum transfer does not contain $1/N$ can indeed be obtained in perturbative calculations. 
First, we derive the Ward identity for the scalar vertex. 
It follows from the fact that the integral equation for the full vertex
\begin{eqnarray}
&&g^{tot} ({\bf k}, \omega, \Omega) = 
g - \frac{3 i {\bar g}}{8\pi^2} \int d^2 q d 
\omega^\prime \nonumber \\
&&\frac{~g^{tot} ({\bf k}+{\bf q}, \omega^\prime, \Omega) 
\chi ({\bf q},\omega - \omega^\prime)}{({\tilde \Sigma}_{k+q} (
\omega^\prime + \Omega) - 
\epsilon_{k+q})({\tilde \Sigma}_{k+q}(\omega^\prime) - \epsilon_{k+q})}
\label{oo2}
\end{eqnarray}
where as before ${\tilde \Sigma}_\omega = \omega + {\tilde \Sigma}_\omega$, 
is equivalent to  the integral equation for ${\tilde \Sigma}_{k}(\omega + \Omega) - {\tilde \Sigma}_{k}(\omega)$:
\begin{eqnarray}
&&{\tilde \Sigma}_{k}(\omega + \Omega) - {\tilde \Sigma}_{k}(\omega)
= \Omega - \frac{3 i{\bar g}}{8\pi^2} \int d^2 q d 
\omega^\prime \nonumber \\ 
&&\frac{({\tilde \Sigma}_{k+q}(\omega^\prime + \Omega) - 
{\tilde \Sigma}_{k+q}(\omega^\prime))~\chi (\omega - \omega^\prime, q)}{({\tilde \Sigma}_{k+q} (
\omega^\prime + \Omega) - \epsilon_{k+q})
({\tilde \Sigma}_{k+q}(\omega^\prime) - \epsilon_{k+q})}
\label{oo3}
\end{eqnarray} 
Comparing (\ref{oo2}) and (\ref{oo3}) we immediately find that
\begin{equation}
g^{tot} ({\bf k}, \omega, \Omega) = g \left( 1 + 
\frac{{\tilde \Sigma}_{k}(\omega + \Omega) - 
{\tilde \Sigma}_{k}(\omega)}{\Omega}\right)
\label{oo4}
\end{equation}
In the limit $\Omega \rightarrow  0$ the r.h.s. of (\ref{oo4}) yields 
$ 1 + \partial \Sigma ({\bf k}, \omega)/\partial \omega$.

The very fact that $g^{tot} ({\bf k}, \omega, \Omega)$ 
is related to  $\partial \Sigma ({\bf k}, \omega)/\partial \omega$ 
implies that 
the large renormalization of the scalar vertex with zero momentum transfer 
is due to the same anomaly which accounts for large $\Sigma (\omega)$. 
Indeed, consider the lowest-order diagram for the vertex renormalization. 
When external bosonic frequency is strictly zero, the two fermionic 
propagators have poles in the same half-plane of frequency, and the 
integration over $\epsilon_k$ does not vanish only due to the fact
 that the spin susceptibility also has poles at 
$|{\bf q}-{\bf Q}| \propto (N \omega)^{1/2}$. 
Obviously, the resulting vertex correction is small in $1/N$. However, at 
finite $\Omega$, the poles in the fermionic propagators are split, 
and there is a nonvanishing contribution from the frequency range where the 
split poles are in different frequency half-planes. 
Evaluating this contribution in the same way as for the 
fermionic self-energy, we indeed find that
\begin{equation}
\frac{\Delta g}{g} =  1 + \lambda \int_0^\Omega~ 
\frac{ d\omega^\prime}{{\tilde \Sigma} (\Omega - \omega^\prime) + {\tilde \Sigma} (\omega^\prime)}~\frac{1}{(1 + |\omega + \omega^\prime|/\omega_{sf})}
\label{oo5}
\end{equation}
It is clear from (\ref{oo5})
 that the large renormalization of $g$ is due to the same anomaly which 
gave rise to a large $\Sigma (\omega)$. Collecting the anomalous
 contributions from the
 higher-order diagrams and summing them up, one should indeed recover Eq. (\ref{oo4}).

Note that this procedure is not as straightforward as one might expect.
Indeed, consider  the limit when 
$\Omega, \omega \ll \omega_{sf}$. In this limit, Eq. (\ref{oo5}) yields 
\begin{equation}
\frac{\Delta g}{g} =  1 + \frac{\lambda}{1 + \lambda}.
\label{oo6}
\end{equation}
As for vanishing $\Omega$, the full $g^{tot} = g (1 + \lambda)$, 
one may suggest that the perturbation series are geometrical, 
i.e. $1 + \lambda/(1 + \lambda) -> 1/(1 - \lambda/(1 + \lambda)) = 1 + \lambda$. However, the actual situation is not like this: 
the explicit calculation of the two-loop vertex correction yields 
\begin{equation}
\frac{\Delta g}{g}_{|2loop} = \frac{2}{\pi} 
\left(\frac{\lambda}{1 + \lambda}\right)^2.
\label{oo7}
\end{equation}
i.e., the next order diagram contains an  extra $\pi/2$ compared to 
 what is would be if the series were geometrical.
 This peculiarity  of the perturbation series emerges  because 
$\Sigma_k (\omega)$ does possess some non-singular 
$k$-dependence along the Fermi surface.
This momentum dependence is indeed fully accounted for 
in the Ward identity but complicates perturbation series in some non-singular way. This is just another example that momentum dependence of the fermionic self-energy complicates calculations but does not affect the physics.  

\subsection{summary of Sec~\protect~\ref{sec_ninf}.}

Let' now summarize what we have at the moment. 

\begin{enumerate}
\renewcommand{\labelenumi}{\roman{enumi}}
\item
We find that at strong coupling, 
one can construct a new, fully self-consistent 
``zero-order'' theory which explicitly includes the
most divergent $O(\lambda)$ piece in the fermionic self-energy, but 
neglects vertex corrections and the renormalization of the Fermi velocity 
which both scale logarithmically with $\lambda$. 
The divergent piece in the self-energy comes from
the second-order diagram. Higher order self-energy and vertex correction
 diagrams only 
account for totally irrelevant, small numerical corrections to this term.
\item  
We  demonstrated that this ``zero-order'' theory  
becomes exact in the formal limit when $N \rightarrow \infty$.  
In this theory, fermions are strongly affected by interaction with 
spin fluctuations and display non-Fermi liquid, quantum-critical 
behavior at frequencies  larger than a typical spin relaxation 
frequency $\omega_{sf}$. In the quantum-critical regime, the fermionic 
self-energy interpolates between linear in $\omega$ and $\sqrt{\omega}$ 
forms. At the same time, there is no feedback from
fermionic incoherence on spin excitations which remain diffusive with 
the same diffusion constant as if they were made from free fermions.  
\end{enumerate}

Our next step is to develop a perturbation theory in $1/N$. Obviously, 
our interest is to understand the role of logarithmical
 vertex and self-energy corrections.   

\section{The perturbation theory in $1/N$}
\label{per_th}

\subsection{one loop RG analysis}

We recall that  the $\log{\lambda}/N$) 
 terms give rise to the two new features in the theory:
 vertex corrections 
which renormalize both fermionic and bosonic self-energies, and  
static fermionic self-energy $\Sigma ({\bf k}, 0)$. 
To the lowest order in $1/N$, we have from (\ref{vert2}) and (\ref{s1}) 
\begin{eqnarray}
\frac{\Delta g}{g}  &=&  \frac{Q(v)}{N}~ \log {\lambda},         \label{vert}\\
\Delta \epsilon_k &=& - \epsilon_{k+Q}~
\frac{12}{\pi N}~ \frac{v_x v_y}{v^2_F}~ \log{\lambda}           \label{se}
\end{eqnarray}
where 
$Q(v) =(4/\pi) {\rm tan}^{-1} (v_{x}/v_{y})$
interpolates between 
$Q=1$ for  $v_x = v_y$, and $Q=2$ for $v_y \rightarrow 0$. 

We see from (\ref{vert},\ref{se}) that the $1/N$ corrections 
to the vertex and to the velocity of the excitations
are almost decoupled from each other: the velocity renormalization does 
not depend on the coupling strength at all, 
while the renormalization of the vertex depends on the ratio of velocities 
only through a non-singular $Q(v)$.
 The absence of the coupling constant in the r.h.s of the Eqs. (\ref{vert},\ref{se}) 
is, we recall,  a direct consequence of the fact that the dynamical part of the 
spin propagator is 
obtained self-consistently within the model. Due to self-consistency,
 the
overall factors in $\Delta\epsilon_k$ and 
$\Delta g/g$ are ${\bar g} (\omega_{sf} \xi^2)$.
 Since the fermionic damping is produced by
the same spin-fermion interaction as the fermionic self-energy, 
$\omega_{sf}$ scales as $1/{\bar g}$, 
and the coupling constant disappears from the Eqs. (\ref{vert},\ref{se}). 

The fact that the lowest-order corrections diverge at $\lambda = \infty$
obviously  implies that the $1/N$ expansion 
breaks down near the QCP, and one has to sum up the series of the 
logarithms. We will do this in a standard one-loop 
approximation by summing up the series in $(1/N) \log \xi$ but 
neglecting regular 
$1/N$ corrections to each term in the series (and, indeed, we neglect regular
$1/N^2$ corrections from FLEX-type diagrams). We verified that in 
this approximation,  the cancellation of the coupling constant holds 
even when $g$ is a running, scale dependent 
coupling. This in turn implies that one can separate the velocity 
renormalization 
from the renormalization of the vertex to all orders in $1/N$.

Separating the corrections to $v_x$ and $v_y$ and 
performing standard RG manipulations, we  obtain a set of two RG equations
for  running $v^R_x$ and $v^R_y$
\begin{eqnarray}
\frac{\di v^R_{x}}{\di L}&=&\frac{12}{\pi N} 
\frac{ (v^R_x)^2  v^R_y}{ (v^R_{x})^{2}+ (v^R_{y})^{2}}         \nonumber \\
\frac{\di  v^R_{y}}{\di L}&=&-\frac{12}{\pi N} 
\frac{(v^R_{y})^2  v^R_x}{(v^R_{x})^{2}+(v^R_{y})^{2}}  \label{Renorm-velocity}
\end{eqnarray}
where $L = \log \xi$. 
The solution of these equations is straightforward and yields
\begin{equation}
v^R_x = v_x Z; v^R_y = v_y Z^{-1}; 
Z=\left(1 + \frac{24 L}{\pi N} \frac{v_y}{v_x}\right)^{1/2}
\label{sol}
\end{equation}
where, we recall, $v_x$ and $v_y$ are the bare values of the 
velocities (the ones in the Hamiltonian). 

\begin{figure}[tbp]
\begin{center}
\epsfxsize=2.5in 
\epsfysize=2.5in
\epsffile{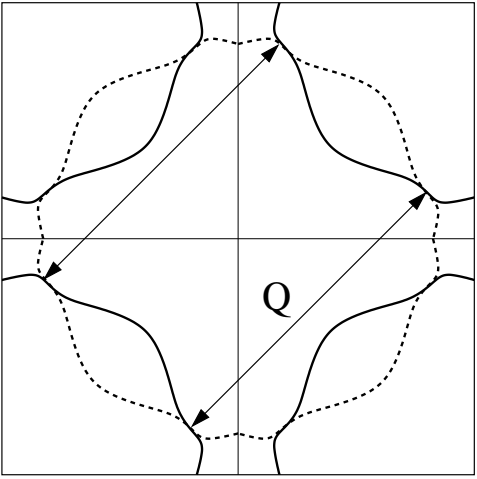}
\end{center}
\caption{ 
The nesting of the Fermi surface 
as the result of the renormalization
of the Fermi velocity. Solid line - the Fermi surface. Dashed line - the
 shadow Fermi surface (shifted by antiferromagnetic ${\bf Q}$). 
 At  $\xi = \infty$,  the velocities at hot spots become perpendicular to the magnetic Brillouin zone boundary. This creates a ``bottle neck effect'': the 
Fermi surface easily disconnects at hot spots immediately below the magnetic transition.}
\label{nesting}
\end{figure}

We see that $v^R_y$ vanishes logarithmically at
$\xi \rightarrow \infty$. This implies that right at the QCP, the
renormalized velocities  at ${\bf k}_{hs}$ and ${\bf k}_{hs}+{\bf Q}$ 
are antiparallel 
to each other, i.e. the Fermi surface becomes nested at hot spots 
(Fig. \ref{nesting}).
This nesting in turn 
creates a ``bottle neck effect'' 
immediately below the criticality as the original and the 
shadow Fermi surfaces approach hot spots with equal 
derivatives (see Fig. \ref{nesting}).
Quite generally, the nesting at hot spots   
is the first step in  the
evolution of the Fermi surface towards hole pockets
~\cite{ch-morr}. 
If the 
nesting occurred at some finite $\xi$, (as one can judge by formally 
extending the lowest 
order result, Eq. (\ref{se}) outside the range of its applicability),
then the system would
 develop strong SDW precursors already in a paramagnetic 
phase~\cite{ks,schr,trembl,ssp}.
It turns out, however, that this process is 
precluded because nesting at finite $\xi$ (i.e.,  $v^R_y =0$ at
finite $v^R_x$)
would imply that  $\omega_{sf}$ renormalizes to zero,
 but renormalized $\omega_{sf}$
 is the  overall factor in the velocity renormalization.
The only way to avoid this negative feedback effect 
is to consider the formal limit of small $N$, 
when the spin damping can be neglected~\cite{cms}. 
In this limit, the feedback effect on the velocity renormalization is 
absent, and the 
Fermi surface evolves towards hole pockets already at a finite $\xi$.  

Another feature of the RG equations (\ref{Renorm-velocity}) is
that they leave the product $v_x v_y$ unchanged. This is a combination 
in which velocities appear in $\omega_{sf}$. The fact that $v_x v_y$ is 
not renormalized  implies that without vertex renormalization, 
$\omega_{sf} \xi^2$ preserves its form, i.e.  spin susceptibility still
 has a simple diffusion pole. 

We next consider vertex renormalization. 
Using again the fact that ${\bar g} \omega_{sf}$ does not depend on 
the running coupling constant, one can straightforwardly extend 
the second-order result for the vertex 
renormalization, Eqn (\ref{vert}), 
to the one-loop RG equation
\begin{equation}
\frac{d g^R}{d L} =  \frac{Q(v)}{N} g^R                        \label{rgg}
\end{equation}
where $g^R$ is a running coupling constant, and 
$Q(v)$ is the same as in (\ref{vert}) but with renormalized velocities 
$v^R_x$ and $v^R_y$ instead of $v_x$ and $v_y$. 
Observe that the coupling constant {\it increases} 
as $\xi \rightarrow \infty$.

At the QCP, the dependence on $\xi$ under the logarithm 
 transforms into the dependence
on frequency and external $q-Q$: 
\begin{equation}
L  \rightarrow (1/2) \log \frac{\bar \omega}{
\mbox{max}(|\omega|,({\bf q}-{\bf Q})^2/\gamma)}.
\end{equation}
Using the fact that for $\xi \rightarrow  \infty$,
 $v^R_y/v^R_x \approx N\pi(v_y/v_x)/(24 L)$  and expanding  $Q(v)$ 
near $v^R_y =0$, we find
 $Q(v) \approx 2 (1 - (2/\pi) v^R_y/v^R_x) = 2 - N (v_y/v_x)/(6 L)$
Substituting this result into 
(\ref{rgg}) and solving the differential equation we obtain in the limit
$L \gg N$
\begin{equation}
g^R = g~ \exp{(2 L/N)} L^{-v_y/6v_x}
 \label{gr}
\end{equation}
\begin{figure}[tbp]
\begin{center}
\epsfxsize=\columnwidth 
\epsffile{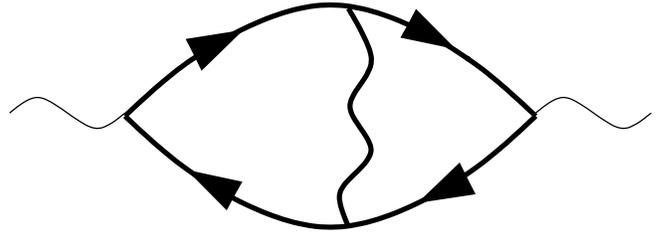}
\end{center}
\caption{ 
The polarization bubble with the vertex correction.}
\label{two-parts}
\end{figure}
We see that the logarithmic factor
 $L$ appears in the exponent, i.e.,  series of logarithmical 
corrections eventually yield power-law dependence of the full vertex on 
frequency and momentum. This result contradicts the 
assertion that the  quantum critical point 
is  described by a purely bosonic ``$\phi^4$'' theory at the upper 
critical dimension $d+z=4$ as in the latter case
all logarithmical corrections  should sum up into the powers of 
logarithms. We explain the origin 
 of the power-law behavior below, and in the next subsection 
compare  our results with the bosonic 
$\phi^4$ theory  in some detail. 
Here we just note that the very fact that $L$ appears in the 
exponent implies that we need to know precisely  in which combination 
$\gamma |\omega|$ and $({\bf q}-{\bf Q})^2$ 
appear in $L$. To address this issue,  
 we explicitly computed
the polarization bubble with a vertex correction 
 at a nonzero momentum and frequency, $\Pi_2 ({\bf q}, \omega)$. 
The corresponding diagram is 
 presented in Fig. \ref{two-parts}. We normalized  
 $\Pi_2 ({\bf q}, \omega)$ such that $\chi ({\bf q}, \omega) = 
\chi_0/(\xi^{-2} + ({\bf q}-{\bf Q})^2 + 
\gamma |\omega| + \Pi_2 ({\bf q},\omega))$.  The computation of the diagram 
 is presented in  Appendix C. The result is
\begin{eqnarray}
&& \Pi_2 ({\bf q}, \omega) = \nonumber \\
&& \frac{Q(v)}{N} (({\bf q}-{\bf Q})^2 + \gamma |\omega|) 
\log \left[\frac{\gamma {\bar \omega}}{
({\bf q}-{\bf Q})^2 + \gamma |\omega|}\right]
\label{dico}
\end{eqnarray}
We see that $\gamma\omega_{sf}$ and 
$({\bf q}-{\bf Q})^2$ are renormalized in exactly the same way. 
This result implies
that  the spin diffusion coefficient  
 {\it does not} undergo logarithmical renormalization, i.e., the 
dynamical exponent $z$ remains equal to two despite singularities. 
Earlier two of us missed the renormalization of the $({\bf q}-{\bf Q})^{2}$ 
term and 
erroneously concluded~\cite{ac_qcp}
 that $z$ becomes larger than $2$. 

That  both 
$\omega$ and $({\bf q}-{\bf Q})^2$ terms in the bosonic propagator 
undergo  logarithmical renormalizations 
requires some extra explanations. 
We argued earlier that to leading order in $1/N$, the spin polarization bubble
has a regular piece, which comes from high fermionic energies and is absorbed into 
a bare  static susceptibility, and an anomalous  $\gamma |\omega|$ term which comes from  fermionic energies which are smaller than $\omega$. 
If we 
linearize fermionic dispersion near $k_F$ and 
integrate first over momentum, and then over frequency, we only obtain 
$\gamma |\omega|$, the high-energy piece
 just does not appear.   
It is then tempting to conclude that the logarithmical vertex 
renormalization should yield  $\omega \log \omega$ term but 
no $({\bf q}-{\bf Q})^2\log |{\bf q}-{\bf Q}|$ term.
One should bear in mind, however, that
for fermions with a linearized dispersion, the polarization operator is 
 linearly  divergent at high energies. Due to divergence, 
 the result of the momentum/frequency 
integration depends on which integration is performed first. 
Indeed, evaluating the particle-hole bubble for free fermions by first 
integrating over frequency and then over momentum, 
we do obtain, in addition to $\gamma |\omega|$ an extra regular piece, 
of order of the upper cutoff in the momentum integral. 

Consider now what happens when we evaluate the same bubble with the 
logarithmical vertex correction.
Since we are interested in the terms which scale either as $\omega L$ 
or $({\bf q}-{\bf Q})^2 L$, we can subtract the result for
$\omega = ({\bf q}-{\bf Q})^{2} =0$. 
 The remaining momentum/frequency integral diverges only logarithmically, and
 hence the result of the integration no longer depends on which integration 
is performed first. 
This implies that while $\gamma \omega$ term in the ``zero-order'' spin propagator is an anomalous piece which cannot be obtained by expanding in $\omega$,
  the logarithmical corrections to the spin propagator 
are {\it not} anomalous and emerge due to the fact that $d=z=2$ case is marginal.
 [Alternatively speaking, these corrections can be found in a standard RG procedure which assumes that typical internal energies are much larger than the external ones.] 
  This reasoning explains why $({\bf q}-{\bf Q})^2$ and $\gamma |\omega|$ 
are renormalized in the same way. 
         
We now discuss why the series of logarithmical vertex corrections yield 
power-law momentum and frequency dependence 
of the renormalized vertex.  
Mathematically, this is a consequence of the fact that 
$\Delta g$ scales as $g^3/\gamma$. 
If $\gamma$ was independent of $g$, then $\delta g$ 
would be 
proportional to  $g^3$, i.e., 
$dg_R^2/dL \sim (g_R^2)^2$ or $1/(g^R)^2 \propto L$.
 This would result in a  logarithmical 
behavior of $g_R$.
In our case, however, bosonic dynamics is made by fermions, 
and $\gamma \propto g^2$. This
 yields $dg_R/dL \sim g_R$, i.e., $\log g_R \propto L$.  

The power-law renormalization of $g_R$ can be understood as follows.
 Suppose we perform calculations at arbitrary $d$.  
The vertex correction contains the product of two fermionic propagators and 
one bosonic propagator. As we discussed above, 
typical momenta in the fermionic propagators are much smaller than in the 
 bosonic propagator. Then one can perform a 2D integration
  over $d\epsilon_k d\epsilon_{k+Q}$ keeping the corresponding momentum dependence only in fermionic propagators.  This momentum integration
 transforms the two fermionic propagators into the constant densities
 of states. The vertex correction then reduces to 
 the evaluation of the frequency integral of $\chi (q, \omega)$
  in dimension $d-2$. The corresponding integral is logarithmically singular
 if $d-2$ is the {\it lower} critical dimension in the problem. As the 
lower critical dimension is the upper critical minus two, the vertex correction becomes logarithmically singular if $d$ is the upper critical dimension in the problem.  This is precisely what we have in our case.  
Now, at the lower critical dimension, there is no restriction as to what the series of logarithms should produce, and they can sum either into the inverse power of a logarithm (as in Heisenberg magnets) or into the exponent of a logarithm (as in XY model). It is therefore not surprising that we did find
 power-law renormalization of $g_R$ at the upper critical dimension. 
    
Note by passing that from general point of view, the spin-fermion model 
is characterized by two different space-time dimensions. 
 One is obviously $d+z_B = d+2$, where $z_B=2$ is 
the bosonic dynamical exponent. 
 The other space-time dimension is associated with the anomalous piece in 
$\Sigma ({\bf k}, \Omega)$. This piece
 is produced solely by fermions (bosons can be treated as static).
 For fermions,  
the dynamical exponent $z_F =1$ (typical fermionic momenta 
and typical fermionic frequencies scale with each other).
 Accordingly, the space-time dimension 
associated with the anomaly is $d+z_F = d+1$. For $d=2$, it is
 equal to $3$ and is below the upper critical dimension. 
 In particular, one can easily check that for arbitrary $d$, 
the anomalous self-energy 
at $\xi = \infty$  scales as $\Sigma (\Omega) \sim \Omega^{(d-1)/2}$ 
and accounts for non-Fermi liquid behavior
for all $d<3$.  
     
We now return to Eq. (\ref{gr}). Substituting 
$L = (1/2) \log({\bar \omega}\gamma/(\gamma |\omega| +({\bf q}-{\bf Q})^2))$, 
we obtain
that  at the QCP, the running coupling constant diverges as
\begin{eqnarray}
g^R &\propto& g~ (\gamma |\omega| + ({\bf q}-{\bf Q})^2)^{-1/N}\nonumber  \\
&\times &|\log{(\gamma |\omega| + ({\bf q}-{\bf Q})^2)}|^{-v_y/6v_x}
 \label{gr1}
\end{eqnarray} 
Substituting this result into 
the spin polarization operator we find that at the QCP, 
\begin{eqnarray}
&&\chi^{-1} ({\bf q}, \omega) \propto \left(\frac{g_R}{g}\right)^2 
(\gamma |\omega| + ({\bf q}-{\bf Q})^2) \nonumber \\
&&
\propto [\gamma |\omega| + ({\bf q}-{\bf Q})^2]^{\frac{N-2}{N}}~ 
|\log(\gamma |\omega| + ({\bf q}-{\bf Q})^2)|^{-v_y/3v_x} 
 \label{chan}
\end{eqnarray}
This result implies that vertex corrections give rise to 
anomalous scaling dimension $\eta$ 
for the  dynamical spin susceptibility, 
\begin{equation}
\chi^{-1} ({\bf q}, \omega) \propto  
(\gamma |\omega| + ({\bf q}-{\bf Q})^2)^{1-\eta}.
\label{eta}
\end{equation}
 Up to logarithmic corrections, $\eta = 2/N$. 
For $N=8$, $\eta =0.25$ and $1-\eta =0.75$. 

We also emphasize that  the renormalization of the bosonic propagator has little effect on fermions as 
 the spin susceptibility appears in the fermionic self-energy in the combination with two side vertices, i.e., as 
$g^2_R \chi ({\bf q}, \omega)$. We see from (\ref{chan}) that this combination retains its bare form 
$(\gamma |\omega| + ({\bf q}-{\bf Q})^2)^{-1}$. Accordingly, the only singular correction to fermionic self-energy comes from the logarithmical
 renormalization of $v_F$: 
\begin{equation}
\Sigma (\Omega) \propto |\Omega|^{1/2}~  
|\log \Omega|^{-1/2}                            \label{sian}
\end{equation}

\subsection{a comparison with $\phi^4$ theory}

The results of the previous subsection 
imply that the  $2D$ fermionic system at the
antiferromagnetic QCP is not described by the effective bosonic theory at the
 upper critical dimension. This  contradicts  
$\phi^4$  theory of the QCP developed mostly by Hertz~\cite{hertz}
 and Millis~\cite{millis} (for recent developments, see~\cite{lp}).  
 Hertz and Millis (HM)  departed from the same spin-fermion model as we study, 
but integrated out fermions in the effective action and obtained the effective Lagrangian for bosonic (spin) degrees of freedom. 
They conjectured that the expansion of the  action in powers of the bosonic variables 
$S_{q,\omega}$ yields  $\phi^4$ theory with the dynamical exponent $z=2$.
\begin{eqnarray}
&&S_{eff} = T \sum_{\omega, q} 
(r + {\bf q}^2 - i\omega) S_{q,\omega} S_{-q,-\omega} \nonumber \\
&& + b \sum_{\omega_i, q_i} S_{q_1,\omega_1}
 S_{q_2,\omega_2} S_{q_3,\omega_3} S_{q_4,\omega_4} 
\delta(\Sigma_i q_i)~
 \delta (\Sigma_i \omega_i) + ....
\label{hm1}
\end{eqnarray}
where $q$ measures a deviation from $(\pi,\pi)$, $r$ is a distance from the critical point, and $b$ is a constant. At $d=2$, the $\phi^4$ theory is at the upper critical dimension ($d+z=4$),
 and the bosonic self-energy is marginally irrelevant. This implies that 
for $r=0$ the self-energy only adds fractional power of 
$\log (\mbox{max}(\omega,q))$  to the bosonic propagator. One can  
check this by explicitly computing the self-energy order by order in $b$.

\begin{figure}[tbp]
\begin{center}
\epsfxsize=\columnwidth 
\epsffile{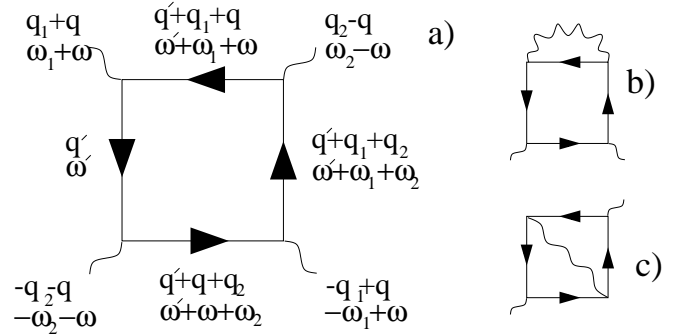}
\end{center}
\caption{ 
a) Diagrammatic representation of the four-fold interaction vertex
 between spin fluctuations. b) and c)
Two ways of contracting the external legs of the 
four-fold  vertex.}
\label{four-vertex}
\end{figure}
The bare bosonic propagator
in the HM theory is the same as our $N=\infty$ 
 result. We, however, dispute the assertion that $b$ can be approximated by a constant. 
Indeed, if this approximation is valid, then the fist-order correction in $b$ involves one bosonic loop and 
is infrared finite. We, however, found in the previous subsection 
that it is  logarithmically singular.
 
The four-boson vertex is  diagrammatically given by the graph 
presented in Fig.~\ref{four-vertex}. 
At external $\omega_i =q_i =0$, it becomes (we skip spin indices for a moment)
\begin{equation}
b = N {\bar g}^2 T \sum_{{\bf q}^\prime, \omega^\prime} 
G^2({\bf q}^\prime,\omega^\prime) 
G^2 ({\bf q^\prime+Q}, \omega^\prime).   
\label{hm2}
\end{equation}
Since 
\begin{equation}
\int d\epsilon_q^\prime G^2 ({\bf q}^\prime, \omega^\prime) =0,
\label{hm3}
\end{equation}
this vertex  vanishes if one linearizes the fermionic dispersion near the Fermi surface.  
The inclusion of the curvature of the fermionic dispersion (i.e., nonuniversal,  lattice effects) makes the momentum 
integral in (\ref{hm2}) and hence $b$ finite~\cite{millis}. 
The
 fact that $b$ is produced by high-energy fermions then  implies 
that it is insensitive to any modifications of the low-energy fermionic dynamics near quantum criticality and can 
be treated as a constant even at the critical point.
     
The potential problem with this reasoning is that although the momentum integral in (\ref{hm3}) indeed vanishes for a linearized dispersion, 
the integrand contains a double pole. 
At finite external $\omega_i$ and $q_i$, 
the double pole splits into two single poles which in a small frequency range of width $\omega_i$ are located in different half-planes. The integration over this range
 does not produce a smallness in $\omega_i$ as the two poles are only separated by a $\mbox{max} (\omega_i, v_F q_i)$. This implies that even for linearized fermionic dispersion,
 the value of $b$ is generally finite and  depends on the ratio of 
$\omega_i$ and $v_F q_i$. 

The evaluation of the four-boson vertex for finite external frequencies 
and momenta is straightforward. For simplicity, we restrict with      
free fermions. One can easily verify 
that the inclusion of the fermionic self-energy does not affect the results below.
Introducing the external momenta and frequencies as in Fig~\ref{four-vertex}a.
we obtain (see Appendix C, Eqn. (\ref{b}))
\begin{equation}\label{hm5}
b=\frac{Ng^{4}}{8\pi v_{x}v_{y}}
\frac{|\omega _{2}+\omega|+|\omega _{2}-\omega |
-|\omega _{1}+\omega |-|\omega _{1}-\omega | }
{[i(\omega _{1}+\omega _{2})-\epsilon _{q_{1}+q_{2}}]
[i(\omega _{1}-\omega _{2})-\epsilon _{Q+q_{1}-q_{2}}]}
\end{equation}
We  see that generally $b$ scales inversely with the incoming frequency,
but the coefficient  depends on the relative magnitudes
of external $\omega_i$ and $v_F q_i$.
This in particular implies that the bosonic self-energy
 depends on which external legs of the four-boson vertex 
are contracted.  There are two 
topologically different self-energy diagrams to first order in $b$. 
They are presented in Fig. \ref{four-vertex} b and c. 
If $b$ was independent of momentum and frequency 
(as is assumed in the HM theory),
 both diagrams would be equivalent and
 yield $b \int d^2 q d\omega \chi ({\bf q},\omega)$. 
This integral is nonsingular in $2D$ and only  
accounts for a nonessential renormalization of $\xi$. 
In our case, the result is different as we already know: 
the first diagram is equivalent to adding fermionic self-energy to the particle-hole bubble and vanishes if we restrict with linearized fermionic dispersion near the Fermi surface. 
 The second diagram is equivalent to adding 
a vertex correction to a 
 particle-hole bubble and is logarithmically singular.

For completeness, in Appendix C we explicitly computed 
both  diagrams by choosing
 the appropriate external frequencies in $b$, and
 convoluting $b$ with the spin propagator.
This computation indeed yields the same results as we just described,
 namely the diagram in Fig. \ref{four-vertex}b vanishes
while that in Fig. \ref{four-vertex}c yields the same logarithmical renormalization as (\ref{dico}).
 
We see therefore that the explicit computation of the {\it first-order}
bosonic self-energy in the effective $\phi^4$ model 
yields a logarithmically singular piece which comes from 
 a singular momentum/frequency dependence of the $\phi^4$ vertex at smallest energies.
 This singularity is missed if one 
 restricts with a constant four-boson vertex, as it is 
{\it assumed} in the $\phi^4$ theory.

A very similar computation has been performed by Lercher and Wheatley~\cite{wheatley}. They also argued that the four-boson vertex is singular at vanishing 
 external momenta and frequency, and depends on the ratio between momentum and frequency. Our result for the vertex slightly differs from theirs, but the conclusions are identical.
       
\subsection{summary of Sec.~\protect\ref{per_th}}

Let us now summarize our $1/N$ results. 

\begin{enumerate}
\renewcommand{\labelenumi}{\roman{enumi}}
\item
We found that 
the singular corrections to the
Fermi velocity cause nesting behavior 
at hot spots at $\xi = \infty$,  but
have little effect on the spin dynamics. These corrections are, however, only 
logarithmically singular at $\xi \rightarrow \infty$.
 The corrections to the vertex on the other hand do not affect Fermi velocity and fermionic self-energy, but give rise to a nonzero 
 anomalous scaling dimension $\eta = 2/N$ of the bosonic propagator. 
\item
We demonstrated that the co-existence  of the logarithmical renormalization of the Fermi velocity  and 
the anomalous dimension in the spin channel is due to the fact that 
the corrections to the spin-fermion vertex are structurally equivalent to
 the renormalization of the order parameter in dimension $d-2$. In the latter case, the logarithmical corrections are not necessary summed up into a series of logarithms.
\item
Our results disagree  with purely bosonic $\phi^4$ theory of the QCP. We argued
 that the difference is due to the fact that in our model, 
the four-boson vertex is not a constant but rather has a complex momentum/frequency dependence at the lowest energies. This is another indication of the presence of anomalies in the spin-fermion model. We explicitly demonstrated how our results can be obtained in the effective $\phi^4$ theory with 
full four-boson vertex made by fermions.    
\item
We also argued that from general perspective, the spin-fermion model 
has two different space-time dimensions, $d+2 =4$ and $d+1=3$. The first is
associated with the RG-type physics, and the second is associated with the anomalous piece in the self-energy which comes from low energies and cannot be obtained within RG procedure. For $2<d<3$, the bosonic dynamics at criticality is
 Gaussian, but fermionic self-energy scales as $\Sigma (\Omega) \propto \Omega^{(d-1)/2}$ and accounts for non-Fermi liquid behavior at the critical point.
\end{enumerate}
\section{Large $N$ theory at finite $T$}
\label{finiteT}
So far, our analysis was restricted to $T=0$. 
We now discuss how the fermionic self-energy and the spin polarization 
operator are  modified at finite $T$.

A conventional wisdom would be that at finite $T$, 
the system in the quantum-critical regime possesses $\omega /T$ scaling, 
i.e., fractional powers of frequency  are 
replaced, at small $\omega$, by the same powers of temperature. We will
see that this effect, indeed, takes place. We will also find, however, that
the fermionic self-energy in the quantum-critical regime 
contains an extra term which does not fit into 
$\omega/T$ scaling. The strength of this extra term depends on
the order of limits $\lambda \rightarrow \infty$ and $N \rightarrow \infty$.   
 
As before, we first present the results of the computations at $N=\infty$ and then extend the theory to finite $N$

\subsection{$N \rightarrow \infty$ limit} 
\subsubsection{spin polarization operator}

We begin with the spin polarization operator $\Pi ({\bf Q}, \omega)$ 
Replacing the integration over Matsubara frequencies in 
Eq. (\ref{Pinew}) by a summation 
as 
\begin{equation}
\int \frac{d \omega }{2 \pi }\rightarrow T \sum_n
\end{equation}
we obtain 
\begin{equation}
\Pi (\omega_m) \!\!=\! \frac{\pi T}{\omega_{sf}}\!\!~\sum_{n}\left( 1\!-\!
\frac{\Omega_n (\Omega_n + \omega_m)}
{\sqrt{(\Omega_n)^2~(\Omega_n + \omega_m)^2}}\right),    
            \label{piT} 
\end{equation}
where, as usual, 
$\omega_m = 2 \pi T m$, $\Omega_n = \pi T (2 n +1)$.
 The summation over $n$ is elementary and yields
\begin{eqnarray}
\Pi _{m}  &=& \frac{\pi T}{\omega_{sf}}~\sum_{n}\left\{ 1- 
{\rm sign} \left(\left[2 n+1\right] 
\left[2(n+m)+1\right]\right) \right\}                            \nonumber \\
& =& \frac{|\omega_m|}{\omega_{sf}}                             \label{piT2}
\end{eqnarray}
We see that at finite $T$, the spin polarization operator preserves exactly
the same form as at $T=0$. In particular, $\Pi ({\bf Q},0) =0$, 
i.e., at finite 
$T$, the spin correlation length 
does not acquire an extra  correction from low-energy fermions. 

\subsubsection{fermionic self-energy}

As the spin polarization operator does not depend on $T$, 
the 
fermionic self-energy is still given by 
Eqs. (\ref{i}) and (\ref{pertnew}).
At $N\rightarrow \infty$ we have 
\begin{eqnarray}
&&\Sigma ({\bf k}_{hs}, \Omega_m) = i \pi T \lambda 
~\sum_{n}~\frac{{\rm sign}\omega _{n}}
{\sqrt{1 + \frac{|\omega_{n-m}|}{\omega_{sf}}}} \nonumber \\
&& = i \pi T \lambda
\left[1  + \sum\limits_{n=1}^{|m|}\frac{1}
{\sqrt{1 + \frac{\omega_{n}}{\omega_{sf}}}}\right]
\mbox{sign}\Omega _{m}.   
\label{seT}
\end{eqnarray}
The extension to  momenta away from 
 a hot spot proceeds in the same way as at $T=0$:
 $\lambda$ and $\omega_{sf}$ are just replaced 
by $\lambda (k)$ and $\omega_{sf} (k)$
 given by (\ref{kdep}).
The full analysis of self energy as function of frequency 
and temperature is performed in the Appendix D. 
 For $|\Omega_m|, T \ll \omega_{sf} (k)$, the 
 self-energy obviously  has a
 Fermi-liquid form 
\begin{equation}
\Sigma ({\bf k}, \Omega_m) = i\lambda (k)~\left(\Omega_m + 
\frac{ (\pi T)^2 -\Omega^2_m}{4 \omega_{sf}(k)}~{\rm sign} \Omega_m \right)                 \label{seT2}
\end{equation}
For $ T \ll \omega_{sf} (k)$ and arbitrary $\Omega_m$, the thermal piece in the
 self-energy preserves $T^2$ dependence, but the prefactor depends 
on frequency:
\begin{equation}
\Sigma_T ({\bf k}, \Omega_m) = i\frac{\pi^2 T^2}{4 \omega_{sf} (k)} 
F\left(\frac{\Omega_m}{\omega_{sf}(k)}\right)~ \mbox{sign} \omega +O(T^4) 
\label{s11-1}   
\end{equation}
where $F(0) =1$, and $F(x \gg 1) \rightarrow 1/3$. The full expression, 
including $T^4$ terms is presented in (\ref{B8}).

In the opposite limit $|\Omega_m|, T \gg \omega_{sf} (k)$, 
the system behavior is quantum-critical.
In this limit, the contributions to the sum in 
 Eq. (\ref{seT}) from all 
$n \neq m$   can be explicitly 
expressed in terms of Riemann Zeta function 
(see Eq. (\ref{B7})). Combining this with the $n=m$  contribution from 
classical, static spin fluctuations, 
we obtain 
\begin{eqnarray}
&&\Sigma ({\bf k}, \Omega_m) = i\pi T \lambda (k) \nonumber \\
&& + i \left(\frac{\pi T {\bar \omega}}{2}\right)^{1/2}
~\left(\zeta\left(\frac{1}{2}\right) - 
\zeta\left(\frac{1}{2}, 1+m\right)\right)                         \label{seT22}
\end{eqnarray}
where $\zeta$ is a Zeta function. For $m \gg 1$, the second term scales as
$\sqrt{T m}$, i.e., as $\sqrt{\omega}$.

In real frequencies, the imaginary part of the self-energy takes the form
(see Appendix D)
\begin{equation} 
\Sigma^{\prime \prime} (\Omega) = \frac{\lambda }{2}Im
 ~\int _{-\infty }^{\infty }
\frac{d\omega  }{\sqrt{1 -i\frac{\omega }{\omega _{sf}}}}
f(\Omega /2T,\omega /2T)
\label{seT2222}
\end{equation}
where $f(x,y) = \tanh (x-y)+1/\tanh (y)$. 
In the Fermi liquid regime, we obviously have $\Sigma^{\prime \prime} (\Omega) \propto T^2$, with $\Omega$ dependent prefactor, as in (\ref{s11-1}). 
In the quantum-critical regime,
\begin{equation}
\Sigma^{\prime \prime} ({\bf k}, \Omega, T) = 
\pi T \lambda(k) + \left(\frac{T\bar{\omega }}{2}\right)^{1/2} 
D\left(\frac{\Omega}{T}\right).                                 \label{seT5}
\end{equation}
The scaling function $D(x)$ is given by (\ref{B9}). In the two limits,
$D(x \gg 1) = \sqrt{x}$, and $D(x \ll 1) = -1.516 + 0.105 x^2$.
 
The real part of $\Sigma ({\bf k}, \Omega)$ is obtained from (\ref{seT5})
by Kramers-Kronig transformation. It contains the same $\sqrt{T}$ dependence
as (\ref{seT5}) with another scaling function of $\Omega/T$, but does not
have a  $\lambda T$ term. This can be easily understood as static
spin fluctuations account for a scattering with a finite 
transferred momentum and zero transferred frequency, and therefore act as 
impurities. Not surprisingly, their contribution to the 
self-energy is purely imaginary.
 
We see from Eqs. (\ref{seT22}), (\ref{seT5})  
that the piece in the self-energy which comes 
  from  nonzero bosonic
Matsubara frequencies (i.e., from $n\neq m$ in (\ref{seT})) 
yields a conventional scaling form of $\Sigma ({\bf k}, \Omega, T)$  in which 
the $\sqrt{\Omega}$ behavior at $\Omega \gg T$ is collaborated by 
$\sqrt{T}$ dependence at $T \gg \Omega$, and the prefactor does not 
depend on the spin correlation length
 
On the other hand, the $n=m$ term in the sum in (\ref{seT}) 
yields a linear in $T$
contribution to the self-energy with the prefactor which scales as $\xi$
and diverges at the QCP. This term 
obviously violates the scaling. 
This violation is a direct  consequence of the fact that in 
our theory $\xi (T) = \xi (T=0)$. For the scaling behavior, one would needed
$\xi^2 (T) - \xi^2 (0) \propto T$.
 We  discuss this issue in a separate subsection below.
\begin{figure}[tbp]
\begin{center}
\epsfxsize=\columnwidth
\epsffile{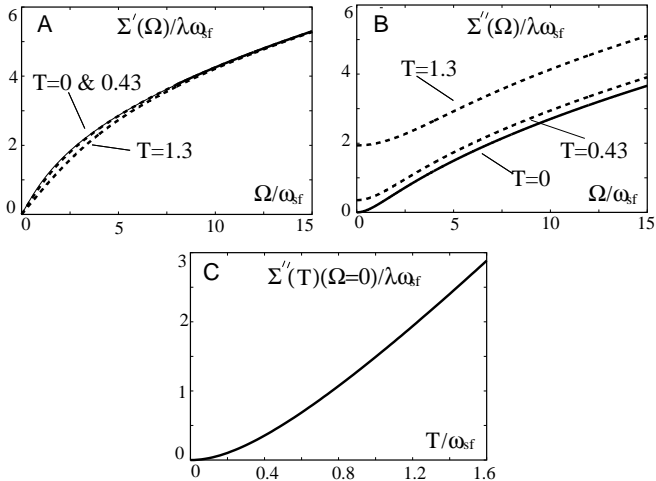}
\end{center}
\caption{ 
The results for the fermionic self-energy at $N\rightarrow \infty $.
}
\label{sigmaT-rob}
\end{figure}

Finally, at $\Omega =0$,
$\Sigma^{\prime \prime} (T)$ has a simple form which 
interpolates between Fermi liquid
and quantum-critical regimes. Taking the $\Omega =0$ limit in 
 (\ref{seT2222}) we obtain
\begin{equation}\label{B81ea} 
\Sigma^{\prime \prime}(\Omega =0) = \pi T \lambda L\left(\frac{T}{\omega_{sf}}\right)
\end{equation}
where
\begin{equation}
L(x) = \frac{1}{\pi x} \int d y Im \frac{1}{\sqrt{1-iy}} \frac{1}{\sinh(y/x)}
\label{B81eea}
\end{equation}
At $x \gg 1$, $L(x) \approx 1$, i.e., $\Sigma^{\prime \prime} (\Omega =0) \approx \pi T \lambda$. In the opposite limit of small $x$, $L(x) = \pi x/4$, 
and we recover the Fermi liquid behavior.

In Fig. \ref{sigmaT-rob} we plot the $N \rightarrow \infty$ result for the 
self-energy at finite $T$. We see
that at small frequencies, ${\rm Im} \Sigma (k, \Omega)$ is linear in $T$
starting already from $T \geq 0.5 \omega_{sf}$. 
At higher frequencies, particularly in the region $\Omega \gg T$,
 ${\rm Im} \Sigma^{ret} ({\bf k}, \Omega)$ is linear 
in $\omega$ with almost $T-$independent slope. A finite  temperature only 
gives rise to  an extra
constant term in ${\rm Im} \Sigma^{ret} ({\bf k}, \Omega)$.    

\subsection{beyond the $N \rightarrow \infty$ limit}

We now consider to which extent we can trust 
the $N \rightarrow \infty$ results for the fermionic self-energy and
the spin polarization operator,
when we apply them to the physical $N=8$. 
We will show that there are two  problems at finite $N$ which are only partly 
 related: 
 one is how strong are vertex corrections, 
the effect of keeping the self-energy of intermediate fermions,
and the momentum dependent piece in the fermionic self-energy (i.e., the renormalization of the Fermi velocity),
 and another one is to what accuracy one can factorize the momentum integration in deriving the equation for the fermionic self-energy. [By factorization we mean that in the diagram for fermionic $\Sigma (k, \Omega)$ one integrates
over the momentum component along the Fermi surface in the bosonic propagator 
and over the momentum component  transverse to the Fermi 
surface only in the fermionic propagator. The resulting
 self-energy is then the convolution of the
effective 1D ``local'' spin susceptibility and the fermionic density of states].

\subsubsection{accuracy of the factorization of the momentum integration
 at finite $T$}

Let's first discuss the second issue. Consider the second-order diagram for the fermionic self-energy.
At $T=0$, the factorization of the momentum integration in this diagram,
 or, equivalently, the
possibility to neglect in the spin susceptibility  the momentum component 
transverse to the Fermi surface,
 is well justified for all relevant frequencies as the correction to the self-energy from keeping the momentum dependence in  the susceptibility has an extra power of $\Omega$ in the Fermi liquid regime and is small by  $1/N^2$ in the quantum-critical regime. 
For $T\neq 0$, analogous result holds for the  $n \neq m$ 
contribution to the self-energy, and the proof almost exactly parallels the one for $T=0$.  The situation is, however,
different for the $n=m$ contribution to (\ref{seT}). 
This term  comes from static spin 
fluctuations, and hence typical $({\bf q}-{\bf Q})_x \sim 
|\omega|/v_F$
has to be compared to $\xi^{-1}$ rather than to 
$(\gamma \omega)^{1/2}$ ($x$ axis is chosen 
transverse to the Fermi surface at ${\bf k}_{hs} +{\bf Q}$). 

The new frequency scale  $v_F \xi^{-1}$ is
related to ${\bar \omega}$ and $\omega_{sf}$ as
\begin{equation}
v_F \xi^{-1} = {\bar \omega} \frac{v^2_F}{v_x v_y} \frac{N}{12\lambda} = 
\omega_{sf} \frac{v^2_F}{v_x v_y} \frac{N \lambda}{3}
\end{equation}
We see that the location of this scale compared to ${\bar \omega}$ 
depends on the 
ratio $N/\lambda$.  In the formal $N \rightarrow \infty$ limit at finite 
$\lambda$, $v_F\xi^{-1} \gg {\bar \omega}$. Then, even at $\omega \sim {\bar \omega}$, $|\omega| \ll v_F \xi^{-1}$, 
and the factorization of momentum integration is well justified. Alternatively, however, 
one can keep $N$ finite and study the system behavior
at $\lambda \rightarrow \infty$ and finite $T$. 
In this limit, $v_F \xi^{-1} \ll {\bar \omega}$, and the corrections to 
the formal $N= \infty$ result for the 
static, thermal self-energy are relevant at frequencies 
$|\omega|\sim  {\bar \omega}(N/\lambda) \ll {\bar \omega}$. 
 In reality, for $N =8$ and $\lambda \geq 1$, 
$v_F \xi^{-1} \leq {\bar \omega}$, i.e., the modification of  
the static piece  in the fermionic self-energy is a 
modest effect for $\omega \leq {\bar \omega}$. 
Note also that this modification is only relevant in the
 quantum-critical regime as in the Fermi liquid regime, $|\omega| \leq \omega_{sf}$, and hence 
$|\omega|/(v_F \xi^{-1}) \leq \omega_{sf}/(v_F \xi^{-1}) = O(1/N)$. 

\begin{figure}[tbp]
\begin{center}
\epsfxsize=\columnwidth 
\epsffile{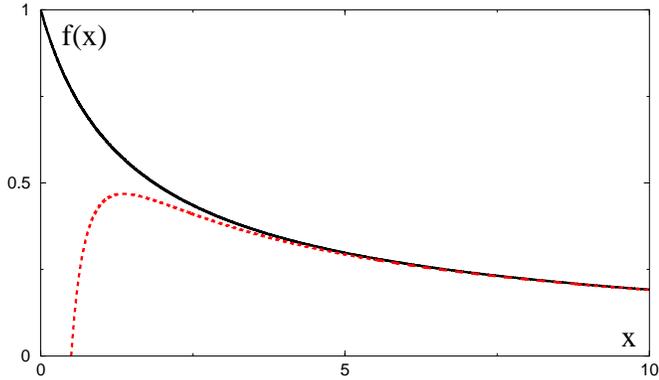}
\end{center}
\caption{ 
Function $f(x)$ defined in (\ref{seT8}). Dashed line is the asymptotic
form $2\log{2x}/(\pi x)$, valid at large $x$
}
\label{f}
\end{figure}

The exact form of $\Sigma (\Omega_m)$  at a hot spot, 
 can be obtained 
from the general formula for the self-energy, 
Eqs. (\ref{i}) and (\ref{pertnew}), extended to a finite $T$.
We separate the contributions from dynamical and static spin fluctuations 
and write $\Sigma (\Omega_m) = \Sigma_{st} (\Omega_m) + \Sigma_{dyn} (\Omega_m)$. As we said, $\Sigma_{dyn}$ can be safely computed at $N = \infty$ as the
 corrections scale as regular powers of $1/N$. 
The result is the second term in Eq. (\ref{seT}):
\begin{equation}
\Sigma_{dyn} (\Omega_m) = i \pi T \lambda 
 \sum\limits_{n=1}^{|m|}\frac{1}
{\sqrt{1 + \frac{\omega_{n}}{\omega_{sf}}}}
\mbox{sign}\Omega _{m}.
\label{ff} 
\end{equation}
In the quantum-critical regime this reduces 
to the second term in Eq. (\ref{seT22})
\begin{equation}
\Sigma_{dyn} (\Omega_m)\! =\! i
\left(\frac{\pi T {\bar \omega}}{2}\right)^{1/2}\!\!\!\!\!\!
~\left(\zeta\left(\frac{1}{2}\right) - 
\zeta\left(\frac{1}{2}, 1+m\right)\right)  
\label{fff}
\end{equation} 
The static contribution has to be evaluated without approximations and reduces to
\begin{equation}
 \Sigma_{st} (\Omega_m) = i~ \lambda T \pi f
\left({\tilde \Omega_m}\right)                                   \label{seT7}
\end{equation}
where ${\tilde \Omega_m}$ = $|\Omega_m|/v_F\xi^{-1}$ and
\begin{equation}
 f\left({\tilde \Omega_m}\right) = \frac{2}{\pi} \frac{
\cos ^{-1} \tilde \Omega_m}
{\sqrt{1 
- {\tilde \Omega_m}^2}}                                           \label{seT8}
\end{equation}
The limiting values of $f(x)$ are
 $f(0) =1$, $f(1) = 2/\pi$, $f(x \gg 1) \approx 
2 \log {2x}/(\pi x)$. The plot of $f(x)$ is presented in Fig. \ref{f}. 
We see that $\Sigma_{st} (\Omega_m)$ coincides with the $N = \infty$ result 
$i \pi \lambda$ only at vanishing ${\tilde \Omega_m}$. At finite 
${\tilde \Omega_m}$, it is smaller than for $N = \infty$, and
 for ${\tilde \Omega_m} \gg 1$, 
$\Sigma_{st} (\Omega_m)$ 
decreases as $1/{\tilde\Omega}_m$, up to logarithmical corrections. In this limit, the 
overall factor in the fermionic self-energy
 scales as $\lambda v_F \xi^{-1} \sim {\bar g}$, i.e., it no 
longer depends on $\xi$. This is indeed what one should expect in the truly 
quantum-critical regime.
\begin{figure}[tbp]
\begin{center}
\epsfxsize=\columnwidth 
\epsffile{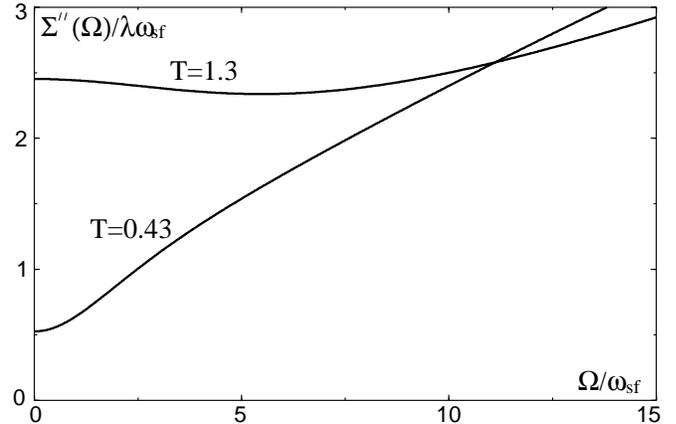}
\end{center}
\caption{ 
The  modified second-order fermionic self-energy - the sum of 
Eqs. (\protect \ref{seT22222}) and (\protect\ref{seT101}). 
It includes the full contribution from  thermal, static 
spin fluctuations, and the $N \rightarrow \infty$ contribution from
dynamical spin fluctuations. 
We used  $v_x = v_y = v_F/\sqrt{2}$, and $\lambda =1$,
in which case $v_F \xi^{-1} = 4 {\bar \omega}/3 = 16 \omega_{sf}/3$.}
\label{sigmaT}
\end{figure}

In real frequencies,  
$\Sigma^{\prime \prime} ({\bf k}_{hs}, \Omega)=
 \Sigma^{\prime \prime}_{st} (\Omega) + 
\Sigma^{\prime \prime}_{dyn} (\Omega)$, where 
$\Sigma^{\prime \prime}_{dyn} (\Omega)$ is the same as in 
(\ref{seT2222}) but with  
$f_1 (x,y) =\tanh (x-y)+1/\tanh (y) -1/y$ instead of $f(x,y)$:
\begin{equation} 
\Sigma^{\prime \prime}_{dyn} (\Omega) = \frac{\lambda }{2}Im
 ~\int _{-\infty }^{\infty }
\frac{d\omega  }{\sqrt{1 -i\frac{\omega }{\omega _{sf}}}}
f_1(\Omega /2T,\omega /2T).
\label{seT22222}
\end{equation}
For $\omega > \omega_{sf}$, this reduces to the second term in 
(\ref{seT5}):
\begin{equation}
\Sigma^{\prime \prime}_{dyn} (\Omega) = 
\left(\frac{T\bar{\omega }}{2}\right)^{1/2} 
D\left(\frac{\Omega}{T}\right).                                 \label{seT55}
\end{equation}
where $D(x)$ is given by (\ref{B9}).

The static contribution is obtained by transferring 
(\ref{seT7}), (\ref{seT8}) onto real axis:
\begin{equation}
\Sigma^{\prime \prime}_{st} (\Omega)= 2 T \lambda {\rm Im} 
\frac{\log\left[{\tilde \Omega}^2 + \sqrt{1 + {\tilde \Omega}^2}\right]}
{ \sqrt{1 + {\tilde \Omega}^2}} 
\label{seT101}
\end{equation} 
For ${\tilde \Omega} \ll 1$, 
$\Sigma^{\prime \prime}_{st} (\Omega) \propto (T/\Omega) 
\log {\Omega/v_F \xi^{-1}}$. We see that the prefactor depends on $\xi$ only in the logarithmical term. This is indeed the same result as
 we found earlier by evaluating the thermal self-energy in Matsubara frequencies.
 
In Fig. \ref{sigmaT}, we plot $\Sigma^{\prime \prime} ({\bf k}_{hs}, \Omega)$
obtained by combining the 
the $N=\infty$ form of $\Sigma_{dyn}$ (Eqn. (\ref{seT22222})) and
 the modified form of $\Sigma^{\prime \prime}_{st}$ (Eqn. (\ref{seT101})).
 Comparing this figure
with Fig. \ref{sigmaT-rob} where we plotted the formal 
 $N =\infty$ result,  
 we observe that the main difference is in the temperature variation of the 
self-energy at high frequencies. 
At $N = \infty$, the plots at different $T$ differ by a constant. 
In the modified expression, the thermal piece in   
$\Sigma^{\prime \prime} ({\bf k}, \Omega)$ 
decreases with frequency, and the curves at different $T$  
progressively come closer to each other.

\begin{figure}[tbp]
\begin{center}
\epsfxsize=\columnwidth 
\epsffile{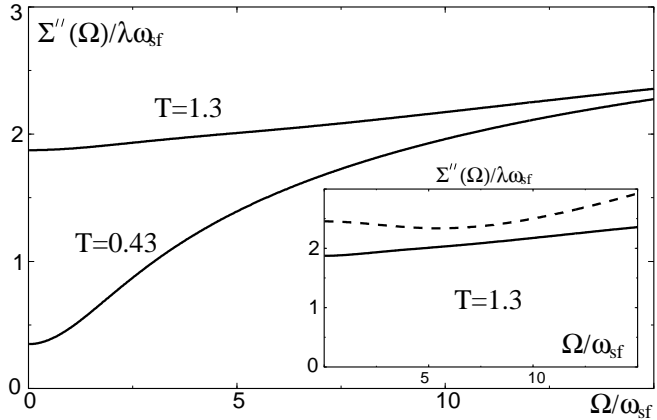}
\end{center}
\caption{ 
The  full second-order result  for the 
$\Sigma^{\prime \prime} (k, \Omega)$ obtained without factorizing the momentum integration in both static and dynamics pieces in the self-energy. 
In the insert the same data for  $T=1.3$ (solid line) is compared with 
the approximate form of the self-energy from
Fig.\protect\ref{sigmaT} (dashed line).
 $v_F \xi^{-1}$ is the same as in Fig.\protect\ref{sigmaT}.}
\label{sigmaT_exact}
\end{figure}

To verify how accurate is 
our ``modified $N = \infty$''  form of the self-energy 
 we present in  Fig.~\ref{sigmaT_exact} the result for the  full  
$\Sigma^{\prime \prime}$,  which we obtained in Appendix D 
(Eqn (\ref{Bbbbb81}) with ${\tilde \Sigma}^2_{R,A} = \Omega - \omega \pm i \delta$)
 without factorizing the momentum integration in both, static and dynamical parts of $\Sigma^{\prime \prime}$. Comparing Figs.~\ref{sigmaT} and 
 ~\ref{sigmaT_exact} we find that the full result is rather close to our approximate one - the only difference is that now he curves at various $T$ progressively come close to each other with increasing frequency, and cross only at very high frequencies.
  
\begin{figure}[tbp]
\begin{center}
\epsfxsize=\columnwidth 
\epsffile{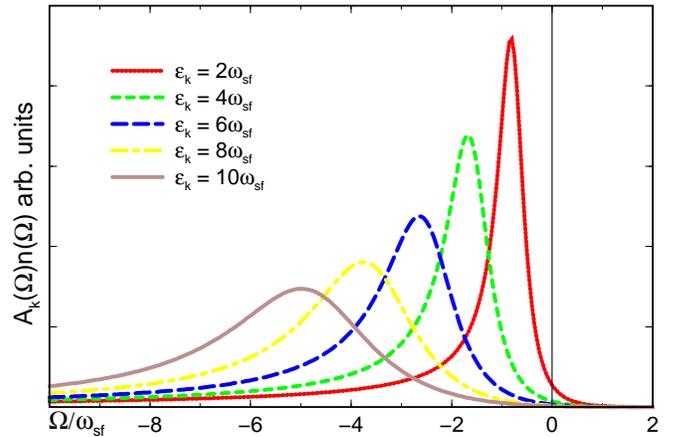}
\end{center}
\caption{ 
The  photoemission intensity $I (\Omega) \propto Im G({\bf k}, \Omega) n_F (\Omega)$ for various $\epsilon_k$. For definiteness we used 
$\lambda =1.7$ and $T = 0.43 \omega_{sf}$.
}
\label{spectr_T}
\end{figure}

In Fig.~\ref{spectr_T} we present the set of the results for the 
photoemission intensity
$I (\Omega) \propto Im G({\bf k}, \Omega) n_F (\Omega)$ for various $\epsilon_k$. They were 
obtained using the spectral 
functions from Fig. \ref{sigmaT_exact}. 
For definitness, we used $\lambda =1.7$ and $T = 0.43 \omega_{sf}$.
 Comparing  these forms with the corresponding results at $T=0$ (see Fig.~\ref{spectr_T0}), we
 see that a finite temperature gives rise  to an additional broadening of the photoemission intensity at small $\Omega \sim \omega_{sf}$, and plays virtually no role for $\Omega \gg \omega_{sf}$. 

\subsubsection{the role of the self-energy of intermediate fermions}

In the above analysis we restricted with the second order self-energy diagram. 
Now we consider what  happens if we keep the self-energy of intermediate fermions (i.e., evaluate the self-energy diagram as in Fig. (\ref{Ninfinity})).
Recall that at $T=0$, the self-energy of intermediate fermions is not small in $1/N$ but nevertheless doesn't play any role
 as the momentum integration can be factorized, and   
$ \Sigma (\omega)$ is expressed  in terms of the  density of states of intermediate fermions.  For the linearized fermionic dispersion, the
 latter is independent on the self-energy and is the same as for free fermions. This implies that the full expression
 for the self-energy coincides with the second-order result.
 At finite $T$, this is no longer the case 
 as the momentum integration in the static piece in $\Sigma (\Omega)$ cannot 
 be factorized. 

In principle, the self-energy of intermediate fermions should include
both $T=0$ and temperature contributions.   
We will show below that the temperature  piece in the self-energy for intermediate fermions becomes relevant for $T \gg N \omega_{sf}$ when also vertex 
corrections and the renormalization of $\epsilon_k$ cannot 
 be neglected. To separate thermal and quantum effects, 
we focus first on $T \ll N \omega_{sf}$ when 
the vertex corrections and the corrections to $\epsilon_k$ are the same as at $T=0$, i.e., are small in $1/N$. 
At these temperatures, the
 self-energy of intermediate fermions 
can be approximated by its $T=0$ value.
 The latter, in turn, is 
well approximated by  the $N = \infty$ result, 
$\Sigma_{T=0} (\Omega) = 2 \lambda 
\Omega/(1 + (1 - i \Omega/\omega_{sf})^{1/2})$.  

\begin{figure}[tbp]
\begin{center}
\epsfxsize=0.9 \columnwidth 
\epsffile{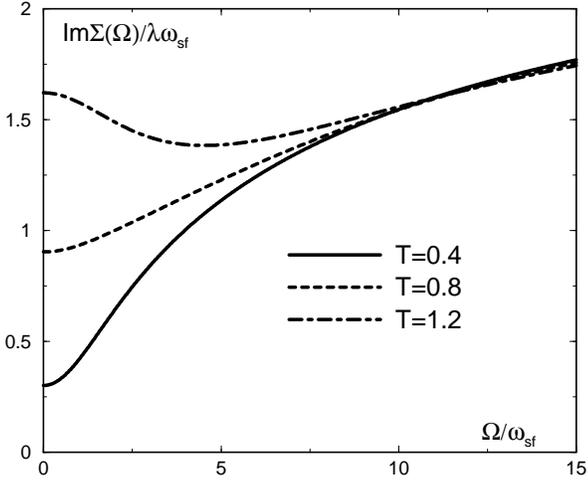}
\end{center}
\caption{ 
The  full result  for the 
$\Sigma^{\prime \prime} (k, \Omega)$ (Eqn. \protect\ref{Bbbbb81}).
The form of $\Sigma^{\prime \prime} (\Omega)$  at low temperatures
 is close to the second-order result plotted in Fig.\protect\ref{sigmaT_exact}.
 The initial decrease of $\Sigma^{\prime \prime} (\Omega)$ at higher $T$ 
is probably the artifact associated with the neglect of vertex corrections. 
 $v_F \xi^{-1}$ is the same as in Fig.\protect\ref{sigmaT}.}
\label{sigmaT_exact_2}
\end{figure}

The full expression for the static part of $\Sigma^{\prime \prime}$ is the same as in Eq. (\ref{seT101}), but now ${\tilde \Omega} = 
(\Omega + \Sigma_{T=0} (\Omega))/v_F \xi^{-1}$. 
In the quantum-critical regime, $\Omega < {\bar \omega}$,
 $\Sigma_{T=0} (\Omega)$ exceeds $\Omega$, and
$\tilde{ \Omega}$ becomes $i \Omega/{\bar \Omega}$ where ${\bar \Omega} = 
(v_F \xi^{-1})^2/{\bar \omega} = N^2 \omega_{sf} (v^2_F/v_x v_y)/36$. 
Then, when ${\bar \Omega} < \Omega < {\bar \omega}$, the static piece 
 in the self-energy behaves as 
 $\Sigma^{\prime \prime}_{st} (\Omega) \propto T N \sqrt{{\bar \omega}/\Omega} \log {\Omega/{N^2 \omega_{sf}}}$. We see that up to a logarithmical prefactor,
$\Sigma^{\prime \prime}_{st} (\Omega)$ acquires the same scaling form as 
the dynamical piece in the fermionic self-energy. In other words, at high enough frequencies, we eventually recover $\Omega/T$ scaling.

In Fig.~\ref{sigmaT_exact_2} we plot 
the full expression for $\Sigma^{\prime \prime}$ obtained in the Appendix D,
 (Eqn. (\ref{Bbbbb81})) 
 without factorizing the momentum integration in both static and dynamic parts 
and keeping the $T=0$, $N= \infty$ self-energy for intermediate fermions. 
 Comparing this figure with the second-order result, Fig.\ref{sigmaT_exact},
 we see that at low $T$, the two forms of 
$\Sigma^{\prime \prime} (\Omega)$ are close to each other, as they indeed 
should be, as at $T=0$ keeping the self-energy of intermediate fermions 
only yields regular $1/N$ corrections. At higher temperatures, $\Sigma^{\prime prime}$ in Fig.~\ref{sigmaT_exact_2} differs from the second-order result 
 at high frequencies, as we found analytically, but it also 
 initially decreases with frequency due to  a stronger downturn renormalization of the static piece then in the second order expression. 
      
This initial decrease  poses a dilemma in selecting the theoretical form of $\Sigma^{\prime \prime}$ for experimental comparisons. From theory perspective, the $T=0$ self-energy of intermediate fermions is not small in $1/N$ and therefore cannot be
 neglected in evaluating $\Sigma^{\prime \prime}_{st} (\Omega)$ for which 
 the momentum integration is not factorized. The restoration of 
 the scaling behavior at finite $T$ due to this self-energy is also a good sign. On the other hand, one can include the self-energy of intermediate fermions 
 but neglect  vertex corrections and the momentum dependence of the self-energy
 only as long as $T < N \omega_{sf}$ (see below). Up to which temperature
 this can be safely done for a physical $N = 8$ is a'priori unclear.
 It looks to us that the initial drop in 
$\Sigma^{\prime \prime} (\Omega)$ in Fig. ~\ref{sigmaT_exact_2} at $T > \omega_{sf}$ is an artifact of overestimating the 
reduction of $\Sigma^{\prime \prime}_{st} (\Omega)$, and should be compensated by vertex corrections. This reasoning implies that for $T > \omega_{sf}$ and $N=8$, 
the second-order result for the self-energy, 
 Fig.~\ref{sigmaT_exact}, may be closer to reality than 
  Fig.~\ref{sigmaT_exact_2}, which explicitly includes the self-energy of intermediate fermions but neglects vertex corrections. In Sec~\ref{summary} we will
 be using the second-order result for the comparisons with the photoemission and optical data. We stress, however, that for $T \leq \omega_{sf}$, the difference between the two results is minor for all experimentally relevant 
 frequencies. In particular, we verified that optical conductivities 
obtained using the second-order and the full self-energy 
 are almost identical for $T \leq \omega_{sf}$.

We now return to the theoretical discussion and 
 consider what happens at
$T \gg N \omega_{sf}$. As we already said, this temperature range is of little use for experimental comparisons with cuprates, at least near optimal doping. 
However, it is important for the understanding of what is the correct 
form of the self-energy at a finite $T$ at the QCP, when 
$\omega_{sf} \rightarrow 0$. Recall in this regard that so far we still have
$\Sigma^{\prime \prime}_{st} (\Omega =0) = \pi T \lambda \propto \xi$. 
On the other hand, from general scaling arguments~\cite{NFL_at_crit_general}
we expect that $\Sigma^{\prime \prime}_{st} (\Omega)$ in the 
quantum-critical regime should be independent on $\xi$. 

\subsubsection{the renormalization of the Fermi velocity}

We begin with the velocity renormalization. 
To estimate it,  we need to analyze the
strength of the static
piece in the self-energy $\Sigma_{k, \Omega_m}$ at  
${\bf k} \neq {\bf k}_{hs}$ and 
$\epsilon_{k+Q} \gg \Omega_m$. Performing the same computation as 
in the previous subsection we obtain
 after simple manipulations, 
\begin{eqnarray}
&& \Sigma ({\bf k}, \Omega_m) - \Sigma ({\bf k}_{hs}, \Omega_m)
 =  \nonumber \\
&&  2 T \lambda (k)
\log \left[\left(\epsilon_{k+Q} + \sqrt{\epsilon^2_{k+Q} + 
v^2_F \xi^{-2}}\right)/(v_F \xi^{-1})\right]                \label{seT9}
\end{eqnarray}
 At small $\epsilon_{k+Q}$,
expanding under the logarithm and using the fact that 
$\epsilon_k = v_x k_x + v_y k_y$ and $\epsilon_{k+Q} = - v_x k_x + v_y k_y$,
 we obtain 
\begin{eqnarray}
v^{R}_{x}&=&v_{x}(1 + \delta)       \nonumber \\
v^R_y &=& v_y (1-\delta)            \nonumber \\
\delta &=& \frac{2 v_xv_y}
{3 v^2_F}~\frac{T}{N \omega_{sf}}~                      \label{seT10}
\end{eqnarray}
We see therefore 
that velocity renormalization becomes relevant at 
$T  \sim N \omega_{sf}$, as we anticipated.

\subsubsection{vertex corrections}

The result for the vertex renormalization parallels the one for the 
self-energy. We can split $\Delta g/g$ into the contributions from static 
and dynamic
spin fluctuations. The contribution from dynamical spin fluctuations scales
as $(1/N) \log T$ and is small for $T \leq {\bar \omega}$ that we are 
interested in. The contribution from static spin fluctuations is larger 
due to the absence of the large $\gamma \omega$ term in the spin 
susceptibility.  Integrating over
momenta of two fermionic propagators in the vertex correction diagram 
 in the same way as at $T=0$, we obtain 
\begin{equation}
\left(\frac{\Delta g}{g}\right)_{st} = 
\frac{2T}{N \omega_{sf}}~{\rm tan}^{-1}  \frac{v_x}{v_y}                                                   \label{vT1}
\end{equation}
This expression is valid for small external $\omega < v_F \xi^{-1}$. At
larger frequencies, $(\Delta g/g)_{st}$ decreases and eventually scales
as $({\bar g} T/ \omega^2) |\log T|$.

We see from (\ref{vT1}) that vertex corrections also become relevant at 
  $T > N \omega_{sf}$.

\subsubsection{the thermal self-energy of intermediate fermions}

Finally, we show that at $T > N \omega_{sf}$ 
 we also have 
 to keep the thermal piece in the self-energy for intermediate fermions,
The easiest way to estimate this temperature is to consider the 
limit $\Omega =0$. At vanishing frequency, the
 $T=0$ piece in the fermionic self-energy vanishes, and we are 
left with only 
$\Sigma_{st} (T)$. At $N = \infty$, $\Sigma_{st} (T) = \pi T \lambda$. 
At finite $N$, this expression is valid if $\Sigma_{st} (T)
 < v_F \xi^{-1}$, or $T < N \omega_{sf}$. At larger $T$, the thermal piece
in the self-energy for intermediate fermions cannot be neglected, at least at 
 small frequencies.

\subsection{self-consistent theory}

We are not familiar with any controlled way to compute the static 
piece in the self-energy at finite $N$,  finite $T$, 
and $\xi \rightarrow \infty$ such that $T \gg N \omega_{sf}$. 
 As an exercise, 
we performed the FLEX-type computation of $\Sigma_{st} (\Omega, T)$ at vanishing $\Omega$. Specifically,  we 
included the full static fermionic self-energy into the 
 self-energy diagram, but neglected vertex corrections and the renormalization of the Fermi velocity. We solved the 
self-consistent equation, obtained $\Sigma_{st} (T)$, and 
 then re-calculated the vertex and velocity corrections 
with the new fermionic propagators. We found that the self-energy no 
longer diverge at the critical point, and that the vertex and velocity 
corrections are at most logarithmical
even when $T \gg N \omega_{sf}$. 
While this analysis is by no means a controllable one, it describes 
the physically appealing effect that the divergence in the self-energy 
and in the vertex are eliminated by thermal dressing of intermediate 
fermionic quasiparticles. 

 The FLEX-type computation of $\Sigma_{st} (T)$ reduces to the solution of the self-consistent equation in the form  
\begin{equation}
{\tilde \Sigma}_{st} (T) = \frac{3 T}{ N {\omega_{sf}}} \int_0^{\infty} 
\frac{dx}{\sqrt{x^2 +1}} \frac{1}{\sqrt{x^2 +1} + {\tilde \Sigma}_{st} (T)}
\label{sT1}
\end{equation}
where ${\tilde \Sigma}_{st} = \Sigma_{st}/(v_F \xi^{-1})$
The renormalization of the Fermi velocity and the vertex 
correction depend on ${\tilde \Sigma}_{st} (T)$ as  
\begin{eqnarray}
\frac{\Delta \epsilon_k}{\epsilon_{k+Q}} 
&=& -\frac{3 T}{ N \omega_{sf}} \int_0^{\infty} 
\frac{dx}{\sqrt{x^2 +1}} \frac{1}{(\sqrt{x^2 +1} + 
{\tilde \Sigma}_{st} (T))^2}                               \nonumber \\
\frac{\Delta g}{g} &=& \frac{T}{N \omega_{sf}} \int_0^{\infty}
 \frac{dx}{\sqrt{x^2 +1}} 
\frac{ {\tilde \Sigma}_{st} (T)}{(x^2 +{\tilde \Sigma}^2_{st} (T))}\nonumber \\
&\times &\frac{1}{
(\sqrt{x^2 +1} + {\tilde \Sigma}_{st} (T))} 
\label{sT2}
\end{eqnarray}
For simplicity, we set $v_x = v_y = v_F/\sqrt{2}$. The integral in (\ref{sT1})
is explicitly evaluated in the Appendix D (see Eqn (\ref{Bd81})).
 The integration in (\ref{sT2})
can also be performed exactly, but the results are
 too lengthy to present here.

At $T \ll N \omega_{sf}$, the self-energy in the denominator can 
be neglected, and we obtain $\Sigma_{st} (T) = \pi T \lambda$. For these $T$, 
\begin{equation}
\Delta \epsilon_k = - \frac{3 T}{N\omega_{sf}}~ \epsilon_{k+Q}; ~~
 \frac{\Delta g}{g} = \frac{\pi T}{2 N \omega_{sf}}.
\end{equation} 
In the opposite limit, solving (\ref{sT1}) with logarithmical 
accuracy, we obtain 
\begin{equation}
\Sigma_{st} (T) = {\bar \omega}
 \left(\frac{T N}{6 {\bar \omega}}\right)^{1/2}
 \left(\log{\frac{3 T}{N \omega_{sf}}}\right)^{1/2}
\label{sT3}
\end{equation}
At these temperatures
\begin{eqnarray}
\frac{\Delta g}{g} &=& \frac{T}{ N \omega_{sf}} 
\frac{\log {{\tilde \Sigma}_{st} (T)}}{{\tilde \Sigma}^2_{st} (T)} = \frac{1}{3} \nonumber \\
\Delta \epsilon_k &=& - \epsilon_{k+Q} 
\left(1 - \frac{1}{\log {{\tilde \Sigma}_{st} (T)}}\right)
\label{sT4}
\end{eqnarray}

Analyzing these results, we find that self-consistent solution of 
the thermal problem yields the results which are qualitatively 
similar to that at $T=0$. 
Namely, the 
 thermal piece in the fermionic self-energy scales as $T^{1/2}$, up to
a logarithmical prefactor, the velocities 
at ${\bf k}_{hs}$ and ${\bf k}_{hs} +{\bf Q}$ logarithmically 
tend to become antiparallel to each other, and vertex correction 
remains finite at criticality. The self-consistent solution indeed 
does not capture full 
logarithmical physics, nevertheless it clearly 
demonstrates that higher order diagrams eliminate the 
unphysical $O(\lambda)$ divergence in the fermionic 
self-energy and yield $\sqrt{T}$ behavior at the QCP.
 Alternatively speaking, the self-consistent, finite $N$
 solution recovers $\omega/T$ scaling in the fermionic variables. 
From this perspective, the only real difference 
between  the self-energy at $T=0$ and $T \neq 0$ in our 
large $N$ analysis   
is that at $T \neq 0$, the width of the intermediate regime where 
$\Sigma^{\prime \prime} \propto T$ scales with  $N$ 
($\omega_{sf} < T < N \omega_{sf}$), while at $T =0$, the 
width of the analogous linear regime 
in $\Sigma^{\prime \prime} (\omega)$ is, 
strictly speaking, of order 1
 around $T \sim \omega_{sf}$. 
Curiously enough, for physical $N=8$, the widths of the linear regimes 
in  $\omega$ and in $T$ are comparable as the one in $\omega$ 
 occurs in between $\omega_{sf}$ and 
$6-8 \omega_{sf}$.
Whether this is more than just a coincidence is unclear to us.
In any event, a more accurate treatment of thermal fluctuations 
is clearly called for. 

Note, however, that while $\omega/T$ scaling in the fermionic self-energy is
 probably recovered in the self-consistent solution, still, there is no
 $\omega/T$ scaling in the dynamical spin susceptibility. Indeed, the thermal
 correction to the spin-fermion vertex remains finite at criticality, while
 at $T=0$, the corrections were logarithmical and eventually yielded an anomalous dimension $\eta$. While, as we said, logarithmical physics is not precisely captured by self-consistent solution, there are no reasons to expect $T^{1-\eta}$ behavior  as the anomalous power at $T=0$ emerged due to the presence of
 anomalies which are purely {\it quantum} effects.

\subsection{a temperature variation of $\xi$}

Finally, we  discuss is some more detail the 
temperature variation of the correlation length.
The issue is whether there exist a universal temperature correction 
to the correlation length.
We recall that in our theory 
 $\xi (T) = \xi (T=0)$. 
This independence of $T$ 
is the consequence of the fact that 
 in the computation of the spin polarization operator, 
we linearized the fermionic dispersion near the Fermi surface 
(see Appendix C). 
Expanding fermionic energies beyond the linear order in $k-k_F$, one 
indeed obtains  thermal correction to $\xi^{-2}$, but the latter is nonuniversal and is of order 
 $(T/\Lambda)^2$,
where $\Lambda$ (our upper cutoff) is comparable to the fermionic bandwidth. 
The  nonlinearity of the
 fermionic dispersion also  gives rise to a constant term in 
 four-boson vertex $b \propto 1/\Lambda$.
For a constant $b$, the first-order bosonic self-energy in 2D, 
$\Sigma_b \sim b T \sum_\omega \int d^2 q \chi ({\bf q}, \omega)$ 
is logarithmically singular at $T \neq 0$~\cite{millis}:  
$\Sigma_b \sim b T \int d^2 q /({\bf q}^2 + \xi^{-2}) \sim b T \log \xi$. 
Cutting the divergence at criticality by 3D effects, we do find that  
 at the QCP, $\xi^{-2} (T) \propto b T$. This correction is larger than $T^2$ term, but still it is nonuniversal and scales inversely with $\Lambda$.  

Another widely used  approach  to the quantum-critical behavior near 
the antiferromagnetic instability explores the assumption that 
 $\chi ({\bf q}, \omega)$ satisfies a temperature independent
 sum rule constraint $
 \int d^2 q \chi ({\bf q}, \omega) = const$~\cite{scs}.
 One can easily demonstrate that
for $z =2$ dynamics at low-energies, the constraint implies that at the 
critical point, 
$\xi^{-2} = a T$ with a model independent, universal $a$~\cite{scs,lp}. 
This  
contradicts the  analysis based on the spin-fermion model. 
The most likely reason for the discrepancy is that the spin-fermion model is
a finite $U$ version of the Hubbard model, while 
the hard constraint is valid only at infinite $U$. Indeed, if $U$ is finite,
the no-double occupancy constraint is not exact, and there 
is a finite probability for two 
electrons to occupy the same site. The spin of this  doubly occupied site 
can be either zero or one, i.e., the ``length'' of the spin at a given site 
is not fixed. 
From this perspective, it is likely that the sigma-model approach
captures the physics at very large $U$, while our theory 
describes the physics at moderate $U \leq \Lambda$, when lattice effects
 (such as $b T$ correction to $\xi^{-2}$) are irrelevant.

\subsection{summary of Sec.~\protect\ref{finiteT}}

We  now summarize what we obtained in this section
\begin{enumerate}
\renewcommand{\labelenumi}{\roman{enumi}}
\item
We found that the spin polarization operator at finite $T$ has 
precisely the same form as at $T=0$. There is no universal $T$ 
dependent correction to the spin correlation length.
\item
We found that the $N=\infty$ theory  for electronic self-energy 
at finite $T$ is rather peculiar: the $T$ dependent corrections from  
scattering on dynamical
 spin fluctuations (the ones with finite Matsubara frequency) have 
the same dependence on $T$ as of frequency, i.e., they obey $\Omega/T$ 
scaling. In the quantum-critical regime, this correction scales as 
$\sqrt{T {\bar \omega}}$. 
At the same time, the piece in the self-energy from thermal scattering 
on static spin fluctuations yields impurity-like contribution 
$\Sigma (T) = i \pi \lambda T$ and still scales with the correlation 
length $\xi$. The still presence of $\xi$ in the quantum-critical 
form of the self-energy is a serious artifact of the $N = \infty$ theory
at a finite $T$.
\item
We demonstrated that this artifact disappears when calculations 
are performed at a finite $N$. First, we obtained the full form 
of the self-energy in the second-order perturbation theory, when intermediate 
 fermions are treated as free particles. We found that at large 
frequencies, the static piece in the self-energy becomes 
independent on $\xi$, up to logarithmical factors. 
Then we demonstrated that for $T \ll N \omega_{sf}$,
 one can obtain in a controllable way
 the full  $\Sigma^{\prime \prime} (\Omega)$, by explicitly 
including the $T=0$ self-energy of intermediate fermions. 
  At frequencies which exceed ${\bar \Omega} = N^2 \omega_{sf} (v^2_F/v_x v_y)/36$, this self-energy obeys $\Omega/T$ scaling. 
We also found that for $T \geq \omega_{sf}$, this  $\Sigma^{\prime prime} (\Omega)$ has a dip at low frequencies which is likely an artifact of overestimation of the reduction of the static part of the self-energy at small but finite $\Omega$. We argued that the 
second-order result for the self-energy may be closer to reality at these $T$. \item
We found that at small $\Omega$, the controllable calculations 
of fermionic $\Sigma (\Omega)$
definitely break down when  $T$ exceeds $N \omega_{sf}$. 
For these temperatures, we performed FLEX-type calculations
  and found that the self-energy recovers 
$\sqrt{T {\bar \omega}}$ form, up to logarithmic corrections. 
These calculations are
  not controlled as the vertex renormalization and the 
renormalization of the Fermi velocity are $O(1)$. We 
argued, however, that these renormalizations affect 
logarithmical factors but not $\sqrt{T}$ dependence. 
\item
Our finite $N$ analysis indicates that at finite $N$, 
the fermionic self-energy likely possesses $\Omega/T$ 
scaling. At $T, \Omega > \omega_{sf}$, 
$\Sigma (\Omega, T)$ is linear in $\Omega$ and $T$ at intermediate energies
 and crosses over to $\mbox{max}(\Omega^{1/2}, T^{1/2})$ behavior 
at high energies. The peculiarity of large $N$ is 
that the width of the intermediate range along
 $T$ axis 
is by a factor of $N$ larger than that along frequency axis.
\item
We found, however, that there is no $\omega/T$ scaling in the spin susceptibility, i.e., $\omega^{1-\eta}$ behavior at the QCP at $T=0$ is not collaborated by
the same temperature behavior of the spin correlation length. We argued that this difference is due to the fact that the anomalous dimension $\eta$ emerges due to the presence of anomalies which are quantum effects.    
\end{enumerate}

\section{optical conductivity}
\label{opt_cond}
In this section we use the results for the self-energy 
 and study in detail the behavior of the optical conductivity.

The diagonal component of the optical conductivity is given by~\cite{mahan}
\begin{equation}
\sigma_{ii} (\omega )= \sigma_1 (\omega) + i \sigma_2 (\omega) 
= \frac{\omega^2_{pl}}{4 \pi} \frac{1}{2\pi} \int d \theta
 \frac{\Pi
_{\sigma }(\theta ,\omega )}{i\omega - \delta}  \label{rs}
\end{equation}
where $\omega_{pl}$ is the plasma frequency, $\theta$ 
is the angle along the Fermi surface,  and $
\Pi _{\sigma }(\theta ,\omega )$ is fully renormalized
current-current correlator at zero momentum transfer. 

Without vertex corrections,  $\Pi_{\sigma} (\theta, \omega)$ is given by  
\begin{equation}
\Pi _{\sigma }\left(\theta, i\omega _{n}\right) =\frac{Q}{\beta }%
\sum_{m}\int \frac{d^{2}k}{\left( 2\pi \right) ^{2}} G \left({\bf k},
i\omega _{n}+i\omega _{m}\right) \text{ }G \left({\bf k}, i\omega
_{m}\right) 
\label{o7}
\end{equation}
The normalization factor $Q$ has to be chosen such that in an ideal Fermi gas 
$\Pi (\omega) =1$.
The vertex corrections to $\Pi _{\sigma }$ 
 in our model are due to irreducible corrections to the 
current vertex. These corrections obviously scale with
 $d\Sigma /d\epsilon_k$~\cite{mahan} and can be safely neglected. 
 There are also corrections to  $\Pi _{\sigma }$
 due to extra interactions which are neglected in the spin-fermion model. Of particular relevance here is the residual $p-$wave type four-fermion  
interaction. This interaction   
 give rise to RPA-type series of particle-hole bubbles~\cite{ioffe_larkin,im2}.
 How strong these corrections are is a'priori unclear and has to be determined by experimental comparisons. We discuss this point in some detail in Sec~\ref{summary}).

For a conventional dirty metal with $\Sigma ({\bf k}, \omega) = i 
\mbox{sign} \omega/2\tau$,
 Eq. (\ref{o7}) yields the Drude formula
 $\sigma (\omega) = (\omega^2_{pl}/4\pi)\tau/(1 - i\omega \tau)$.
 In our case, however, the fermionic self-energy strongly depends on 
frequency, and the behavior of  $\sigma(\omega)$ is more complex.

For experimental comparisons, it is convenient to express the conductivity in 
terms of the generalized Drude model~\cite{timusk} 
\begin{equation}
\sigma (\omega) = \frac{\omega^2_{pl}}{4\pi} \frac{\tau (\omega, T)}{
m^*(\omega, T)/m - i\omega \tau (\omega, T)}
\label{o80}
\end{equation}
where the effective  relaxation rate $\tau^{-1} (\omega)$ 
and the effective electron 
mass $m^* (\omega, T)$ depend on temperature and frequency.
 Both $\tau$ and $m^*$ can be expressed in terms of $\sigma_1$ and $\sigma_2$ as 
\begin{eqnarray}
 \tau^{-1} (\omega) &=&  
\frac{\omega^2_{pl}}{4\pi} \left[{\rm Re} \sigma^{-1} (\omega)\right]
\label{o8} \\
\frac{m^*}{m} &=&-\frac{1}{\omega } 
\frac{\omega^2_{pl}}{4\pi} \left[{\rm Im} \sigma^{-1} (\omega)\right].
\label{o81}
\end{eqnarray}

Sometimes, the effective relaxation rate is also defined as 
\begin{equation}
\frac{1}{\tau^*} = \omega \frac{\sigma_1}{\sigma_2}
\label{o82}
\end{equation}

We now obtain the explicit expression for $\sigma (\omega, T)$.
 Substituting the fermionic Green's functions into (\ref{o7}), using the 
spectral representation $G({\bf k}, i\omega
_{m}) = (1/\pi) \int d x~Im G^R ({\bf k}, x)/(x - i \omega_m)$ to convert to real frequencies and retarded variables, and
   integrating over $\epsilon_k$, we obtain after a simple 
algebra~\cite{im3,shulga}
\begin{equation}\label{o9}
\Pi^R _{\sigma }\left(\theta,\Omega\right) =
\int d \omega  \frac{n_F (\omega +\Omega) - n_F(\omega )}
{\Omega +\Sigma^A_{\omega +\Omega} 
- \Sigma^R_{\omega }}
\end{equation}
where $\Sigma_{\omega } = \Sigma ( \theta, \omega )$. 

The relaxation rate is related to $\Pi^R_{\sigma}$ by
\begin{equation}
\tau^{-1} (\omega) = 
\frac{\omega {\rm Im} 
\left\langle \Pi^R _{\sigma}(\theta,\omega)\right\rangle}
{({\rm Im} \left\langle \Pi^R _{\sigma}(\theta,\omega)\right\rangle)^2 
+ ({\rm Re} \left\langle\Pi^R _{\sigma}(\theta,\omega)\right\rangle)^2} 
\label{o10}
\end{equation}
where $\left\langle\dots 
\right\rangle = \int \frac{d\theta}{2\pi } \dots$ means
averaging over the Fermi surface. 

The angular dependence of the fermionic self-energy is due to 
the fact that both the coupling $\lambda$ and spin relaxation frequency $\omega_{sf}$ depend on $\theta$ 
(see Eqn. (\ref{kdep}); for small $\theta$, $\tilde k$ introduced in 
(\ref{kdep}) is related to $\theta$ as $\tilde k = k_F \theta$). 
One can easily make sure that the integral over $\theta$ is determined by 
$\theta = O(1)$, i.e., the 
conductivities  at low $\omega$ and $T$ 
predominantly come from the momenta away from hot spots.
Strictly speaking, this implies that very close to criticality, 
our forms of the self-energy, obtained by expanding near hot spots, are
 inapplicable for the computations of the conductivity at small $\omega$ and $T$. Still, however, even at $\xi = \infty$, our formulas for the self-energy 
 can be used to compute conductivity at frequencies above 
$\omega_{sf} (\theta_{max})$, as at these frequencies, the 
whole Fermi surface behaves as a hot spot. At small frequencies and $\xi \rightarrow \infty$, the critical behavior is affected by the instability in the 
 pairing channel (see Sec.~\protect\ref{pairing_vertex} below), and has to be modified anyway.

Below we 
 focus on the forms of conductivities at finite $\xi$ when pairing fluctuations can be neglected. As we already said and will discuss in more detail in Sec~\protect\ref{summary}, at these $\xi$, the      
self-energy dependence on $\theta$ is modest and $\omega_{sf} (\theta_{max})$
 is comparable to $\omega_{sf} (\theta =0)$. 
 In this situation, the $\theta$ dependence of the self-energy affects 
 the numbers but not the functional forms of the conductivities. In the 
 calculations below, we just neglect the $\theta$ dependence of $\Sigma$ without further discussions. 
We now consider the behavior of conductivities at various 
frequencies and temperatures.  

\subsection{$T=0$}
  
At $T=0$, there are three different frequency regimes: 
i) $\omega \ll \omega _{sf}$, ii)  $\omega \geq {\bar \omega}$ 
 and iii)  $\omega \gg \omega_{sf}$.
 In the first regime, the fermionic self-energy has a Fermi-liquid form, 
$\Sigma ^{R}(\omega ) = \lambda \omega +i\lambda \omega ^{2}/4\omega _{sf}$,
and conductivities indeed also behave as in a Fermi liquid
\begin{eqnarray}
\sigma_1 (\omega) &=& \frac{\omega^2_{pl}}{24\pi}~
\left\langle
\frac{\lambda}{(1 + \lambda)^2 \omega_{sf}}\right\rangle \nonumber \\
\sigma_2 (\omega) &=& \frac{\omega^2_{pl}}{4\pi}\frac{1}{\omega }
\left\langle \frac{1}{1 + \lambda} \right\rangle\nonumber\\
\tau^{-1} (\omega) &=& \frac{\omega^2}{6} 
\frac{\left\langle \lambda/[(1 + \lambda)^{2}\omega _{sf}]\right\rangle}
{\left\langle 1/(1+\lambda)  \right\rangle^{2}} \nonumber \\
\frac{m^*}{m} &=& \frac{1}{\left\langle 1/(1 + \lambda)\right\rangle}
\label{o11}
\end{eqnarray}
\begin{figure}[tbp]
\begin{center}
\epsfxsize=\columnwidth 
\epsffile{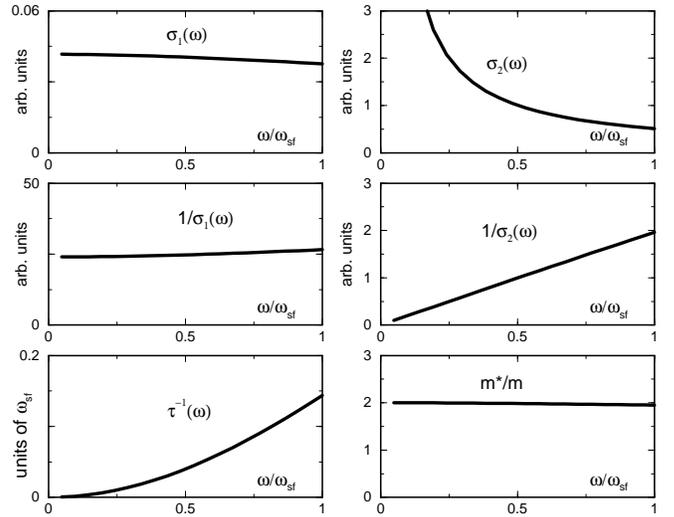}
\end{center}
\caption{ 
Real and imaginary parts of conductivity, effective relaxation rate, 
and effective mass as functions of $\omega$ in the Fermi-liquid regime.
 For definiteness, we used $\lambda =1$.
}
\label{zeroT_ReIm_Tau_effMass_smallOmega}
\end{figure}

In Fig. \ref{zeroT_ReIm_Tau_effMass_smallOmega} 
we plot the two conductivities, the 
effective relaxation rate and $m^*/m$ at $\omega \leq \omega_{sf}$, 
obtained by solving (\ref{o9}) with some angular-independent, ``average'' 
$\lambda$ and $\omega_{sf}$. The Fermi-liquid forms are clearly visible at 
the lowest frequencies. We see, however, that the deviations from the 
Fermi-liquid behavior become visible already at $\omega \sim \omega_{sf}/2$. This
is a consequence of the fact that the fermionic self-energy rapidly deviates
from Fermi-liquid form at $\omega > \omega_{sf}/2$.
\begin{figure}[tbp]
\begin{center}
\epsfxsize=\columnwidth 
\epsffile{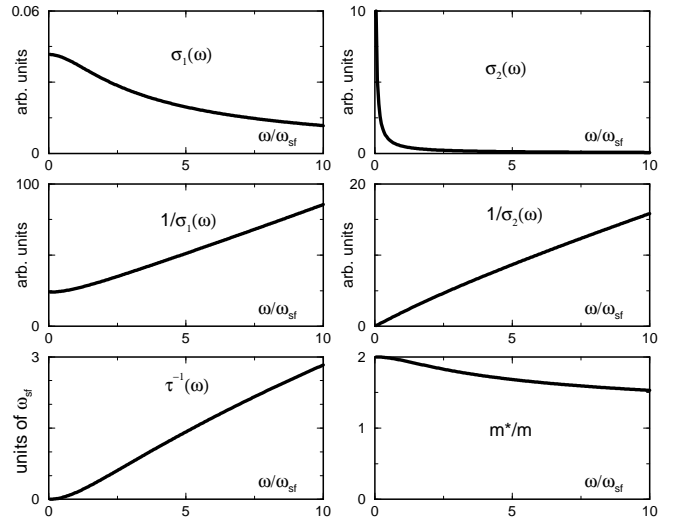}
\end{center}
\caption{Frequency dependence of $\sigma _{1}$, $\sigma _{2}$, $1/\tau $, 
and $m^{*}/m$  at $T=0$ at intermediate frequencies for $\lambda =1$. 
}
\label{zeroT_ReIm_Tau_effMass}
\end{figure}

At intermediate ${\bar \omega} \gg \omega \geq \omega_{sf}$, 
the fermionic $\Sigma (\omega)$ can be
 approximated by a linear function of frequency up to 
$\sim 6-8 \omega_{sf}$. This results in 
the linear in $\omega $ behavior of $1/\tau$ and $1/\sigma_1$ 
In Fig. \ref{zeroT_ReIm_Tau_effMass}  
we show the  results for the conductivities, 
$1/\tau$ and $m^*/m$ in this region. The linearities are 
clearly visible. 
\begin{figure}[tbp]
\begin{center}
\epsfxsize=\columnwidth 
\epsffile{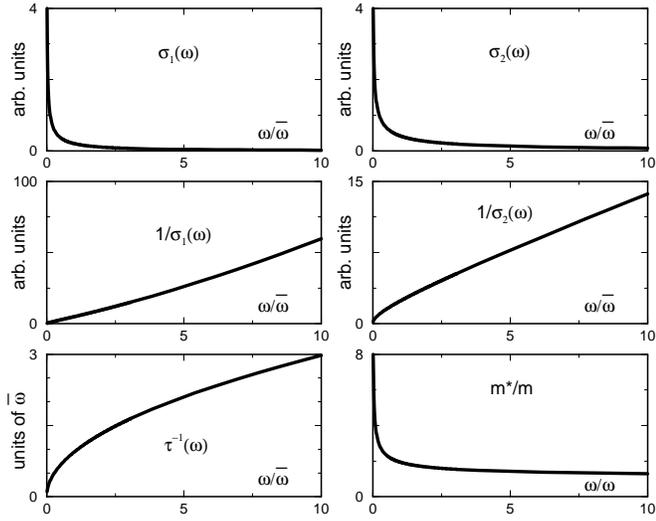}
\end{center}
\caption{Frequency dependences of $\sigma _{1}$, $\sigma _{2}$, $1/\tau $, 
and $m^{*}/m$  at $T=0$ as functions of $\omega/{\bar \omega}$ at $\omega \gg \omega_{sf}$. For simplicity, we set $\omega _{sf} = 0$. 
}
\label{zeroT_0wsf_ReIm_Tau_effMass}
\end{figure}
Finally, at $\omega \gg \omega_{sf}$, 
$\Sigma^R (\omega) = \sqrt{{\bar\omega}/2|\omega|} ( \omega + i |\omega|)$ 
becomes
independent of $\theta$ i.e., the whole Fermi surface acts as a hot spot.
In this limit, we obtained the exact expression for $\Pi_\sigma (\omega)$.
\begin{equation}\label{exactPi}
\Pi^R _{\sigma }(\omega ) = 2-a\frac{\pi }{2}+4\frac{a^{2}-1}{\sqrt{a^{2}-2}}
\tan^{-1}\frac{(\sqrt{2}-1)\sqrt{a^{2}-2}}{a+\sqrt{2}}
\end{equation}
where $a = \sqrt{(i\omega -\delta)/\bar{\omega }}$.
This expression can be simplified at 
$\omega \ll \bar{ \omega}$ and $\omega \gg \bar \omega$ (i.e., at $a\ll 1$ and $a\gg 1$, respectably).  
At $\omega \ll {\bar \omega}$ the bare $\Omega$ term in 
the denominator of (\ref{o9}) is small compared to the fermionic 
 self-energy, while at $\omega \gg {\bar \omega}$,
 the  self-energy is 
smaller than $\Omega$.  At $\omega \ll \bar \omega$,  i.e., 
$a \ll 1$, we have  
\begin{eqnarray} 
 \sigma_1 (\omega) &\approx& \frac{\omega^2_{pl}}{4 \pi }
 \frac{A}{\sqrt{2{\bar \omega} \omega}} \nonumber \\
 \sigma_2 (\omega) &\approx& \sigma_1 (\omega)   \nonumber \\
\tau^{-1} (\omega) &\approx& \frac{1}{A}
\sqrt{\frac{\omega \bar{\omega }}{2}}            \nonumber \\
\frac{m^*}{m} &\approx& \frac{1}{\omega \tau}  
\label{o12a}
\end{eqnarray} 
where $A= 2 - \sqrt{2} \log{(\sqrt{2}+1)} 
\approx 0.754$.
In the opposite limit, $\bar{\omega }\ll \omega$, i.e., $a \gg 1$, we have
\begin{eqnarray} 
 \sigma_1 (\omega) &\approx& \frac{\omega^2_{pl}}{4 \pi} 
 \frac{2\sqrt{2}}{3}\frac{\bar{\omega }^{1/2}}{\omega ^{3/2}} \nonumber \\
 \sigma_2 (\omega) &\approx& \frac{\omega^2_{pl}}{4 \pi} 
\frac{1}{\omega }                                             \nonumber\\
\tau^{-1} (\omega) &\approx& \frac{4}{3}\sqrt{
\frac{\omega {\bar \omega}}{2}}                                     \nonumber \\
m^*/m &\approx& 1  
\label{o12}
\end{eqnarray} 
The behavior of the conductivities, the 
relaxation rate and the effective mass at
$\omega  \gg \omega_{sf}$ and arbitrary $\omega/{\bar \omega}$ 
is shown in Fig.\ref{zeroT_0wsf_ReIm_Tau_effMass}. To emphasize the dependence on $\omega/{\bar \omega}$, we plotted the results right at $\omega_{sf} =0$. Two features in Fig.\ref{zeroT_0wsf_ReIm_Tau_effMass} worth mentioning. First, 
$1/\tau (\omega)$ scales as $\sqrt{\omega}$ and virtually does not change between $\omega< {\bar \omega}$ 
and $\omega > {\bar \omega}$ (the two limiting results for $1/\tau$ 
are  amazingly close to each other as to a good accuracy $1/A = 4/3$!).
 Second, 
$1/\sigma_{1} (\omega)$ initially (at $\omega \leq \bar \omega$) increases as
$\sqrt{\omega}$, but then remains linear in $\omega$ over a substantial 
frequency range and crosses over to $\omega^{3/2}$ behavior only at 
very large $\omega/{\bar \omega}$.

\begin{figure}[tbp]
\begin{center}
\epsfxsize=\columnwidth 
\epsffile{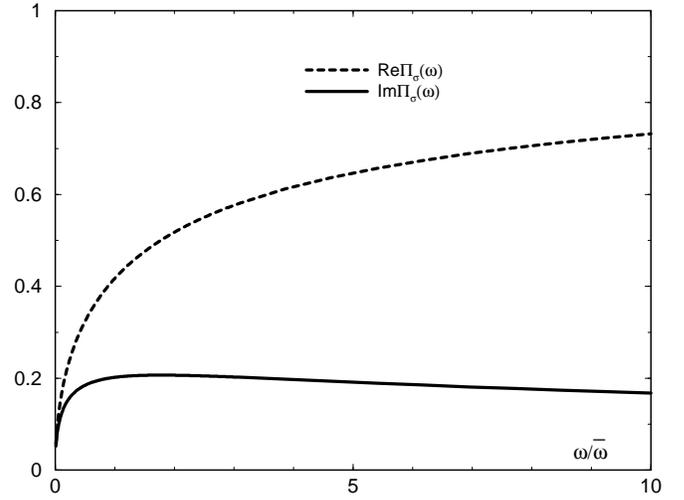}
\end{center}
\caption{
Frequency dependence of 
$Im\Pi _{\sigma }(\omega )$ and $Re\Pi _{\sigma }(\omega )$
at $T=0$ and $\omega _{sf} = 0$. 
}
\label{zeroT_0wsf_Polariz}
\end{figure}

This behavior can also be extracted from the form of the 
polarization bubble presented in Fig. \ref{zeroT_0wsf_Polariz}: we see that 
$Im \Pi_\sigma (\omega)$ first increases as $\sqrt{\omega}$, 
but then  nearly saturates at a value
near $0.2$ and very slowly decreases with increasing $\omega$. 
It is amazing that $Im \Pi_\sigma (\omega)$ remains flat up to 
frequencies which by an order of magnitude exceed the lower boundary
 for $\sqrt{\omega}$ behavior in the fermionic self-energy.

\begin{figure}[tbp]
\begin{center}
\epsfxsize=\columnwidth 
\epsffile{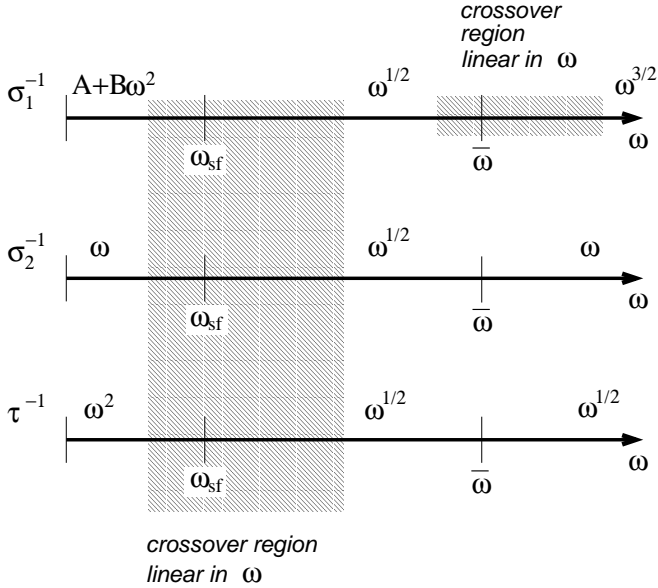}
\end{center}
\caption{
A sketch of the sequence of crossovers for the two conductivities 
and the relaxation 
rate for $\lambda \gg 1$. The shaded regions are the extended crossover regimes. For $\lambda \sim 1$, the crossover region at $\omega \geq \omega_{sf}$ extends over ${\bar \omega}$. In this situation, the two regions of linear behavior of $\sigma^{-1}_1$ merge. We found numerically that the slope of $\sigma^{-1}_1 (\omega)$ does not change much between the two crossover regimes, i.e., $\sigma^{-1}_1 (\omega)$ remains nearly linear in $\omega$ over a very wide frequency range.}
\label{conductivity_schematic}
\end{figure}

We now assemble the $T=0$ results. 
For $\lambda \gg 1$, 
${\bar \omega} = 4 \lambda^2 \omega_{sf} \gg \omega_{sf}$, and the scales
${\bar \omega}$ and $\omega_{sf}$ are well separated in the sense that 
the upper boundary of the linear behavior of the fermionic self-energy 
($6-8 \omega_{sf}$) is located below ${\bar \omega}$. 
In this situation, 
$\sigma^{-1}_1(\omega)$ is quadratic in $\omega$ at the lowest 
frequencies,  becomes linear in $\omega$ at $\omega \sim \omega_{sf}/2$, 
then crosses over to 
$\sqrt{\omega}$ behavior at $\omega \sim 6-8 \omega_{sf}$, then again becomes 
linear at $\omega \sim {\bar \omega}$,  and eventually behaves as
 $\omega^{3/2}$
at the highest $\omega \geq {\bar \omega}$. The last crossover is, however, only suggestive as it occurs at very high $\omega \sim 10^2 {\bar \omega}$ when our low-energy theory is clearly inapplicable.  
The inverse $\sigma_2$ behaves as $\omega$ in the Fermi-liquid regime,
changes slope but remains linear at $\omega >  \omega_{sf}$
 then crosses over to $\omega^{1/2}$ behavior at $6- 8 \omega_{sf}$,
 and at higher frequencies
 behaves as $\omega$. In the ``$\sqrt{\omega}$'' regime, 
real and imaginary components of the conductivity are close to each other.
 The relaxation rate $1/\tau (\omega)$ is quadratic 
in frequency at the smallest frequencies, 
becomes linear in $\omega$ at  $\omega \sim \omega_{sf}/2$, and 
 crosses over to $\sqrt{\omega}$ form at $\omega \sim 6-8 \omega_{sf}$.
This behavior is schematically 
shown in Fig. \ref{conductivity_schematic}.
We emphasize that that the behavior of conductivities 
at $\omega \gg \omega_{sf}$ is fully universal and 
depends only on the ratio $\omega/{\bar \omega}$ but not on the coupling $\lambda$ (see Eq. (\ref{exactPi})). In 
particular, at, e.g.,  $\omega \sim {\bar \omega}/2$, 
\begin{equation}
\sigma_1 \approx 0.37 \frac{\omega_{pl}^{2}}{4\pi {\bar \omega}}~~
\sigma_2 \approx 1.76 \sigma_1~~ \frac{1}{\tau} \approx 0.66 {\bar \omega}~~
\frac{m^*}{m} \approx 2.32
\label{o12b}
\end{equation}

 For $\lambda \sim 1$, the upper cutoff for the linear behavior of fermionic self-energy exceeds 
$\bar \omega$. In this situation, there should be no intermediate 
$\sqrt{\omega}$ regime. Numerically, we found that $\Pi (\omega)$
 does not change much with frequency between $\omega_{sf}$ and
$\omega \geq {\bar \omega}$, i.e.,  the inverse conductivities 
and the relaxation rate remain  linear in 
$\omega$ with little change of the
 slope over a wide frequency range.

\begin{figure}[tbp]
\begin{center}
\epsfxsize=\columnwidth 
\epsffile{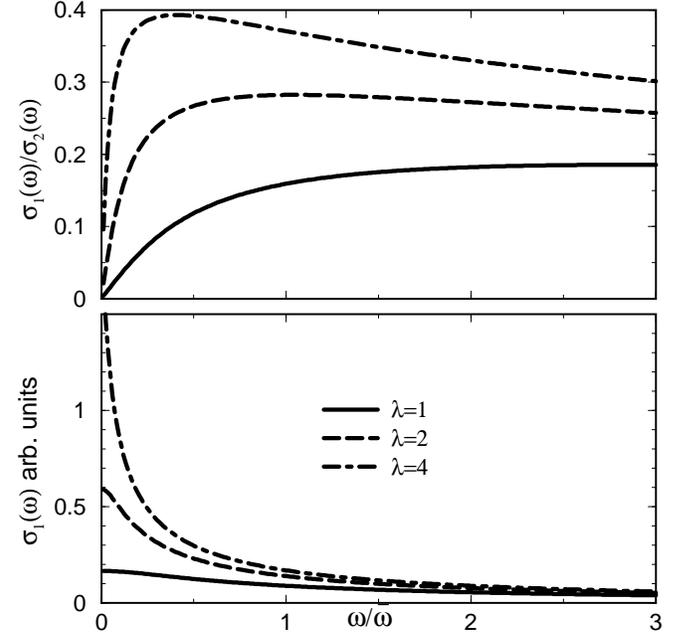}
\end{center}
\caption{Real and imaginary parts of the conductivity for 
various $\lambda$.}
\label{zeroT_ReIm_Tau_effMassLambda.eps}
\end{figure}

For completeness, in Fig.~\ref{zeroT_ReIm_Tau_effMassLambda.eps} we 
present the results for the two conductivities  for various $\lambda$. 
We see that while the functional forms of the conductivities do not
change much with $\lambda$, 
 the ratio of $\sigma_2/\sigma_1$ indeed depends on $\lambda$. This dependence 
may be used as a guide for selecting $\lambda$ which best fit the data. 

\subsubsection{Sum rule for optical conductivity}

The optical conductivity satisfies so-called $f$-sum rule~\cite{sum_rule}
associated with the conservation of the number of carriers
\begin{equation}\label{sumR}
\int \limits_{0}^{\infty }\sigma _{1}(\omega )d\omega = 
\frac{\omega _{pl}^{2}}{4\pi }
\end{equation}
This sum rule has a practical application as it is used to extract 
 the plasma frequency $\omega _{pl}$
from the experimental data.
The issue we consider is up to which frequency one has to integrate to satisfy the sum rule. We will demonstrate that the integral in (\ref{sumR}) 
is poorly convergent, and one has to integrate up to an unexpectedly high frequency to get the right value of $\omega_{pl}$.

\begin{figure}[tbp]
\begin{center}
\epsfxsize=\columnwidth 
\epsffile{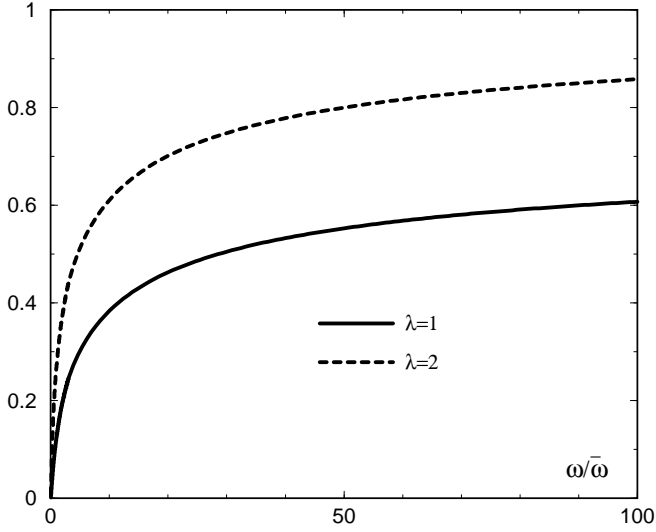}
\end{center}
\caption{ 
The sum rule (\ref{sumR}) as a function of the upper limit of integration.
We used $T=0$, $\lambda =1$.
}
\label{sumRule}
\end{figure}

The results of the numerical integration of the conductivity $\sigma_1 (\omega)$ up to some frequency $\omega$ are presented in 
Fig. \ref{sumRule}. We see that the convergence is very slow. 
This is related to the fact that over a very substantial frequency range, 
the imaginary part of the polarization bubble remains almost a constant (see Fig.~\ref{zeroT_0wsf_Polariz}), and hence $\sigma_1 (\omega)$ scales as $1/\omega$ in which case the integral logarithmically depends on the upper cutoff in frequency integration. Only at very high $\omega \gg {\bar \omega}$, the conductivity crosses over to $\omega^{-3/2}$ behavior, and the integral over $\sigma_1$ converges.  

In cuprates, the value of the plasma frequency is normally obtained by 
 integrating the measured $\sigma_1 (\omega)$ up to 
about $2-2.5 eV$~\cite{timusk,homes,basov,tu,qui}.
 We will argued below that this scale roughly corresponds to 
about $10 {\bar \omega}$ 
Fig \ref{sumRule} then indicates that the value of the plasma frequency extracted from these measurements is somewhat smaller than the actual value of $\omega_{pl}$, the difference decreases with increasing $\lambda$. 

\subsection{finite $T$}

We next consider the behavior of the optical conductivity at finite $T$. 
We begin with the limit $\omega \rightarrow 0$.

\subsubsection {finite $T$, $\omega \rightarrow 0$}

At finite $T$, $Im \Sigma (\omega)$ becomes finite at $\omega =0$, and 
 $\Pi_\sigma (\omega)$ at the lowest frequencies can be obtained by 
just expanding in $\omega$ in  (\ref{o9}). Performing the expansion, we
obtain 
\begin{eqnarray}
  \sigma _{1}(T)&=&\frac{\omega^{2} _{pl}}{4\pi }\frac{1}{8T} 
\int ^{\infty }_{-\infty }\frac{1}{{\rm Im}\Sigma ^{R}_{\omega }}
\frac{d\omega }{\cosh ^{2}(\omega /2T)}              \nonumber \\
\tau ^{-1}(T) &=& \frac{\omega^{2} _{pl}}{4\pi }\sigma ^{-1}_{1} \nonumber\\
m^{*}/m &=& \frac{\sigma ^{-2}_{1}}{16T}\int ^{\infty }_{-\infty }
\frac{1+\partial {\rm Re}\Sigma ^{R}_{\omega }/\partial \omega }
{[{\rm Im}\Sigma ^{R}_{\omega }]^{2}}  
 \frac{d\omega }{\cosh ^{2}(\omega /2T)}                 \label{condT}
\end{eqnarray}
Obviously, the  typical frequency in these integrals
 is of order of temperature.
At small $T\ll \omega _{sf}$,  
the  self energy has the Fermi-liquid form (see Eq. (\ref{seT2})).
The straightforward calculations then yield
\begin{eqnarray}
\sigma_{1} (T) &=& \frac{\omega^{2}_{pl}}
{4\pi }\frac{\left\langle \omega _{sf}\right\rangle}
{6T^{2}}                                                     \nonumber\\
\tau ^{-1}(T) &=& \frac{6T^{2}}
{\left\langle \omega _{sf}\right\rangle}                      \nonumber\\
m^{*}/m &=& \frac{6}{\pi ^{2}}
\frac{\left\langle (1+\lambda)\omega^{2} _{sf}\right\rangle}
{\left\langle \omega _{sf}\right\rangle^{2}}
\left(1+\frac{6}{\pi ^{2}}\zeta(3)\right)
\end{eqnarray}
where $\zeta(z)$ is Riemann zeta function.
\begin{figure}[tbp]
\begin{center}
\epsfxsize=\columnwidth 
\epsffile{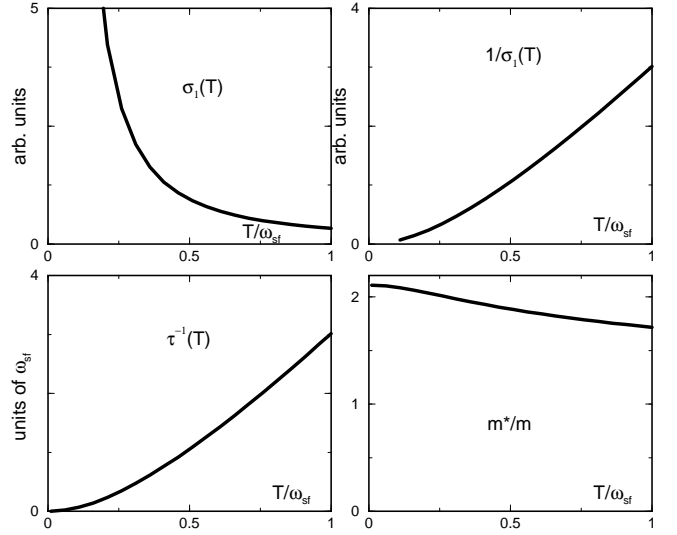}
\end{center}
\caption{Temperature dependences for $\sigma _{1}$, $\sigma _{2}$, $1/\tau $, 
and $m^{*}/m$ at finite $T<\omega _{sf}$ 
and $\omega =0$. We used $\lambda =1$. 
}
\label{cond_ReIm_1overTau_effMass_vsT_omega0_smallT}
\end{figure}
To obtain the conductivity at $T \gg \omega _{sf}$ we need as an input 
the self energy at $\omega \gg \omega_{sf}$. 
The dominant contribution to its imaginary part comes 
from static spin 
fluctuations. At $T \ll N \omega_{sf}$ which we assume to hold, this 
static piece is independent of frequency and just reduces to  $\lambda \pi T$.
Substituting this result into (\ref{condT}), we immediately obtain
\begin{eqnarray}
\sigma _{1}(T)  &\approx& \frac{\omega^2_{pl}}{8\lambda \pi^2 T} \nonumber \\
\tau^{-1}(T) &\approx& 2\lambda \pi T  
\label{condT-inter}
\end{eqnarray}
We see that both $\sigma^{-1}_1$ and $\tau^{-1}$ are 
linear in $T$, and the slopes scale with the coupling constant. 
The computation of the mass ratio requires more care as it 
involves $Re \Sigma^R_\omega$. For intermediate 
$\omega_{sf} \ll \omega \ll {\bar \omega}$,  
 $\partial {\rm Re}\Sigma ^{R}/\partial \omega \gg 1$. Substituting 
$Re \Sigma^R_\omega$ into (\ref{condT})
 and evaluating the 
frequency integral we obtain
\begin{equation}
m^{*}/m \approx  0.48 \sqrt{\bar{\omega }/T}     
\end{equation}
In the opposite limit $\omega \gg {\bar \omega}$, 
 $\partial {\rm Re}\Sigma ^{R}/\partial \omega \ll 1$. 
The integral in (\ref{condT}) is then trivially evaluated and yields
\begin{equation}
m^{*}/m \approx 1    
\end{equation}

\begin{figure}[tbp]
\begin{center}
\epsfxsize=\columnwidth 
\epsffile{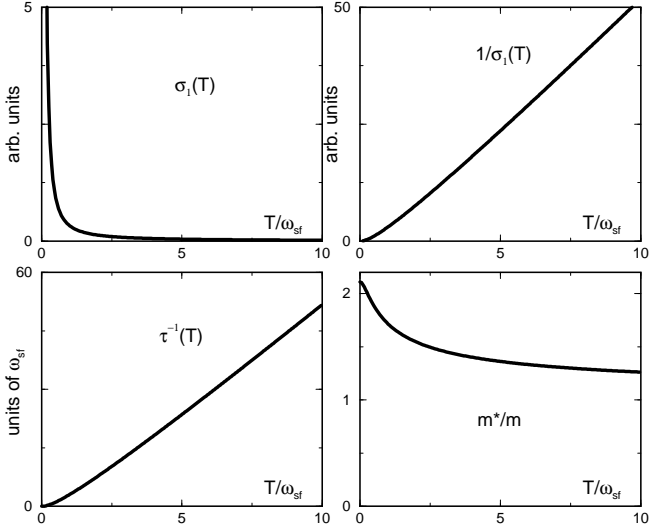}
\end{center}
\caption{Temperature dependences for 
$\sigma _{1} (T)$, $\rho (T) = 1/\sigma_{1} (T) $, $1/\tau $, 
and $m^{*}/m$ at $\omega =0$ and $T \geq \omega_{sf}$. 
We used $\lambda =1$. 
}
\label{cond_ReIm_1overTau_effMass_vsT_omega0}
\end{figure}
In Fig.\ref{cond_ReIm_1overTau_effMass_vsT_omega0} we present the results for 
the dc conductivity, resistivity, relaxation rate and the mass ratio
 at $T \geq \omega_{sf}$ . We see that although the 
asymptotic ``high temperature''  behavior
 of the conductivity and the relaxation rate, Eq. (\ref{condT-inter}) 
is reached only at rather high temperatures, the linear in $T$  behavior of  
$\sigma^{-1}_1$ and $\tau^{-1}$ (and, accordingly, the linear in $T$ 
behavior of the resistivity) begins already at a low $T \sim \omega_{sf}$ and
over a wide temperature range can be well approximated by $AT -B$ where 
 $A$ is somewhat smaller than in  (\ref{condT-inter}), and $B$ 
is a small positive number.  

\subsubsection{finite $T$, finite $\omega$}
\begin{figure}[tbp]
\begin{center}
\epsfxsize=\columnwidth 
\epsffile{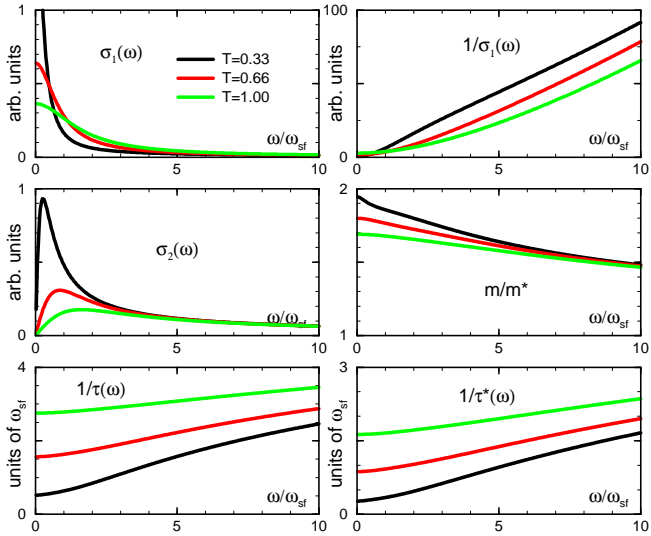}
\end{center}
\caption{Frequency dependence of the conductivities, 
inverse $\sigma_1$, mass ratio and the relaxation rates $1/\tau $ 
and $1/\tau^*$ at different temperatures. We used $\lambda =1.7$. 
The results were obtained with the second-order self-energy (see Fig\protect\ref{sigmaT_exact}).}
\label{cond-tau-all}
\end{figure}

Finally, in Fig. \ref{cond-tau-all}  we present 
the results for the conductivities, the mass ratio, and the 
relaxation rates $1/\tau$ and $1/\tau^*$
 when both frequency and temperature are finite. We see 
that for any $T$, the inverse
conductivity $\sigma_1$ 
is still linear in $\omega$ over a substantial frequency 
range, and the slope is almost independent of $T$.
 Observe also that the conductivity curves for various $T$ cross each other such that at small $\omega$, conductivity 
decreases with $T$, while at large $\omega$ it increases with the temperature.
 The conductivities in Fig.  \ref{cond-tau-all} are 
obtained with the 
 second-order fermionic self-energy (its imaginary part is plotted in Fig.
~\ref{sigmaT_exact}). We verified that if we use the ``full''
 self-energy (\ref{Bbbbb81}), we obtain almost identical 
results for the conductivities, mass ratio and relaxation rates.

\subsection{summary of Sec.~\protect\ref{opt_cond}}

We now summarize the main results of this section

\begin{enumerate}
\item
At $T=0$, real and imaginary parts of the optical conductivity, 
$\sigma_1 (\omega)$ and
$\sigma_2 (\omega)$, and the effective scattering 
rate $1/\tau (\omega)$ undergo a series
 of crossovers schematically shown in 
Fig~\ref{conductivity_schematic}. In particular,
 for moderate couplings, $\sigma_1^{-1}$ and $\sigma_2^{-1}$ 
are linear in $\omega$ over
a substantial frequency range. The relaxation rate $1/\tau$ 
is also linear in $\omega$ but
 over a smaller range of frequencies, and then crosses over 
to $\sqrt{\omega}$ behavior.
\item
We found that the $f-$sum rule for $\sigma_1$ is recovered only 
if the integration in $I(\omega) =\int_0^{\omega} \sigma_1 (x) dx$ 
is extended up to  extremely high frequencies. If the integration 
is performed up to frequencies about 10 times
 larger than ${\bar \omega}$,  $I(\omega)$ saturates 
(see Fig.~\ref{sumRule}) at some intermediate value 
which depends on $\lambda$. 
The implication of this result for experiments is that  
$\omega_{pl}$ extracted from integrating up to a particular cutoff is
 somewhat smaller than the actual plasma frequency.
\item
At finite $T$, the most important features are the development 
of the peak in $\sigma_2 (\omega)$ at $\omega$ comparable to $T$, and
the crossing between $\sigma_1 (\omega)$ curves at different $T$: 
the  conductivity at the lowest $T$ 
is the largest at the smallest $\omega$, but
 become the smallest at larger $\omega$  (see Fig. \ref{cond-tau-all}). 
The relaxation rate on the other hand
 do not display this kind of behavior.
\item
The effective mass $m^*$ is by a factor $1+\lambda$ larger than the 
bare electron mass, but the ratio $m^*/m$ decreases with
increasing frequency and temperature.
\end{enumerate}

\section{other observables}
\label{other_observables}

In this section, we 
briefly discuss the behavior of other observables - 
 the fermionic
density of states, the NMR relaxation rate and the Raman intensity.

\subsection{fermionic density of states}

The fermionic density of states (DOS) is given by
\begin{equation}
N(\omega) = \frac{1}{\pi} \sum_k |{\rm Im} G ({\bf k},\omega)| 
\label{o1}
\end{equation}
Assume  for simplicity that the  Fermi surface is circular. Then 
 we can split momentum integration into the integration over $d\theta$ along the Fermi surface, and over
$d\epsilon_k =v_F d k_\perp$, where $k_\parallel$ is a  momentum component 
along the Fermi surface.
Substituting  $G^{-1} ({\bf k}, \omega) = 
\omega + \Sigma (k_\parallel, \omega) - \epsilon_k$ 
and integrating over $d \epsilon_k$ 
first we obtain that
\begin{equation}
N(\omega) = const 
\label{o2}
\end{equation}
independently of the strength of the fermionic self-energy. It is only essential that the self-energy does not depend on $\epsilon_k$.
 
For more complex Fermi surface, $N(\omega)$ does acquire some  
frequency dependence which, however, is completely unessential from physics perspective. Also, Eq. (\ref{o2}) is indeed only valid at frequencies much smaller
 than the fermionic bandwidth, when one can linearize the fermionic 
dispersion near the Fermi surface. At frequencies comparable to bandwidth, $N(\omega)$ curves down and eventually vanishes.

\subsection{neutron scattering and NMR relaxation rate}

The neutron scattering measures the  dynamical structure factor $S({\bf q},\omega) = (1+ 2 n_B (\omega)) 
\chi^{\prime \prime} ({\bf q}, \omega)$.
Substituting the expression for the susceptibility, Eqs (\ref{chi}), (\ref{Pi1}),
we obtain
\begin{equation} 
S({\bf q},\omega) = \frac{\chi_0 \xi^2}{\omega_{sf}}~
\frac{\omega(1+ 2 n_B (\omega))}
{(1 + \xi^2 ({\bf q}-{\bf Q})^2)^2 + (\omega/\omega_{sf})^2}
\label{o3}
\end{equation}
At $\omega \gg T$, the Bose factor is irrelevant, 
$S({\bf q},\omega)$ at a given $q$ has a simple $x/(1+x^2)$ form where $x= \omega/(\omega_{sf} (1 + \xi^2 ({\bf q}-{\bf Q})^2))$ is a rescaled frequency.

The NMR relaxation rate $1/T_1$ is proportional to the momentum integral of 
 $S({\bf q},\omega)$ 
weighted with the momentum dependent formfactor $F({\bf q})$~\cite{nmr}:
\begin{equation}
T^{-1}_1 (\omega) \propto \sum_q F({\bf q}) S({\bf q},\omega)
\label{o4}
\end{equation}
The typical frequencies in the NMR experiments are in the range of few MHZ ($\sim 10^{-4} K$), hence  $T \gg \omega$ 
for all reasonable $T$. Then $1 + 2 n_B (\omega) \approx 2 T/\omega$, and
\begin{equation}
T^{-1}_1  \propto 2T \lim\limits_{\omega \rightarrow 0}
  \sum_q F({\bf q}) \frac{\chi^{\prime \prime}({\bf q},\omega)}{\omega}
\label{o5}
\end{equation}
The form factor $F({\bf q})$ depends on the local environment of the 
isotope atom. For $Cu$ NMR, $F({\bf q})$ does not vanish
 at ${\bf Q} = (\pi,\pi)$.
 Substituting $\chi^{\prime \prime} 
({\bf q}, \omega)$ into (\ref{o5}),
one can easily check that the momentum integration in (\ref{o5}) is 
 then confined to small $|{\bf q}-{\bf Q}| \sim \xi^{-1}$, and
hence 
\begin{equation}
\frac{1}{T_1 T} \approx \frac{F({\bf Q}) \chi_0}{4\pi \omega_{sf}}
\label{o6}
\end{equation}
As the product $F({\bf Q}) \chi_0$ can be extracted independently from 
Knight shift data~\cite{nmr}, the measurements of $1/T_1$ allow one to 
determine $\omega_{sf}$ which, we remind, is one of the two
 input parameters in our theory.

\subsection{Raman intensity}

Another observable sensitive to the form of the fermionic self-energy 
is the intensity of the Raman absorption.  
In a Fermi gas, a transferred photon energy
$\omega$ can be absorbed by a metal only
 if $\omega$ is smaller than $v_F |q|$ when $q$ is a transferred 
momentum (this is the Fermi golden rule)~\cite{a_ge}.
As photon momentum is  vanishingly small, the 
 absorption is possible only for extremely small 
$\omega \sim \omega_{in} v_F/c$ where $\omega_{in}$ is the incident 
photon frequency, and $c$ is the velocity of light.
However, if fermions have a nonzero ${\rm Im} \Sigma$, the absorption 
is possible for all frequencies: the energy extracted from  photons is dissipated due to a finite lifetime of fermionic excitations.

The intensity of the (non-resonance) Raman absorption  
$R(\omega)$ is generally proportional to the 
imaginary part  of the particle-hole bubble at 
zero external momentum and finite frequency, weighted with the form-factors
which depend on the scattering geometry~\cite{klein,dev}. For the most studied 
$B_{1g}$ scattering, the form-factor 
$F_R ({\bf k}) \propto \cos k_x - \cos k_y$
 is at maximum near hot spots, and can be approximated by a 
constant. The Raman intensity is then simply given by 
\begin{equation}
R(\omega) \propto {\rm Im} \Pi_R (\omega) 
\label{o14}
\end{equation}
The Raman bubble $\Pi_R (\omega)$ differs from $\Pi_\sigma (\omega)$ as
side vertices are now scalars.
As we discussed in Sec.~\ref{sec_ninf},
 a scalar vertex with zero total momentum 
 is subject to strong vertex corrections.
For a density-density correlator with  momentum independent vertices,
vertex corrections fully cancel fermionic self-energy, and as a result 
there is no dissipation~\cite{dev2}. [This cancellation 
 is the consequence of the fact that the number of particles is conserved]. 
However, for $B_{1g}$ Raman scattering, the side vertex
 has $d-$wave symmetry, and the renormalization of the vertex requires a  
$d-$wave component of the interaction. In this situation, there is no 
 exact relation between fermionic self-energy and vertex corrections. 
We explicitly verified that for magnetically-mediated scattering, 
the correction to the $B_{1g}$ Raman vertex 
  is nonsingular even when $\xi \rightarrow \infty$. Neglecting this 
nonsingular vertex renormalization,  
we obtain a
simple relation between the normal state Raman intensity and the normal state
 conductivity:
\begin{equation}
R(\omega) \sim \omega \sigma_1 (\omega).
\label{015}
\end{equation} 
In particular, at $T=0$, 
$R(\omega) \propto \omega$ at small frequencies, 
then saturates, then crosses over to $R(\omega) \propto 
\sqrt{\omega}$, and then (if lattice effects do not interfere) 
first saturates and then begins decreasing as $\omega^{-1/2}$. 
This last regime, however, is
likely already  masked by the development of a ``two-magnon''
 peak from short-range spin fluctuations~\cite{bl}.

\section{pairing vertex}
\label{pairing_vertex}

\begin{figure}[tbp]
\begin{center}
\epsfxsize=\columnwidth 
\epsffile{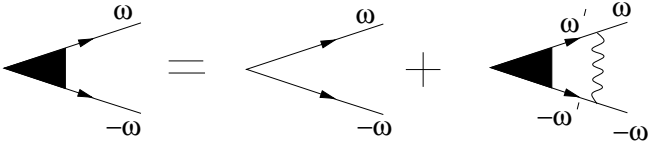}
\end{center}
\caption{
The diagram for the full vertex in the particle-particle channel.
 The bare (unshaded) vertex is infinitesimally small. The wavy line is the spin propagator. Since the spin-mediated interaction is retarded, the full vertex
 depends on frequency $\omega$.
}
\label{pairing}
\end{figure}

Finally, we briefly discuss the development of the pairing instability
in the spin-fermion model. 
It is customary for
 the analysis of the pairing problem
  to introduce an infinitesimally small particle-particle vertex $g^0_{pp} (k,-k) \equiv g^0_{pp} (k)$ 
and study its renormalization by magnetically mediated interaction. 
The corresponding diagram is presented in Fig \ref{pairing}. The 
temperature where the  renormalized vertex diverges 
is obviously the onset of pairing.
The magnetically-mediated pairing interaction is repulsive for $s-$wave pairing, but is  attractive in the $d_{x^2-y^2}$ channel~\cite{david_review,scal_pr}. 
Accordingly, we will be searching for a $d-$wave solution for the full pairing vertex $g_{pp} (k)$. For momenta near hot spots, the $d-$wave constraint implies that $g_{pp} (k+Q) = - g_{pp} (k)$. 

We demonstrate in this section that
 the renormalization of the pairing 
vertex is not a $1/N$ effect, and hence the  
pairing instability is possible even in the 
limit when $N \rightarrow \infty$, when our Eliashberg-type theory becomes exact.

For a  particle-hole vertex with a nonzero momentum 
transfer, the $1/N$ smallness of the vertex renormalization comes from the 
fact that for large $N$, the momentum dependence of the spin susceptibility 
can be neglected, and 
the  vertex correction is a convolution of the two fermionic 
densities of states and the dynamical spin 
susceptibility at momentum ${\bf Q}$. The latter contains 
$N$ in the denominator, and this gives rise to the overall 
$1/N$ factor in the  correction to a particle-hole vertex. 
For zero momentum transfer, the $1/N$ does not appear, 
but as we discussed in Sec~\ref{vert_zero}, 
this is the consequence of the fact that the momentum integral is singular and has to be properly regularized.
  
For a particle-particle vertex, the situation is  different as now 
 vertex renormalization  involves fermions with opposite momenta and 
frequencies, and hence at a given frequency, the poles in the two 
fermionic propagators are located in different half-plains of $\epsilon_k$. 
As a result, there is no double pole in the momentum integral. The latter in turn implies that at $N \rightarrow \infty$, one can again  
factorizing the momentum integration i.e., neglect in the 
spin susceptibility the momentum component transverse to the Fermi surface.
Still, however, one {\it has to} integrate the susceptibility over the momentum component along the Fermi surface. This last procedure changes the power of $1/N$ in the denominator such that eventually $1/N$ is eliminated from the equation for $g_{pp} (k)$ (see below).  

Two of us and Finkel'stein have demonstrated explicitly~\cite{acf} that 
for  relevant $k$ near each of hot spots, the momentum variation of  
  $g_{pp} (k)$ is irrelevant and can be neglected.  
The momentum integration then can be performed explicitly, and at $T=0$ we obtain
\begin{equation}
\frac{\delta g_{pp}}{g^{0}_{pp}} \propto  
 \frac{3 g^2}{8\pi^2 v_F} ~\int \frac{d \omega_m}{
|\omega_m -i \Sigma (\omega_{m})|}~\chi_L (\omega_m - \Omega_m)    
\label{p1}
\end{equation}
where $\Omega$ is the external frequency ($g_{pp} = g_{pp} (\Omega, -\Omega)$),
 and $\chi_L (\omega)
 = \pi \chi_0 \xi/(1 + |\omega|/\omega_{sf})^{1/2}$
 is the ``local'' 1D susceptibility 
obtained by integrating over momentum component along the Fermi 
surface and neglecting ${\bf q}-{\bf Q}$ transverse to the Fermi surface .
 Substituting this susceptibility into (\ref{p1}),
 we  obtain 
\begin{eqnarray}
\frac{\delta g_{pp}}{g^{0}_{pp}} &=& \frac{1}{4}  
\int \frac{d \omega_m}{\omega_m} \frac{\sqrt{\omega_{sf}} + 
\sqrt{\omega_{sf} + |\omega_m|}}{\sqrt{\omega_{sf} + |\omega_m - \Omega_m|}}
\nonumber \\
&&
~\frac{2 \lambda}{1+ (2\lambda) + \sqrt{1 + |\omega_m|/ \omega_{sf}}}
 \label{p2}
\end{eqnarray}
We see that the vertex correction is positive, i.e., 
the addition of a spin-fluctuation exchange 
enhances the particle-particle vertex.
 Rescaling frequency as $x = \omega/\omega_{sf}$ we immediately find that 
${\delta g_{pp}}/{g^{0}_{pp}}$ depends only on $N-$independent effective
 coupling $\lambda$.  The origin of this result is rather transparent 
in the quantum-critical regime $\omega_m \gg \omega_{sf}$. 
Here $\chi_L \propto 
\sqrt{\omega_{sf}/\omega_m}$ scales as $(1/N)^{1/2}$. Simultaneously, 
\begin{equation}
\omega_m - i \Sigma (\omega_m) \approx \sqrt{\omega_m {\bar \omega}} 
\sim \lambda \omega_m  \sqrt{\frac{\omega_{sf}}{\omega_m}}
\end{equation}
 also scales as $(1/N)^{1/2}$. 
Taking the ratio of the two expressions we obtain $N-$independent result 
\begin{equation}
\frac{g^2}{v_F} ~\frac{\chi_l (|\omega_m-\Omega_m|)}
{\omega_m -i\Sigma (\omega_m)} 
\sim \frac{1}{\omega_m}~ 
\left(\frac{\omega_m}{|\omega_m-\Omega_m|}\right)^{1/2}.
\end{equation} 

Another issue is how large is the pairing instability temperature.
 For 
$\lambda \leq 1$, the answer is rather obvious as the pairing problem is very similar to the BCS phonon problem with $\omega_{sf}$ playing the same role as 
 the Debye frequency for phonons. Indeed, when $\lambda \leq 1$, $\omega_{sf}$ 
in (\ref{p2}) provides a natural upper cutoff for the frequency 
 integral. Extending Eq. (\ref{p2})  to finite  $T$ in a standard way and 
evaluating the frequency sum with logarithmical accuracy, we obtain 
\begin{equation}
\frac{\delta g_{pp}}{g^{0}_{pp}} = \frac{\lambda}{1 + \lambda} 
\log {\frac{\omega_{sf}}{T}}
\label{bcs}
\end{equation}
Obviously then,
 $T_{ins} \propto \omega_{sf} \exp^{-(1+\lambda)/\lambda}$. This is similar to  the McMillan formula for phonons~\cite{mcmillan}. The $1+\lambda$ factor 
in the numerator accounts for the mass renormalization at frequencies below $\omega_{sf}$.

Eq. (\ref{bcs}) is formally valid for all couplings $\lambda$ provided that
 the pairing problem is {\it confined} to a region $\omega \leq \omega_{sf}$ 
where the system has a Fermi-liquid behavior. 
However, we see from (\ref{p2}) that the pairing kernel preserves 
$1/\omega$ form also at frequencies larger than $\omega_{sf}$ and 
crosses over to $1/\omega^{3/2}$ form only at $\omega > {\bar \omega}$. 
For ${\bar \omega} \gg \omega_m \gg \omega_{sf}$, we have from (\ref{p2})
\begin{equation}
\frac{\delta g_{pp}}{g^{0}_{pp}} = \frac{1}{4}  
\int \frac{d \omega_m}{\sqrt{|\omega_m (\omega_m - \Omega_m)|}}
\label{pv_1}
\end{equation}
Alternatively speaking, the kernel for the pairing problem remains $O(1/\omega)$ not only in the Fermi-liquid regime, but also in the quantum-critical regime where the system does not behave as a Fermi-liquid. 
This implies that 
$T_{ins}$ in principle 
can be of order $\bar \omega$, i.e., substantially higher than in the 
McMillan formula. This is not guaranteed, however, as the pairing 
kernel in (\ref{p2})  depends 
on both internal and external frequency, and the logarithm in the r.h.s. of
 (\ref{pv_1}) is cut by $\Omega_m$. In the latter case, the ladder series are not geometrical,  and
 there is no a'priori 
guarantee that they yield an instability at $T \sim {\bar \omega}$.

The full analysis of the pairing problem requires extra care
and is beyond 
the scope of the present paper. We just 
cite the result~\cite{acf}:
the instability temperature $T_{ins}$ does scale with ${\bar \omega}$ at 
strong coupling, and saturates at $T_{ins} \approx 0.17 {\bar \omega}$ 
at $\xi = \infty$. Numerically, 
the saturation of $T_{ins}$ was detected in Ref.~\cite{ml}.
The behavior below $T_{ins}$ is rather involved, as we discussed in 
the Introduction (see Fig. \ref{phase-diagram-normalSC}). 

Note in passing that the studies of the pairing instability at the 
verge of the magnetic transition yield rather nontrivial results 
also when the pairing is mediated by ferromagnetic spin 
fluctuations~\cite{ferro}.

The fact that the 
 pairing instability temperature scales with ${\bar \omega}$ which, we remind,
 is 
the upper cutoff frequency for the quantum-critical 
behavior, implies that our normal state analysis which neglects pairing channel is valid only in a restricted frequency and temperature range. Still, however, numerically $T_{ins} \ll \bar \omega$, and hence there exists a wide region of frequencies/temperatures where normal state quantum-critical analysis is valid. A related issue is how strong are the pairing fluctuations above $T_{ins}$. 
We  show below that the pairing susceptibility has an overall factor $1/N$ and hence pairing fluctuations 
only affect fermionic self-energy in a narrow range near $T_{ins}$. Outside this region, they contribute only $1/N$ corrections to the fermionic self-energy
 and hence do not affect our $N = \infty$ analysis.  

\begin{figure}[tbp]
\begin{center}
\epsfxsize=\columnwidth 
\epsffile{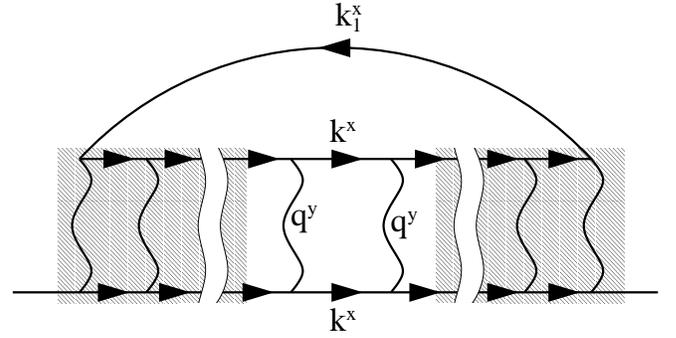}
\end{center}
\caption{
The diagram for the correction to the electronic self-energy due to fluctuations in the pairing channel.
}
\label{SEpairingCorrection}
\end{figure}

The corrections to the electronic self-energy due to pairing fluctuations
are  shown diagrammatically in Fig \ref{SEpairingCorrection}.
Consider first the shaded regions in 
Fig. \ref{SEpairingCorrection}. Each time we add an extra bosonic line 
to the ladder, we also add two fermionic lines. 
As we demonstrated above, this addition  
does not contribute extra powers of $1/N$. The $1/N$ factor comes from the 
 unshaded part of this diagram. It contains a pair of fermionic propagators, 
and a {\it pair} of bosonic propagators, i.e., it contains an extra bosonic propagator compared to the building block in the shaded part of the diagram.  
The bosonic propagator scales as $1/N$ as there are $N$ channels for spin decay.
We explicitly verified that this $1/N$ appears as the overall factor in the diagram, i.e., 
\begin{equation}
\Sigma_{pp} (\Omega) \propto \sqrt{\omega {\bar \omega}} \frac{\Gamma}{N}
\end{equation}
where $\Gamma$ 
is the enhancement factor from the shaded part of the diagram. 
The divergence in $\Gamma$ implies pairing instability and 
obviously affects  fermionic self-energy. However as long as $\Gamma \leq N$, 
 $\Sigma_{pp} (\Omega)$ has a prefactor of $1/N$, and can be neglected 
in the limit $N\rightarrow \infty $.  

A final comment. The pairing instability at a finite $T_{ins}$ at $\xi = \infty$ obviously implies that there exists a dome on top of the magnetic quantum critical point. Inside this dome, 
 the quantum-critical theory has to be reconsidered as the fermions near hot spots become gaped. 
In other words, the quantum-critical behavior which causes the 
 pairing instability at a finite $T_{ins}$ by itself gets affected as a feedback from the pairing. In particular, we explicitly verified that the correction
 to the spin-fermion vertex is saturated at finite value below $T_{ins}$, i.e., it is no longer logarithmically singular. This, in particular,  
 implies that the anomalous magnetic response in the quantum-critical regime,
 which we found in Sec.\protect~\ref{per_th} 
 exists only at intermediate energies and cannot be extended to
 the lowest $\omega$ (which fall inside the dome). The quantum-critical behavior inside the dome is a subject of a separate study and we will not discuss it here.
 
\section{Summary and experimental comparisons}
\label{summary}
In this paper, we presented the full scale calculations for the 
normal state properties of the spin-fermion model. The model 
involves low-energy fermions 
 interacting with their own spin collective excitations which 
are assumed to be peaked at the antiferromagnetic momentum 
${\bf Q}=(\pi,\pi)$. We argued that this model is the 
low-energy version of the lattice, Hubbard-type models 
for strongly interacting fermions, provided that the 
spin fluctuations are the only low-energy collective 
bosonic excitations. The model indeed
 makes sense only if the spin-fermion coupling 
does not exceed fermionic bandwidth which we assume to hold. If this
 condition is not satisfied, the separation between low-energy and high-energy
 excitations becomes problematic.   

We demonstrated that near magnetic 
instability, the model falls into the strong coupling 
limit, where the conventional perturbation theory does 
not work. The corrections to the Fermi-gas behavior are 
the strongest for fermions near hot spots - the points 
at the Fermi surface separated by antiferromagnetic ${\bf Q}$. 
The presence of hot spots is essential
 for our study - without them the critical theory at $T=0$ 
would have had a different dynamical exponent $z=1$.
 
We developed a controlled way to perform calculations in the strong coupling 
 limit by expanding in $1/N$ where $N (=8$ for the physical case) 
is the number of hot spots in the Brillouin zone.  Equivalently, 
one can extend the model to a large number of electron flavors $M$ 
and expand in $1/M$.

Our major results are summarized in Figs. ~\ref{sigmaT0} and \ref{sigmaT_exact} 
for the fermionic spectral function, and Fig. ~\ref{cond-tau-all}
for optical conductivity. For the dynamical structure factor, 
our major result is Eqn (\ref{chan}). 
We demonstrated that near the QCP, the 
dimensionless spin-fermion coupling $\lambda \propto \xi$ 
is large, i.e., the system falls into the strong coupling regime. 
 This strong coupling behavior can also be reached at intermediate 
$\xi$ when the spin-fermion interaction ${\bar g}$ increases.

At strong coupling, the region near the QCP
 is divided into the
  Fermi liquid regime and the quantum-critical regime where the system behavior
is the same as at criticality.  The quantum-critical behavior extends  
roughly up to frequencies comparable to the spin-fermion coupling constant 
${\bar g}$. The crossover from the Fermi liquid to the 
quantum-critical behavior on the other hand occurs at energies of order
 $\omega_{sf} \sim \lambda^{-2} {\bar g} \ll {\bar g}$.
We found that there is a single crossover energy for both electronic 
and magnetic properties of the system. 

We now list the catalog of our key results. We first list the results 
at $T=0$ and then show how they are modified at finite $T$.

\subsection{$T=0$}

\subsubsection{Fermi-liquid regime}

In the Fermi-liquid regime, we found
for the fermionic self-energy
\begin{equation}
\Sigma (k, \Omega) = \Omega\lambda (k) + i \lambda (k)
 \frac{\Omega |\Omega|}{4 \omega_{sf} (k)} 
\label{s1}
\end{equation}
 The momentum-dependent coupling constant $\lambda (k)$ and 
and $\omega_{sf} (k)$ are given by  
\begin{equation}
\lambda (k) = \lambda/(1 + ({\tilde k} \xi)^2)^{1/2}, \,\,\,\,
\omega_{sf} (k) = \omega_{sf} (1 +  ({\tilde k} \xi)^2)        \label{s2}
\end{equation}
where ${\tilde k}$ is the deviation from a hot spot along the Fermi surface
(${\tilde k} = \epsilon_{k+Q}/v_F$). These Fermi liquid forms are valid
when $\Omega < \omega_{sf} (k)$.  Obviously, the self-energy corrections 
are the largest for fermions near hot spots. For the same frequencies, 
we found that the spin susceptibility near  the 
antiferromagnetic momentum ${\bf Q}$ can be 
well approximated by its static form
\begin{equation}
\chi ({\bf q}, \omega) 
\approx \frac{\chi_0 \xi^2}{1 + ({\bf q}-{\bf Q})^2 \xi^2}
\label{s3}
\end{equation}

In this limit, $\sigma_1 (\omega)$ weakly depends on $\omega$,
and $\sigma_2 (\omega)$ scales as 
 $\omega^{-1}$.

\subsubsection{quantum-critical regime, intermediate frequencies}

At intermediate frequencies 
$ \omega_{sf} (k) \leq \Omega \leq 6-8 \omega_{sf}$,
we found that to a surprisingly good accuracy, 
the imaginary part of the fermionic self-energy is 
 linear in frequency: 
\begin{equation}
\Sigma^{\prime \prime} (k, \Omega) \approx 0.3 \lambda (k) ( \Omega - 0.7 \omega_{sf})
\label{s6}
\end{equation}  
Note, however, that this behavior is {\it not} an 
intermediate asymptotic which width 
could be controlled by some parameter of a system, but rather a peculiar 
behavior in a 
wide crossover region between truly Fermi-liquid and truly 
quantum-critical (high frequency) regimes. The real part of the self-energy
 $\Sigma^{\prime} (k, \Omega)$ roughly follows $\omega \log |\omega|$ behavior
 which is the Kramers-Kronig transform of (\ref{s6}), but this form 
is only approximate as the linearity in $\Sigma^{\prime \prime}$ 
exists only in a limited frequency range. 

At frequencies larger than $\omega_{sf}$,
 one also has to use the full expression for the 
dynamical spin susceptibility as neither the gap in the static 
susceptibility nor the dynamical piece can be neglected
\begin{equation}
\chi ({\bf q}, \omega) = \frac{\chi_0 \xi^2}
{1 + ({\bf q}-{\bf Q})^2 \xi^2 - i |\omega|/\omega_{sf}}
\label{s7}
\end{equation} 
Finally, at these frequencies, both $\sigma_1 (\omega)$ and 
$\sigma_2 (\omega)$ scale as $1/\omega$.

\subsubsection{quantum-critical regime, high frequencies}

At high frequencies, $\Omega \sim {\bar \omega}
 \gg \omega_{sf} (k)$,
 the whole Fermi surface
behaves as a hot spot. We found that in this regime, 
the fermionic self-energy scales as $\sqrt{\omega}$: 
\begin{equation}
\Sigma (\Omega) = e^{i\frac{\pi}{4}} 
\sqrt{{\bar \omega} |\Omega|} \mbox{sign} {\Omega}
\label{s4}
\end{equation}
This form of the self-energy implies that fermionic Green's function 
$G^{-1}(k, \Omega) \propto (i\Omega - \epsilon^2_k/{\bar \omega})$. 
Comparing this with the Fermi liquid result 
$G^{-1}(k, \Omega) \propto \Omega - \epsilon_k/(1 + \lambda (k))$, we see
that as the system crosses over to the quantum-critical regime, the 
 pole in the fermionic propagator gradually moves from the real frequency 
axis onto the imaginary axis. The pole along imaginary axis implies that 
fermions are completely overdamped and can only propagate diffusively. 

Simultaneously, the dynamic spin susceptibility acquires a strong 
frequency dependence  which also gives rise to the diffusive 
behavior near $q=Q$:
\begin{equation}
\chi ({\bf q}, \omega) 
\approx \frac{\chi_0}{({\bf q}-{\bf Q})^2 - i |\omega| \gamma}
\label{s5}
\end{equation}
where $\gamma = (\omega_{sf} \xi^2)^{-1}$ is independent of $\xi$.
Alternatively speaking, both fermionic and spin excitations are diffusive 
at $\omega \sim {\bar \omega} \gg \omega_{sf}$.

Finally, in the true quantum-critical regime
 $\sigma_1  (\omega)$ interpolates between
$\omega^{-1/2}$  at $\omega < {\bar \omega}$, 
and  $\omega^{-3/2}$ at the highest frequencies. 
The imaginary part of conductivity interpolates 
between $\omega^{-1/2}$ for 
$\omega < {\bar \omega}$ and $\omega^{-1}$ for $\omega > {\bar \omega}$. 
We found that  the crossover region 
for $\sigma_2 (\omega)$ is rather narrow, 
but that for $\sigma_1 (\omega)$ is very wide, and in a wide range of frequencies $\omega > {\bar \omega}$, $ 4\pi/\omega^2_{pl}~
 \sigma_1 (\omega)$ is close to 
$0.2 \omega^{-1}$. 

\subsubsection{quantum-critical point}

Finally, we analyzed the forms of the self-energy and the spin 
susceptibility when  $\omega_{sf} \rightarrow 0$, and for fermions near
 hot spots 
the quantum-critical behavior extends down to the lowest frequencies.
In this limit, there exists a wide
 frequency range where on one hand $\omega \gg \omega_{sf}$, and on 
the other hand $\omega \ll {\bar \omega}$, i.e.,  a truly 
{\it low-energy} quantum-critical regime. 
We found that in this regime, 
Eqs (\ref{s4}) and (\ref{s5}) are modified by subleading, 
$\log (\omega/{\bar \omega})$ terms in the perturbation series.  
We studied the effects of 
extra logarithms in the one-loop RG theory assuming that the 
number of hot spots $N$ is large and neglecting nonlogarithmic corrections in $1/N$ ($N=8$ in real situation). 
We found that the 
spin dynamics is  still described by the dynamical exponent $z=2$, but
the dynamical susceptibility acquires an anomalous dimension:
\begin{eqnarray}
\chi ({\bf q}, \omega_m) &\propto& 
( \gamma |\omega_m| + ({\bf q}-{\bf Q})^2)^{-1+\eta} \nonumber \\
&&  [\log (\gamma |\omega_m| + ({\bf q}-{\bf Q})^2)]^{-v_y/3v_x}
\label{s8}
\end{eqnarray}          
where $\eta = 2/N = 0.25$,  and 
and $v_y$ and $v_x$ are
the components of the bare Fermi velocity near hot spots: 
$\epsilon_k = v_x k_x + v_y k_y; \epsilon_{k+Q} = -v_x k_x + v_y k_y$.
The appearance of a finite $\eta$ is the consequence  of the 
fact that the spin-fermion model contains  anomalies which give rise to a complex structure of the four-boson vertex at the QCP. In this respect, our
theory  differs from the effective bosonic $\phi^4$ theory, which is 
marginal in $d=2$ since  $z+d =4$. 

The fermionic self-energy is not affected by the anomalous dimension $\eta$ 
and differs from (\ref{s4}) only by a logarithmical factor:
\begin{equation}
\Sigma (\Omega_m) \propto |\Omega_m|^{1/2}~  
|\log \Omega_m|^{-1/2}                            \label{s9}
\end{equation}   
Finally, the Fermi velocity is also logarithmically renormalized in the 
quantum-critical region:
\begin{equation}
 v^R_x = v_x \left(1 + \frac{24 L}{\pi N} \frac{v_y}{v_x}\right)^{1/2};
 v^R_y = v_y \left(1 + \frac{24 L}{\pi N} \frac{v_y}{v_x}\right)^{-1/2} 
\label{s10}
\end{equation}
where $L = |\log[{\mbox{min} (\xi^{-1}, \gamma \omega)}]|$
This renormalization implies that as $\xi$ diverges, $v^R_x \rightarrow 0$,
and the velocities at ${\bf k}_{hs}$ and ${\bf k}_{hs} +{\bf Q}$ become 
antiparallel to each other, i.e., the Fermi surface becomes ``nested'' at
hot spots.  This ``nesting'' is 
the first step in the transformation from a large Fermi 
surface to a small one, 
consisting of hole pockets. If this effect occurred
 at a finite $\xi$, then one might expect a subsequent 
topological transition in which a large Fermi surface disconnects into  
hole pockets and the rest.  We found, however, that the ``nesting'' 
occurs only right at $\xi = \infty$, and is only a weak 
logarithmical effect. This is a consequence 
of the fact that the low frequency 
spin excitations are diffusive rather than propagating.
If the Fermi surface did not contain hot spots or the damping was just 
weaker, one could expect
 more stronger nesting effects and, accordingly, some spin-density-wave precursors in the normal state. 

The logarithmical modification of  the fermionic self-energy 
 gives rise to logarithmical modifications of the conductivities: 
both real and imaginary parts of the conductivity behave 
as $\sigma (\omega) \propto |\log \omega|^{1/2}/\sqrt{\omega}$.
 
We next list the results at finite $T$.

\subsection{finite $T$} 

\subsubsection{Fermi liquid regime}  

The Fermi liquid regime for $T$ dependent terms in the fermionic self-energy
is confined to $\pi T \leq 0.5 \omega_{sf} (k)$. For these 
temperatures we obtained
 $\Sigma (k, \Omega) = \Sigma (k, \Omega, T=0) + \Sigma_T (k, \Omega)$ where
$\Sigma (k, \Omega, T=0)$ is given by (\ref{s1}), and $\Sigma_T (k, \Omega)$
 predominantly affects the imaginary part of the self-energy: 
\begin{equation}
\Sigma_T (k, \Omega) = i~ 
\frac{\pi^2 T^2}{4 \omega_{sf} (k)} \mbox{sign} \Omega~
F\left(\frac{\Omega}{\omega_{sf}(k)}\right)~ +O(T^4) 
\label{s11}   
\end{equation}
where $F(x)$ is a smooth function of $x$ with the limits 
$F(0) =1$, $F(x \gg 1) \rightarrow 1/3$. The full expression, 
including $T^4$ terms, is presented in (\ref{B8}).

We found that the spin susceptibility is not affected by a 
finite temperature except for a regular, $O((T/\Lambda)^2)$ 
correction to the correlation length. The conductivities 
are affected and change to $\sigma_1 (\omega) \propto T^{-2}$ 
and $\sigma_2 (\omega) \propto \omega/T^4$ at $\omega < T^2/\omega_{sf}$. The 
resistivity indeed obeys a Fermi liquid form 
$\rho (T) \propto T^2$. 
   
\subsubsection{quantum-critical regime, intermediate temperatures}

At intermediate temperatures, $0.5 \omega_{sf} < T < N \omega_{sf}$,
 we found that the thermal contribution 
to the self-energy is predominantly linear in $T$
In the formal limit $N \gg 1$, there are two different types of linear 
behavior.  One occurs at relatively high $T$, when on one hand, 
$ T < N \omega_{sf}$, and on the other hand, $ T $ is 
larger than $6-8 \omega_{sf}$. In this regime, the dominant 
piece in $\Sigma_T ({\bf k}, 0)$ is
 the impurity-like contribution from the scattering by
 static spin fluctuations. It yields 
$\Sigma_T ({\bf k}, 0) = i \pi T \lambda_k$. 
The contribution to $\Sigma_T ({\bf k}, 0)$ from dynamical spin fluctuations is
smaller and scales as $\sqrt{T}$. In explicit form (see Eqn (\ref{B8}))
\begin{equation}
\Sigma^{\prime \prime}_T (k, 0) =  \pi T \lambda (k) -
 1.516 \left(\frac{T\bar{\omega }}{2}\right)^{1/2} 
\label{s12}
\end{equation}

At $ T < 6-8 \omega_{sf}$, the scattering from dynamical 
spin fluctuations also yields a
linear in $T$ piece in the fermionic self-energy. In this temperature range, we
found, to a good accuracy
\begin{equation}
\Sigma^{\prime \prime}_T (k, 0) = 0.4~\pi T \lambda (k)   
\label{s121}
\end{equation}

At finite $\Omega$ the self-energy is more involved 
and is generally given by 
Eq. (\ref{Bbbbb81}). The dynamical piece in the self-energy
 is well approximated by the $N = \infty$ result, 
Eqns. (\ref{seT22222}) and (\ref{seT55}). 
In the static piece, the corrections to the $N = \infty$ result $i \pi T \lambda$ become relevant above typical frequency 
evolves is ${\bar \Omega} = N^2 \omega_{sf} (v^2_F/v_x v_y)/36$.
At $N = \infty$ and finite $\lambda$, this ${\bar \Omega}$ exceeds 
${\bar \omega} = 4 \lambda^2 \omega_{sf}$. However, at finite $N$ and large 
$\lambda$, ${\bar \Omega} \ll {\bar \omega}$. 
We found a controllable way to compute the full fermionic $\Sigma (\Omega)$ at $T \ll N \omega_{sf}$ and arbitrary $\Omega/{\bar \Omega}$.
We found that at
${\bar \omega} > \Omega > {\bar \Omega}$, the static
 piece in the fermionic self-energy decreases as 
 $\Sigma^{\prime \prime}_{st} (\Omega) \propto T N \sqrt{{\bar \omega}/\Omega} \log (\Omega/(N^2 \omega_{sf}))$. At these frequencies, the 
 full self-energy (sum of dynamical and static pieces) 
obeys the scaling form $\Sigma (\Omega, T) \propto \sqrt{T} f(\Omega/T)$, 
up to logarithmic corrections.
 For completeness, we presented both the full result for the self-energy, 
Fig.~\ref{sigmaT_exact_2}, and the second-order result, 
Fig~\ref{sigmaT_exact}. We argued that although the derivation 
of the full self-energy is justified for $T < N \omega_{sf}$, 
by numerical reasons,
 vertex corrections may  become relevant
 by numerical reasons  already 
 at $T \geq \omega_{sf}$. We conjectured that at these temperatures,
  the  second-order result may be closer to reality than the ``full'' expression which includes self-energy of intermediate fermions but neglects vertex corrections.

The dynamical spin susceptibility is the same as in (\ref{s7}), and the 
 conductivities scale, at vanishing 
$\omega$, as $\sigma_1 \propto 1/T$, $\sigma_2 \propto \omega/T^2$. 
The behavior of conductivities at finite frequencies is rather 
involved and we refrain from discussing the limits. The full 
result is presented in Fig.~\ref{cond-tau-all}.

\subsubsection{quantum-critical regime, high temperatures}

At high temperatures, $\pi T > N \omega_{sf}$, vertex 
corrections and the corrections to the Fermi velocity 
cannot be neglected, and our controlled computational 
scheme breaks down. In this regime we obtained the results by 
neglecting vertex corrections without justification and
performing
 self-consistent FLEX -type calculations. 
We then verified that the vertex correction, evaluated using the full fermionic 
propagators remain $O(1)$, i.e., at least they do not diverge. 
We found  that the contribution to 
$\Sigma_T ({\bf k}, 0)$ from 
 scattering on static spin fluctuations scales down from $\pi T \lambda$ 
and at the highest $T$ 
behaves, up to logarithms,
 as $\sqrt{T}$ with $\xi-$independent coefficient, i.e., in the same way as the
contribution from the scattering on dynamical spin fluctuations. 
Combining the two contributions, we obtained
\begin{equation}
\Sigma^{\prime \prime}_T (k) = {\bar \omega} \left(\frac{T N}{6 {\bar \omega}}\right)^{1/2}
 \left(\log{\frac{3 T}{N \omega_{sf}}} - 1.516 
\left(\frac{3}{N}\right)\right)^{1/2}
\label{s15}
\end{equation}

This evolution of the thermal piece in the self-energy affects 
the conductivities which scale at $\omega \rightarrow 0$ as 
$\sigma_1 (T) \propto T^{-1/2}, 
 \sigma_2 (T) \propto \omega/T$.
  
\subsubsection{quantum-critical point}

At $\omega_{sf} \rightarrow 0$, the temperature range $\pi T > N \omega_{sf}$ 
extends down to the lowest frequencies.
We found at vanishing $\omega_{sf}$ and a finite $T$ that the fermionic self-energy scales as
\begin{equation}
\Sigma^{\prime \prime}_T  \propto T^{1/2} |\log T|
\label{s16}
\end{equation}
The  
angle between the Fermi velocities at ${\bf k}_{hs}$ 
and ${\bf k}_{hs}+{\bf Q}$  scales as $1/|\log T|$ (see (\ref{sT4})).
 Both of these results are identical to what we obtained at $T=0$, 
 if we substitute $T$  by $\Omega$ 
(see Eqs. (\ref{s9}) and (\ref{s10})). 
Equal powers of logarithms at $T=0, \Omega \neq 0$ and $T \neq 0, \Omega =0$ 
 mean that the system at the QCP 
possesses $\omega/T$ scaling for fermionic variables. This is quite expected, 
in view of the fact that the fermionic self-energy is not 
affected by the anomalous exponent in the spin susceptibility. 
 However, as we said, the self-consistent FLEX solution is uncontrolled,
 and  therefore the equivalence between $\Omega$ and $T$ 
in our solution is not the proof that
 the $\omega/T$ scaling  actually  exists.
       
On the other hand, our theory definitely yields no $\omega/T$ scaling 
for the dynamical spin susceptibility. At $T=0$, we found the 
anomalous exponent $\eta$ (see (\ref{s8})). At finite $T$ and 
$\omega = ({\bf q}-{\bf Q})^{2} =0$, we found 
that the susceptibility behaves regularly, as $\chi \propto 1/T^2$. 
This behavior is certainly modified due to $\phi^4$ interaction 
between bosonic modes. This effect is, however, beyond the scope 
of our theory as the $\phi^4$ interaction is produced by high-energy 
fermions and scales as $1/\Lambda$. The anomalous behavior at $T=0$ on the 
other hand is universal and survives in the 
limit $\Lambda \rightarrow  \infty$. 

\subsection{comparison with experiments}

We now compare our key results for the spectral function and 
conductivity with the experimental data on optimally doped high $T_c$ cuprate 
$Bi2212$  for which 
both photoemission and conductivity data are available.
We remind that the two inputs in our theory are the coupling constant
$\lambda$ and the overall scale ${\bar \omega} = 4 \lambda^2 \omega_{sf}$.
Alternatively, we can reexpress $\lambda$ as 
$\lambda = 3 v_F \xi^{-1}/(16 \omega_{sf})$ and 
use $v_F \xi^{-1}$ and $\omega_{sf}$ as input parameters.

The value of the Fermi velocity can be obtained from the photoemission data at
high frequencies, when the self-energy corrections to the 
fermionic dispersion become relatively minor. 
The three groups which reported the 
MDC data on $Bi2212$ for momenta along zone diagonal~\cite{kaminski,johnson1,bogdanov1}
all agree that that the bare 
value of the Fermi velocity along the diagonal 
is  rather high: $2.5 -3 eV A$, or $0.7-0.8 eV a$
 where $a \sim 3.8 A$ is the $Cu-Cu$ distance. 
We used the $t-t^{\prime}$ model to relate this 
velocity with that at hot spots. Using the value of the velocity and the
 experimental facts that the Fermi surface is located 
at ${\bf k} \approx (0.4 \pi/a, 0.4 \pi/a)$ for momenta along 
zone diagonal and at $ {\bf k} \approx 
(\pi/a, 0.2 \pi/a)$ for $k_x$ along the zone boundary, we found 
$t \sim 0.2-0.25 eV$, 
$t^{\prime} \approx -0.36 t$ and $\mu \approx -1.1 t$. These numbers
 roughly agree with other studies~\cite{dagotto}. 
 The
hot spots are located at ${\bf k}_{hs} = (0.16\pi, 0.84 \pi)$ 
and symmetry related points, and the velocity at a hot spot is approximately 
a half of that at zone diagonal. This yields $v_F \approx 0.35-0.4 eV a$.

The values of $\omega_{sf}$ and $\xi$
 can in principle be extracted from neutron scattering 
data on $S({\bf q},\omega) 
\propto \omega /((1 +({\bf q}-{\bf Q})^2 \xi^2)^2 +
(\omega/\omega_{sf})^2)$, 
 and from NMR data.
We are not familiar with the detailed analysis of the normal state 
neutron and NMR data for $Bi2212$ and will rely on the data for near 
optimally doped $YBCO$.
The NMR analysis~\cite{nmr_extract}
yields $\omega_{sf} \sim 20 meV$ and $\xi \sim 2a$. 
 The  neutron data in the normal state 
are more difficult to analyze because of the background which 
increases the width of the neutron peak and 
masks some frequency dependence. The data  show~\cite{neutron_extract}
 that the dynamical structure factor in the normal state  is indeed 
peaked at ${\bf q}= {\bf Q}=(\pi/a,\pi/a)$, and the  
width of the peak increases with frequency 
and at $\omega =50 meV$ reaches $1.5$ of its value at $\omega =0$.
A straightforward fit to the theoretical formula, Eqn(\ref{s7}), yields  
$\omega_{sf} \sim 35-40 meV$ and $\xi \sim a$ 
which are predictably larger than $\omega_{sf}$ and $\xi$ 
extracted from NMR. We will use NMR values 
$\omega_{sf} \sim 20 meV$ and $\xi =2a$ 
for further estimates. 

Combining the results 
for $v_F$, $\xi$ and $\omega_{sf}$, we obtain $\lambda \sim 2$.
 This in turn yields ${\bar \omega} \sim 0.3 eV$. 
As an independent check of the internal consistency of the 
estimates, note that our recent analysis of the superconducting 
state~\cite{acs} yields the resonance neutron frequency at 
$\omega_{res} \sim 0.25{\bar \omega}/\lambda$, i.e., at 
$\omega_{res} \sim 40 meV$. This is quite consistent with the data. 

Away from hot spots, the effective coupling decreases as 
$\lambda (k) = \lambda (1 + (\epsilon_{k+Q}/v_F\xi^{-1})^2)^{1/2}$. 
This formula is indeed valid only in some vicinity near hot spots as 
we didn't include in the theory the variation of the Fermi velocity 
along the Fermi surface. Nevertheless, experimentally, even for 
$k = k_F$ along zone diagonal (the furthest $k$ point away from a 
hot spot), $\epsilon_{k_F+Q}/v_F (k_F)\xi^{-1} \approx 1.4$~\cite{valla}. 
This shows that the 
coupling does vary along the Fermi surface, 
but this variation is modest: $\lambda$ is reduced 
by at most $1.7$ when we move from hot spots towards zone diagonal. 
Actual reduction can be even smaller as $v_F$ by itself increases as
 one approaches zone diagonal. 
For $\omega_{sf} (k)$, our theory predicts that it increases with the 
deviation from hot spots. Note, however, that 
$\omega_{sf} \propto \sin \phi_0$, where $\phi_0$ is the angle 
between velocities at ${\bf k}$ and ${\bf k}+{\bf Q}$.
In our theory, we assumed that this angle does not change.  
 In reality, $\phi_0$ angle tends to $\pi$ as ${\bf k}$ approaches the 
zone diagonal, and this {\it reduces} $\omega_{sf}$.

\begin{figure}[tbp]
\begin{center}
\epsfxsize=\columnwidth 
\epsffile{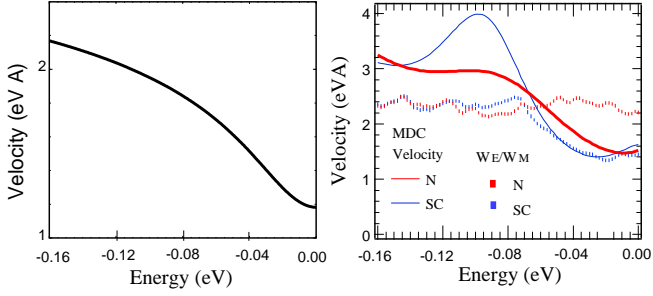}
\end{center}
\caption{a. The theoretical result for the effective velocity
 of the quasiparticles $v^*_F = v_F/(1 +
 \partial\Sigma^{\prime}(\omega)/\partial\omega)$. For definiteness we used
  $\omega_{sf} = 20 meV$, $\lambda = 1.7$ and $v_F = 3 eV A$.  
b. Experimental result for the effective velocity, 
extracted from the MDC dispersion \protect\cite{kaminski} along zone diagonal. 
Observe the bump in the frequency dependence of the 
velocity at $70 -80 meV$ in the data and at 
about $3 -4 \omega_{sf}$ in the theory.}
\label{velocity}
\end{figure}

 Since  for both $\lambda (k)$ and $\omega_{sf}(k)$ there 
are competing effects which 
 we cannot fully control, 
 we believe that effective $\omega_{sf} (k) $ and $\lambda (k)$ 
 should just
 be obtained from the fit to the photoemission data.
 In particular, $\omega_{sf}$ can be extracted from the 
  MDC (momentum distribution curve)  
measurements of the electronic dispersion 
$\omega + \Sigma^\prime (\omega) = \epsilon_k$.  
In Fig~\ref{velocity}
 we compare our $(1 + \partial\Sigma^{\prime}(\omega)/\partial\omega)$ 
 with the measured variation of 
 the effective velocity $v_F (\omega)$ of the electronic dispersion along zone diagonal~\cite{kaminski}. We see that the theoretical 
dispersion has a bump at $\omega \sim 3 \omega_{sf} (k_{diag})$.
 Experimental 
curves look quite similar and show the bump 
at $\sim 70-80 meV$~\cite{kaminski,johnson1,bogdanov1}. 
This yields $\omega_{sf} (k_{diag}) \sim 25 meV$, i.e., 
 almost the same as near hot spots.

 Note in passing that although 
$\epsilon_{k +Q}/v_F \xi^{-1}$ does not vary much when $k$ moves along the 
Fermi surface,  the fact that 
 the Fermi velocity is very high implies that 
 $\epsilon_{k_F +Q}$ is roughly
$v_F \sqrt{2} *0.22 \pi/a \approx  0.8 eV$, i.e., it is comparable to a bandwidth. 
This implies that  one certainly cannot neglect 
the curvature of the Fermi surface in the theoretical analysis. 
In other words,  the Fermi-surface is very different from a near 
square which one would obtain for only nearest neighbor hopping. 
Furthermore,  the fact that the 
 Fermi velocity is large implies 
the physics at energies up to few hundred meV is confined to a 
near vicinity of the Fermi surface, when one can safely 
 expand $\epsilon_k$ to a linear order in $k-k_F$. Finally, 
  van-Hove singularities (which we neglected) do play some role~\cite{onufr,katanin}
 but as $\epsilon_{0,\pi} \approx 0.34 t \sim  85 meV$, we expect that the
 van-Hove singularity  softens due to fermionic incoherence 
 and should not substantially affect the physics.

We now briefly compare the experimental and theoretical 
forms of the fermionic spectral function and optical conductivity.
    
\subsubsection{spectral function}

\begin{figure}[tbp]
\begin{center}
\epsfxsize=\columnwidth 
\epsffile{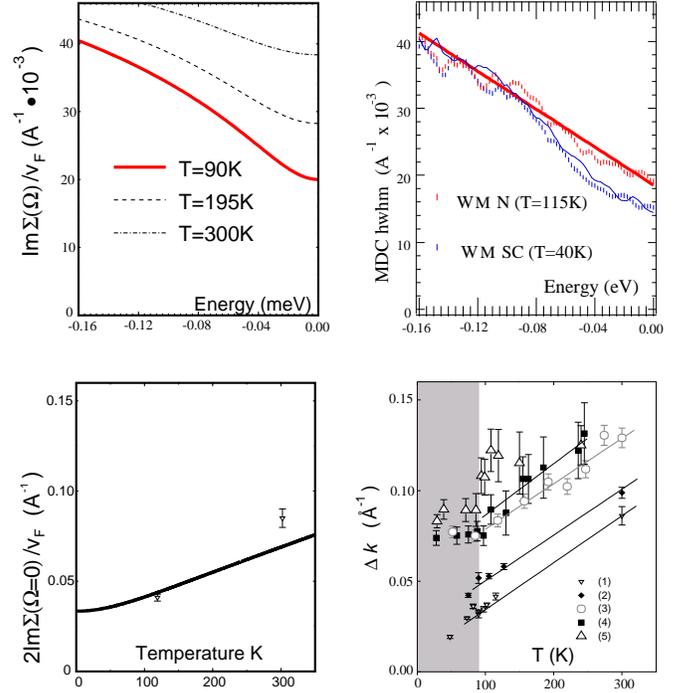}
\end{center}
\caption{A comparison between theoretical results and the photoemission MDC data.   For the Lorentzian lineshape
 of the MDC signal, observed in experiments, the MDC hwhm equals
 to $\Sigma^{\prime \prime}/v_F$.
 Upper panel - the results for the MDC hwhm vs frequency at a given $T$.
 The experimental results are taken from\protect\cite{kaminski}.
 Lower panel - the MDC fwhm vs temperature at $\Omega =0$. 
 The experimental results (right figure and the points on the left figure) are taken from \protect\cite{johnson1}.
}
\label{MDC_exp}
\end{figure}

We first use our form of the fermionic self-energy to fit the 
MDC data which measure the width of the photoemission peak as a function of $k$ at a given frequency. In Fig.~\ref{MDC_exp} we compare our results for $\Delta k =
\Sigma^{\prime \prime} (k, \Omega)/v_F$ with the measured~\cite{kaminski}
 $\Delta k$ vs frequency at $T \sim 100K$ and temperature at $\Omega \rightarrow 0$~\cite{johnson1}. For definiteness 
 we used $\lambda =1.7$ and $\omega_{sf} = 20 meV$. The slope of $\Delta k$ is
 chiefly controlled by $\lambda$. We see that 
$\lambda \leq 2$ yields rather good agreement with the data on  
both, frequency and temperature dependence of the self-energy.
On the other hand, 
the magnitude of our $\Sigma^{\prime \prime}$ is smaller than in the data. 
 To account for the values of $\Delta k$, we had to {\it add} a constant 
 of about $50 meV$ to $\Sigma^{\prime \prime}$.  The origin of this constant term is unclear. It may be the effect of impurities~\cite{va}, and it also may be the effect of other scattering channels which we ignored~\cite{im2}.
 It is essential, however, that 
the functional dependence of  $\Sigma^{\prime \prime} (\Omega T)$ is 
 captured by spin-fluctuation scattering. 

\begin{figure}[tbp]
\begin{center}
\epsfxsize=\columnwidth 
\epsffile{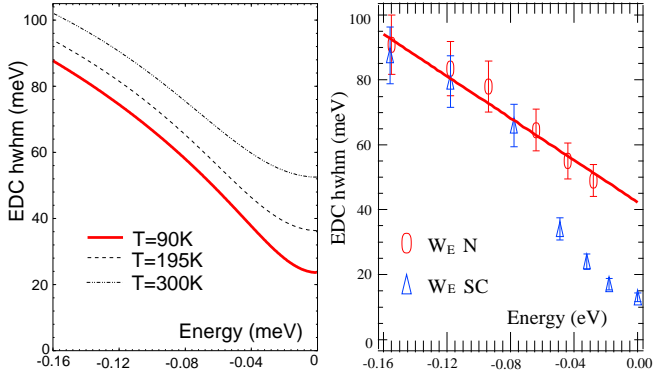}
\end{center}
\caption{The theoretical result for  
$\Sigma^{\prime \prime} (\Omega)/(1 + \Sigma^\prime (\Omega)/\Omega)$. 
The points are the experimental hwhm of the EDC dispersion
 from \protect\cite{kaminski}. 
}
\label{EDC_exp}
\end{figure}

In Fig~\ref{EDC_exp} we present our results for the 
hwhm of the EDC (energy
distribution curve) which measures fermionic 
$I_k(\Omega) = A_k(\Omega) n_F(\Omega)$ as a function of 
frequency at a given $k$ ($A_k (\Omega) = 
(1/\pi) Im G(k, \Omega)$). 
For Lorentzian lineshape, the EDC hwhm is given by $\Sigma^{\prime \prime} (\Omega)/(1 + \Sigma^\prime (\Omega)/\Omega)$~\cite{kaminski}. 
The data are taken at $T =115K$~\cite{kaminski}. 
Again,  the theoretical slope reasonably agrees with the experimental one. 
Some discrepancy is associated with the fact that the experimental lineshape
 is not a Lorentzian and hence the measured hwhm is not exactly 
 $\Sigma^{\prime \prime} (\Omega)/(1 + \Sigma^\prime (\Omega)/\Omega)$.

\subsubsection{conductivity}

\begin{figure}[tbp]
\begin{center}
\epsfxsize=\columnwidth 
\epsffile{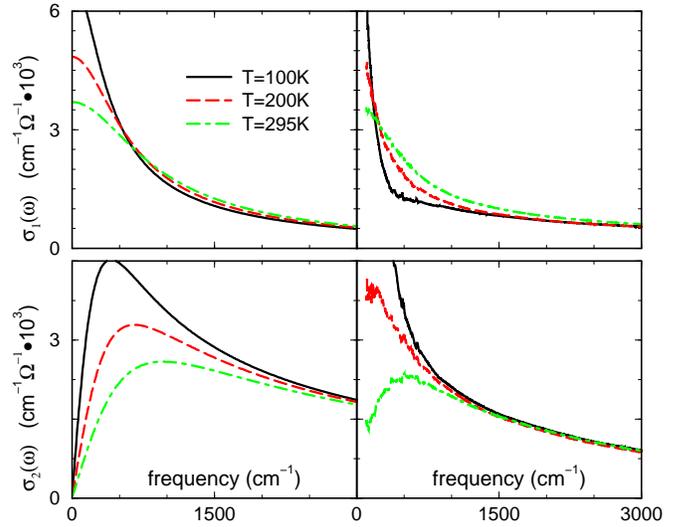}
\end{center}
\caption{The theoretical and experimental results for 
the real and imaginary parts of  optical conductivity. 
The data are from  \protect\cite{tu}.
}
\label{conductivity_exp}
\end{figure}

In Fig. In Fig.~\ref{conductivity_exp} we compare our theoretical results 
for the conductivity with the experimental data for $\sigma_1$ and $\sigma_2$
at different temperatures~\cite{tu}. 
For definiteness we used the same $\lambda =1.7$ and
$\omega_{sf} = 20meV$ as in the fit to the photoemission data along zone diagonals. We checked that the change in $\lambda$ affects the ratio $\sigma_2/\sigma_1$ at high frequencies, but does not change the functional forms of the conductivities. We adjusted the plasma frequency to obtain the agreement between dc conductivity and $Sigma^{\prime \prime}$ extracted from the MDC photoemission data 
using $v_F \sim 3 eV A$. This adjustment yields
$\omega_{pl} \sim 20000 cm^{-1}$. This value is somewhat larger that 
 $\omega_{pl} \sim 16 000 cm^{-1}$ obtained experimentally by integrating 
$\sigma_1$ up to about $2-2.5 eV$~\cite{basov,tu,qui}. This discrepancy is consistent with our theoretical result that the sum rule is satisfied only at extremely high frequencies
 (see Fig.~\ref{sumRule}). 

We see that theoretical $\sigma_1 (\omega)$ and $\sigma_2 (\omega)$ capture
the essential features of the measured forms of the conductivities. In particular, the curves of $\sigma_1$ at different temperatures cross such that at the lowest frequencies, the conductivity decreases with $T$ which at larger frequencies it increases with $T$ (the same behavior has also been detected in ~\cite{timusk,qui}.  The imaginary part of conductivity decreases with $T$ at any frequency, and the peak in $\sigma_2 (\omega)$ increases in magnitude and shifts to lower $T$ with decreasing $T$~\cite{tu,qui,orenst}. At frequencies above $1500 cm^{-1}$ both $\sigma_1$ and $\sigma_2$ weakly depend on $T$ and are comparable in the amplitude.

\begin{figure}[tbp]
\begin{center}
\epsfxsize=\columnwidth 
\epsffile{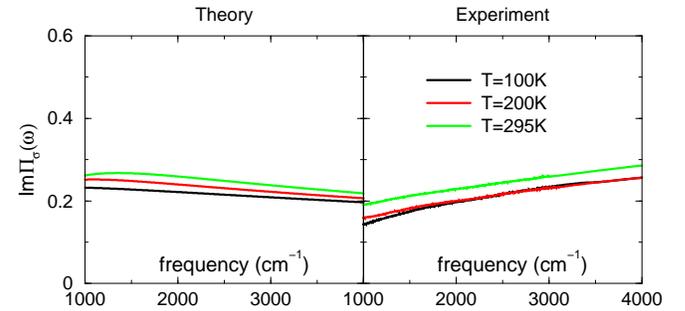}
\end{center}
\caption{The theoretical and experimental results for 
$\Pi_\sigma^{\prime \prime} (\omega) = 4\pi \sigma_1 \omega/\omega^2_{pl}$.
The data are from  \protect\cite{tu}.
}
\label{pol}
\end{figure}

In Fig~\ref{pol} we compare experimental and theoretical results for the 
imaginary part of the full particle-hole polarization bubble 
$\Pi_\sigma^{\prime \prime} (\omega) = 4\pi \sigma_1 \omega/\omega^2_{pl}$
Theoretically, at $T=0$, $\Pi^{\prime \prime}_\sigma (\omega)$ saturates 
at a value of about 0.2 {\it independently} on $\lambda$ 
and remains almost independent on frequency over 
 a very wide frequency range (see Fig. \ref{zeroT_0wsf_Polariz}). 
We see that the theoretical value of $\Pi_\sigma$ does not change much with $T$. Experimental data also clearly show a near saturation of 
$\Pi_{\sigma }^{\prime \prime}$ at a value close to $0.2$. 
We consider this agreement as a strong argument in favor of our theory.

\begin{figure}[tbp]
\begin{center}
\epsfxsize=\columnwidth 
\epsffile{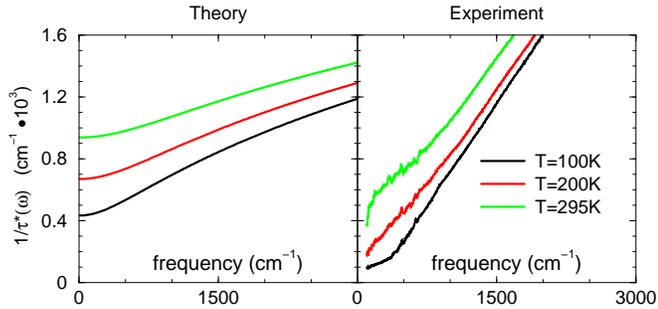}
\end{center}
\caption{The theoretical and experimental results for 
$1/\tau^* = \omega \sigma_1/\sigma_2$. 
The data are from  \protect\cite{tu}.
}
\label{tau_star}
\end{figure}

The agreement between our theory and the experiment is, however, not
a perfect one. In Fig.~\ref{tau_star} we show theoretical and experimental results 
 for $1/\tau^* = \omega \sigma_1/\sigma_2$. The advantage of 
comparing $1/\tau^*$ 
is that this quantity
 does not require one to know what the plasma frequency is. 
We see that while both experimental and theoretical curves  are linear in
 frequency, the slopes are off roughly by a factor of $3$.
This discrepancy is related to the fact that in our theory, at high enough 
frequencies, $Pi^{\prime}_\sigma (\omega)$ is roughly 3 times larger than 
$\Pi^{\prime \prime}_\sigma (\omega)$ (see Fig. \ref{zeroT_0wsf_Polariz})
 and hence $\sigma_2/\sigma_1 \sim 3$, whole experimentally $\sigma_1$ and 
$\sigma_2$ nearly coincide at high frequencies. Also,  at zero frequency, the
 theoretical value of $1/\tau^* (\omega =0)$ at, say, 
 $T =300K$ is larger
 than in the data~\cite{tu,qui}. From Fig~\ref{tau_star} 
the ratio of theoretical and experimental $1/\tau^* (\Omega =0)$ 
is roughly 1.5. 

 The discrepancy in $1/\tau^*$ 
is important for the theory as it indicates 
that either the averaging over the Fermi surface and vertex correction within a bubble, or the RPA-type corrections to conductivity~\cite{ioffe_larkin,im2} 
 do play some role. Still, however, Figs. \ref{conductivity_exp} 
 and \ref{pol} indicate
that the general trends of the behavior of the conductivities, at least near optimal doping, are reasonably well captured by the spin-fluctuation theory.     

In this paper we collected all what we know at the moment 
about the normal state properties 
of the spin-fermion model. The next step is to consider what happens 
below the pairing instability. We plan to 
present the detailed account of these results in the near future. 

\section{acknowledgements}

It is our pleasure to thank A. M. Finkel'stein for stimulating
discussions on numerous aspects of strong coupling effects in cuprates. We
are also thankful to E. Abrahams, A.A. Abrikosov, B.L. Altshuler, 
D. Basov, G. Blumberg, J.C. Campuzano, P. Coleman, L.P.
Gor`kov, M. Grilli, L. Ioffe, P. Johnson, R. Joynt, B. Keimer, R. Laughlin, 
 M. Lavagna, 
D. Khveschenko, G. Kotliar,  H. von L\"{o}hneysen, 
A. Millis, M. Norman, C.  P\'{e}pin, D. Pines, 
A. Rosh, S. Sachdev, Q. Si, O. Tchernyshyov, A.Tsvelik, J. Tu and J.
Zasadzinski for useful conversations. We are also thankful to D. Basov and
J. Tu for sharing their unpublished results with us and to M. Norman for bringing ref.~\cite{wheatley} to our attention. The research
was supported by NSF DMR-9979749 (Ar. A and A. Ch.), and by the Ames
Laboratory, operated for the U.S. Department of Energy by Iowa State
University under contract No. W-7405-Eng-82 (J.S). This work was 
 partly done while A.C. was on a sabbatical leave at 
 the Rutgers University and the NEC Research Institute. The hospitality of both  places is acknowledged with thanks. 

\section{Appendix A}

In this Appendix, we evaluate 
several integrals presented in Sec.~\ref{n_s}. 

\subsection {direct perturbation theory}

We first compute the second order 
fermionic self-energy and the vertex correction in a
direct perturbation theory for the spin-fermion model. 
``Direct'' means that the calculations are performed with the 
bare value of the spin susceptibility.

We begin with the perturbative 
fermionic self-energy $\Sigma ({\bf k}, \Omega_m) = 
(\Omega_m - \epsilon_{k+Q})~I({\bf k}_{hs}, 0)$, where $I ({\bf k}_{hs}, 0)$
is given by (\ref{pertnew0})
\begin{eqnarray}
I({\bf k}_{hs}, 0) &=&  \frac{3 ~{\bar g}}{(2\pi)^3} 
\int d^2 {\tilde {\bf q}} d \omega_m
~\frac{1}{\xi^{-2} + {\tilde {\bf q}}^2 + (\omega/v_s)^2}    \nonumber \\ 
&\times&\frac{1}{( i \omega_m  - v_F {\tilde q}_x)^2}
\label{ssss}
\end{eqnarray}
Here ${\tilde {\bf q}} = {\bf q}-{\bf Q}$, 
and $x-$axis is chosen along ${\bf v}_F$ at ${\bf k} + {\bf Q}$.   
Rescaling the momentum and frequency ${\tilde q}_{x}=\xi ^{-1}x$, 
${\tilde q}_{y}=\xi ^{-1}y$,
and $\omega =\xi ^{-1}v_{s}z$, transforming to the spherical coordinates
and integrating over $r=\sqrt{x^{2}+y^{2}+z^{2}}$,
we obtain ($\eta = v_{F}/v_{s}$)
\begin{eqnarray}
I(k_{hs}, 0) &=& -\frac{\lambda }{2\pi \eta }
\int\limits_{0}^{\pi /2}\frac{d\theta }{\sin \theta }
\int \limits_{0}^{2\pi }  
\frac{d\phi }{\left(i\eta ^{-1}\cot\theta -\cos\phi\right)^{2}}  \nonumber\\
&=&-\lambda \eta\int \limits_{0}^{\pi /2}
\frac{\sin \theta \cos \theta d\theta }
{\left[1 +(\eta ^{2}-1)\sin ^{2}\theta \right]^{3/2}} =
- \frac{\lambda }{1+\eta },                                    \label{as2}
\end{eqnarray}

Similar analysis works also for the vertex correction. 
We have from (\ref{vert1}) 
\begin{eqnarray}
\frac{\Delta g}{g} &=& -\frac{\bar g}{8\pi^3} 
\int d^2 {\tilde q} d \omega
~\frac{1}{\xi^{-2} + {\tilde {\bf q}}^2 +\left(\omega/v_s\right)^{2}}\nonumber \\
&\times&\frac{1}{i \omega - (v_x {\tilde q}_x + v_y {\tilde q}_y)}
~\frac{1}{i \omega + v_x {\tilde q}_x - v_y {\tilde q}_y}~         \label{av1}
\end{eqnarray}
Rescaling the momentum and frequency in the same way as before,
 we obtain from (\ref{av1})
\begin{eqnarray}
\frac{\Delta g}{g}&=& -\frac{\lambda }{6\pi ^{2}}\frac{v_{F}}{v_{s}}
\int \frac{dxdydz}{1+x^{2}+y^{2}+z^{2}}
\frac{1}{(iz-\alpha y)^{2}-\beta ^{2}x^{2}}                 
\end{eqnarray}
where $\alpha = v_{y}/v_{s}$ and $\beta =v_{x}/v_{s}$ ($\alpha^2 + \beta^2 = \eta^2$). Introducing spherical coordinated and integrating over $r$ 
we  obtain
\begin{eqnarray}
&&\frac{\Delta g}{g} = - \frac{ \lambda v_{F}}{6\pi v_{s}}\int_{0}^{\pi /2}\!\!\!
\frac{d\theta }{\cos \theta }
\int _{0}^{2\pi }\!\!\!\!\!\frac{d\phi }
{(i\tan \theta -\alpha \sin \phi )^{2}-\beta ^{2}\cos ^{2}\phi } \nonumber \\
&&=\frac{ \lambda v_{F}}{3v_{s}}\!\int _{0}^{\pi /2}\!\!\!\!\!
\frac{d\theta \cos \theta \sin \theta }
{\left[\sin ^{2}\theta +\alpha ^{2}\cos ^{2}\theta  \right]
\sqrt{\sin ^{2}\theta +(\alpha ^{2}+\beta ^{2})\cos ^{2}\theta}} \nonumber\\
&&=\frac{\lambda}{3}\frac{v_{F}}{v_{s}}\frac{1}{\beta \sqrt{1-\alpha ^{2}}}
\ln \frac{\alpha \left(\sqrt{1-\alpha ^{2}}-\beta  \right)}
{\sqrt{\left(\beta ^{2}+\alpha ^{2} \right)
\left(1-\alpha ^{2}\right)}-\beta}                                \label{av3}
\end{eqnarray}

At $v_s\rightarrow\infty$, i.e., for a purely static bare spin susceptibility,
 $\Delta g/g$ becomes
\begin{equation}
\frac{\Delta g}{g} =  
\frac{\lambda}{3}\frac{v_{F}}{v_{x}}\sinh^{-1}\frac{v_{x}}{v_{y}} \label{av4}
\end{equation}
For $v_x = v_y$, we have $\frac{\Delta g}{g} = \frac{\lambda \sqrt{2}}{3}\ln (1+\sqrt{2})
\approx 0.415 \lambda$.

In the opposite limit $v_s \ll v_F$, we obtain from (\ref{av3})
\begin{equation}
\frac{\Delta g}{g} = 
\frac{\lambda}{3}~\frac{v_s v_F}{v_{x}v_{y}}~ \tan^{-1}\frac{v_x}{v_y}   \label{av5}
\end{equation}

\subsection{renormalized perturbation theory}

We next repeat these computations using the
 relaxational form of the spin susceptibility. 
We first re-evaluate
$I({\bf k}_{hs}, 0)$. 
From Eq. (\ref{pertnew}) we have 
\begin{eqnarray}
I({\bf k}_{hs}, 0) &=&  3 {\bar g}\xi^2 
\int \frac {d^2 {\tilde q} d \omega_m}{(2\pi)^3}
~\frac{1}{1 + ({\tilde {\bf q}} \xi)^2 + \frac{|\omega_m|}
{\omega_{\rm sf}} + \frac{\omega^2_m}{v^2_s \xi^{-2}}}        \nonumber \\ 
& & \times \frac{1}{(i\omega_m - v_F {\tilde q}_x)^{2}}~      \label{pertnew1}
\end{eqnarray} 
The key difference with the direct perturbation theory is that now
the $\omega^2$ term in the susceptibility becomes subleading at low frequencies.
Let's first neglect this term. 
 Introducing, as before, 
${\tilde q}_x \xi = x,~{\tilde q}_y \xi =y$, and also 
$\omega_m = \omega_{sf} t$,
 we then rewrite
Eq. (\ref{pertnew1}) as
\begin{eqnarray}
I({\bf k}_{hs}, 0) &=&  \frac{ 3 v_x v_y}{2 \pi^2 N v^2_F}
~\int d x d y d t 
\frac{1}{1 + x^2 + y^2 + |t|}                       \nonumber \\ 
&\times& ~ \frac{1}{(x  - i t/a)^2}~               \label{pertnew2}
\end{eqnarray} 
where 
$a = (v_F \xi^{-1}/\omega_{sf}) = (N \lambda/3)~(v^2_F/v_x v_y)$.
One can easily make sure that the neglect of the 
 $\omega^2$ term in the susceptibility is justified when $a \gg 1$
 (see also below). The integration over $y$ is straightforward and yields
\begin{eqnarray}
I({\bf k}_{hs}, 0) &=&  \frac{ 3  v_x v_y}{2 \pi N v^2_F}
~\int \frac{ d x d t}
{(1 + x^2 + |t|)^{1/2}}                            \nonumber \\ 
&\times& ~ \frac{1}{(x  - i t/a)^2}                      \label{pertnew3}
\end{eqnarray} 
The last term in (\ref{pertnew3}) is a double pole. The integral 
over $x$
then does not vanish only because the first term contains a branch cut. 
Deforming the contour of integration over $x$ to include the 
branch cut along imaginary  $x$,
 and integrating over the branches of the branch cut, we obtain
\begin{equation}
I({\bf k}_{hs}, 0) = -\frac{12 v_x v_y}{\pi N v^2_F}~\int_0^{\sim 1/a}
 d t \frac{1}{1 + t}.                      \label{pertnew4} 
\end{equation}
Evaluating the integral with logarithmical accuracy, we obtain  
\begin{equation}
I_{reg} ({\bf k}_{hs}, 0) =  -\frac{12 v_x v_y}
{\pi N v^2_F}~\log\lambda                                     \label{Iregapp}
\end{equation}
This is the result we cited in the Eq. (\ref{Ireg}).

With little more efforts, one can evaluate $I({\bf k}_{hs}, 0)$
 for arbitrary $a$ and $v_s/v_F$ which are two parameters for $I({\bf k}_{hs}, 0)$. To avoid presenting very long formulas, we only consider 
 the two limits $v_s \rightarrow \infty$ (i.e., a purely static bare 
spin susceptibility), and $v_F = v_s$. 
In the first case, performing the same computations as above but keeping 
 $a$ arbitrary, 
 we obtain after some algebra
\begin{equation}
I({\bf k}_{hs}, 0) = - \frac{\lambda}{\pi} \int_0^{\infty}  
\int_0^{\infty}~\frac {y dy dx}
{( 1 + x + a y)(x + y^2)^{3/2}}                              \label{pertnew5}
\end{equation}
The straightforward integration yields 
\begin{equation}
I_{reg} ({\bf k}_{hs}, 0) = - \lambda~ \frac{1 + \frac{a}{\pi} 
\log{\frac{a}{2}}}{1 + \frac{a^2}{4}}                        \label{pertnew6}
\end{equation}
This is the result we cited in the Eq. (\ref{Iint}). 

Similar considerations for $v_s = v_F$ yield
\begin{equation}
I({\bf k}_{hs}, 0) = - \frac{4 \lambda}{\pi a^2}\!\! 
~\left(\pi\!\! -a\! -\!\! \sqrt{4 -a^2} \cot^{-1} 
\frac{a}{\sqrt{4-a^2}}\right)                               \label{pertnew7}
\end{equation}
At $a \rightarrow 0$, $I({\bf k}_{hs}, 0) \rightarrow - \lambda/2$
in agreement with (\ref{as2}).

We next show how to compute the full (regular plus anomalous) second-order 
$\Sigma ({\bf k}_{hs}, \Omega) $ at
 arbitrary $\Omega$. We have
\begin{eqnarray}
&&\Sigma ({\bf k}_{hs}, \Omega_m) = 3 {\bar g} \xi^2 
\int \frac {d^2 {\tilde q} d \omega_m}{(2\pi)^3}\nonumber \\
&&~\frac{1}{1 + ({\tilde {\bf q}} \xi)^2 + |\omega_m|/\omega_{\rm sf} +\frac{\omega^2_m}{v^2_s \xi^{-2}}}~\frac{1}{i\Omega_m  + 
i\omega_m  - v_F {\tilde q}_x}            \label{pertnew8}
\end{eqnarray}
Introducing the same variables $x, y ,t$ as before, and 
also $\Omega = s\omega _{sf}$ and using the
spectral representation of the Green's functions
 we obtain \[
Im\Sigma (\Omega ) = \frac{\lambda \omega _{sf}}{\pi }  
Im\!\!\int _{0}^{s }\!\!dt\int\!\!\frac{dy}
{y^{2}\!+\!1\!-\!it +((s \!+\!t)^{2}\!-\!\eta^2 t^{2})/a^2}
\]
The integration is straightforward, and the result is
\begin{equation}
Im\Sigma (\Omega ) =  Im\frac{\lambda \omega _{sf} a}
{\sqrt{1-\eta^{2}} } 
\ln \frac{\sqrt{s +t_{1}}+\sqrt{s +t_{2}}}
{\sqrt{t_{1}}+\sqrt{t_{2}}}
\label{b101} 
\end{equation}
where $t_{1}$ and $t_{2}$ are the roots of the quadratic equation
\begin{equation}
(1+ it)a^2 = (s +t)^{2}-\eta^2 t^{2} = 0
\end{equation}
For purely static bare susceptibility ($v_{s} = \infty $, 
or $\eta =0$), Eqn (\ref{b101}) is simplified to 
\begin{equation}
\Sigma (k_{hs},\Omega ) = 2\lambda \omega _{sf} a~
\ln \frac{i\sqrt{K_{\Omega }-1}+\sqrt{K_{\Omega }+1}}
{i\sqrt{K_{\Omega }\!\!-\!1\!\!+\!A_{\Omega }} + 
\sqrt{K (\Omega) + 1\!\! -\!A_{\Omega }\!}}\label{sigma-full}
\end{equation}
where
\begin{equation}
K_{\Omega }^{2} = 1+\frac{4}{a^2}\left(1-i\frac{\Omega }{\omega _{sf}}\right);
\,\,\,\,A_{\Omega } = \frac{2i}{a^2} \frac{\Omega}{\omega _{sf}}
\end{equation}
(we used Kramers-Kronig relation to obtain $Re \Sigma$)

 We next evaluate the 
lowest order  correction to the spin-fermion vertex at 
the bosonic momentum ${\bf Q}$. 
 The expression for $\Delta g$ 
in terms of fermionic Green's functions is given by  Eq. (\ref{vert1}).
Neglecting the subleading 
$\omega_m^2$ in the spin propagator, introducing 
$x = {\tilde q}_x \xi,~ y = {\tilde q}_y \xi$,
 and $t = \omega/\omega_{sf}$, and 
expanding
 the quasiparticle energies to linear order in deviations
from hot spots 
 we obtain from (\ref{vert1})
\begin{eqnarray}
\frac{\Delta g}{g} &=& 
\frac{v_x v_y}{2 \pi^2 N}~
 \int d x dy dt
~\frac{1}{1 + x^{2}+y^{2} + |t|}           \nonumber \\
&\times&\frac{1}{(v_{x} x)^{2}-
\left(v_{y} y - i v_F t /a \right)^{2}}            \label{verta1}
\end{eqnarray}
we remind that $a = (N \lambda/3)~(v^2_F/v_x v_y) \gg 1$.
Let's perform the integration over $y$ first. 
As both poles in the last term in 
(\ref{verta1}) are in the
same half-plane, it is convenient to close the 
integration contour over a different half-plane 
where only the  spin susceptibility has a pole. 
Evaluating the integral  we obtain, 
\begin{eqnarray}
&&\frac{\Delta g}{g} =
\frac{v_x v_y}{\pi N}~
 \int_0^{\infty} \di t~ \int_{-\infty}^{\infty}~
\frac{\di x}{\sqrt{1 + x^{2}+ |t|}}       \nonumber \\
&\times& \Big[(v_F x)^{2} + v^2_y (1 + t) + 
v_{F}^{2}\left(t/a\right)^2                           \nonumber \\
&+& 2 v_y v_{F}\frac{t}{a} 
~\sqrt{1 + x^{2}+ |t|}\Big]^{-1}.            \label{verta2}
\end{eqnarray}   
To a logarithmical accuracy, we can neglect $1/a$ terms in the 
integrand of (\ref{verta2}) and set the upper limit of frequency integration 
to be 
$t_{max} \sim a^{2}$.
Introducing next 
$x = (1 + t)^{1/2} p$, and performing the integration over $p$ first,
 we obtain from (\ref{verta2})
\begin{equation}
\frac{\Delta g}{g}=\frac{1}{2N}~Q\left(\frac{v_{x}}{v_{y}}\right)
~\int_0^{a^{2}}\frac{\di t}{1+ t}    \label{verta3}
\end{equation}
where
\begin{eqnarray}
Q\left(\frac{v_{x}}{v_{y}}\right) &=& \frac{4}{\pi}
~\frac{v_x v_y}{v^2_F}~\int_{0}^{\infty} ~
\frac{\di p}{\sqrt{1 + p^2}}~\frac{1}{p^2  + (v_y/v_F)^2}         \nonumber \\
&& =  \frac{4}{\pi}~{\rm tan}^{-1} 
\frac{v_x}{v_y} = \frac{4 \phi_0}{\pi}                         \label{verta4}
\end{eqnarray}
 and $\phi_0$ is the angle between the directions of velocities at hot spots
separated by ${\bf Q}$.
 The integration over $t$ in (\ref{verta3}) is elementary,
 and we obtain
\begin{equation}
\frac{\Delta g}{g} = 
\frac{1}{N}~Q\left(\frac{v_x}{v_y}\right)~\log \lambda         \label{verta5}
\end{equation}
This is the result we quoted in the text.
For $\lambda \rightarrow \infty$, the logarithmic dependence on $\lambda$ 
transforms into the logarithmic dependence on the external 
frequency: $\log \lambda \rightarrow (1/2) 
\log{\omega_{max}/|\omega|}$ where $\omega_{max} \sim v^2_F/{\bar g}$.       

It is  instructive to observe  that the same result can also 
be obtained by noticing that 
the $\omega$ terms 
in the fermionic propagators are much smaller than the $\sqrt{\omega}$ term
from the spin susceptibility, Eq. (\ref{verta1}). These $\omega$ terms can then be neglected, and 
the vertex correction can be straightforwardly 
simplified to
\begin{equation}
\frac{\Delta g}{g} = \frac{2 v_x v_y}{\pi^2 N} I \log \xi,    \label{verta51}
\end{equation}
where
\begin{equation}
I = \int \frac{d {\tilde x} d{\tilde y}}{1\!\! +\! {\tilde x}^2 \!\!+\! {\tilde y}^2}
~\frac{1}{({\tilde x} v_x\!\! +\! {\tilde y} v_y \!\!-\! i0)
({\tilde x} v_x\!\! -\! {\tilde y} v_y\!\! +\! i0)}                             \label{verta52}
\end{equation}
and compared to our previous notations, 
we introduced ${\tilde x} = x/t^{1/2},~{\tilde y} = y/t^{1/2}$. 
Naively, one might expect that the 2D 
integral in (\ref{verta52}) is determined by the two 
poles at vanishing ${\tilde x}$ and ${\tilde y}$. However, the contribution 
to the integral from small ${\tilde x}, {\tilde y}$ diverges as $\log^2$,
and has to be regularized. The extra 
$1 + x^2 +y^2$ in the denominator of (\ref{verta52}) provides such a regularization. We checked that for
 $v_x = v_y$, the regularization is irrelevant, and 
the result for $\Delta g/g$ comes from the two poles at vanishing
${\tilde x}, ~{\tilde y}$.   For
other ratios of velocities, the regularization  yields an extra
contribution to the integral 
from the range where ${\tilde x}, {\tilde y} = O(1)$. A straightforward 
calculation then yields $I = (2\pi/(v_x v_y)) {\rm tan}^{-1} v_x/v_y$, and 
hence the same $\Delta g/g$ as in (\ref{verta5}).  
Physically, this result implies that except for the case when 
$v_x = v_y$, a part of the  vertex correction comes from small 
momenta (in reality, 
of order $\omega/v_F$), 
and a part comes from much larger momenta, of order 
$(\omega/(\omega_{sf} \xi^2))^{1/2} \gg \omega/v_F$.       
 
 Above we computed $\Delta g/g$ with the logarithmical accuracy. 
One can easily check, however,  that
 the 3D integration over momentum and frequency in 
(\ref{vert1}) is convergent, i.e., the subleading, non-logarithmical term can 
 also be   determined  within the low-energy model.
 This calculation is more
convenient to perform in polar coordinates,  by expressing
$\epsilon_k = v_F {\tilde q} \cos (\phi + \phi_0/2)$, $\epsilon_{k+Q} = 
v_F {\tilde q} \cos (\phi - \phi_0/2)$. 
The integration over ${\tilde q}$ is straightforward, 
and performing it we obtain
\begin{eqnarray}
\frac{\Delta g}{g} &=& - \frac{2 |\sin \phi_0|}{\pi^2 N}
Re \int_0^{\pi}~d\phi~
\frac{\log[\sin (\phi/2)]}{\cos \phi + \cos \phi_0}            \nonumber \\
&\times& \left( 2\log a + 
\log{\sin \phi/2} \right)                                      \label{verta6}
\end{eqnarray}
where, we remind, 
$a = v_F \xi^{-1}/\omega_{sf} = (N \lambda/3) v^2_F/(v_x v_y)$. The $\log a$
 term in (\ref{verta6})  is indeed 
the same as in (\ref{verta5})
 The second  piece is the regular contribution to $\Delta g/g$. Evaluating this piece numerically we find that for $\phi_0 = \pi/2$ i.e., $v_x = v_y$, 
which is close to the actual situation in cuprates,
$\Delta g/g = (1/N) (\log a - 0.93)$. This result 
implies  that for moderate $\lambda$, vertex correction is rather small numerically: 
 $\Delta g/g \approx 0.09$ for e.g., $\lambda =1$, and $\Delta g/g \approx 0.18$ for $\lambda =2$. 

\section{Appendix B}
\label{a_formal}
In this Appendix we discuss the general structure of the  $1/N$ expansion.
Suppose for definiteness that we keep the number of hot spots finite and 
extend the theory to a large number of fermionic flavors.
As usual, the extension of the theory to large $N$ implies an appropriate rescaling of the coupling constant ${\bar g}$. In our case, however, we also have
 to rescale the Fermi velocity in the same way as ${\bar g}$.
Indeed, consider the lowest order fermionic loop 
(the particle- hole bubble). We found above that it
 is proportional to $\overline{g}N/v_{F}^{2}$. 
The power of the Fermi velocity in the denominator accounts for the number of
fermion propagators in the diagram. Evidently we would like this diagram
 to be of order one. Then we will have a ``zero-order'' theory with no factor of $N$.  Usually this is achieved by
assuming that the coupling constant is inversely proportional to $N$.
However if we do this extension, we find that the anomalous 
fermionic self
energy, which is also a part of the ``zero-order'' theory 
scales as $1/N$ as it is proportional to 
 $\overline{g}/v_{F}$. (recall that the fermionic self-energy 
contains  only
one fermionic line and no summation over the flavor index).
A proper extension which makes both particle-hole bubble and anomalous 
 fermionic self-energy of order 1 is
\begin{equation}
\overline{g} \rightarrow \overline{g} N  \label{scal1}
\end{equation}
and 
\begin{equation}
v_{F} \rightarrow v_{F} N.
  \label{scal2}
\end{equation}
We now consider the general structure of the Feynman diagrams not included into the ``zero-order'' theory. Since
$v_{F}$ scales with $N$, each running fermionic line acquires 
 a factor $1/N$. Similarly, since $\overline{g}$ scales with $N$,
 a  diagram with $n$ vertices acquires a prefactor $%
N^{n/2}$ (recall that ${\overline g} \propto g^2$). 
Finally, if a given diagram to the thermodynamic potential has $m$
closed fermion loops, a summation over fermionic flavors yields an
 additional prefactor $N^{m\text{ }}$ results. As a result,  a
diagram with $n$-vertices and $m$ closed fermionic loops behaves as 
\begin{equation}
D\left( n,m\right) \propto N^{m-n/2},  \label{diagram}
\end{equation}
We used the fact that the number of internal fermionic lines in the diagram is
$n$.

\begin{figure}[tbp]
\begin{center}
\epsfxsize=0.8\columnwidth 
\epsffile{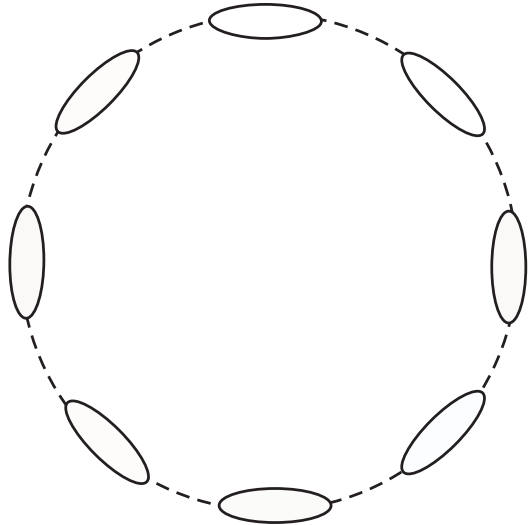}
\end{center}
\caption{ 
A diagram for the thermodynamic potential to zero order in $1/N$.}
\label{dia1}
\end{figure}

The  thermodynamic potential for the ``zero-order'' theory 
 is given by the diagram shown in Fig.\ref{dia1}. 
In this diagram, $m=n/2$ and hence $D = O(1)$. The fermionic and bosonic self
energies are obtained by cutting one fermionic (or bosonic) line, and
 are also  of order 1 as they indeed should be in the ``zero-order'' theory. 
 Note, however, that this independence of $N$ is the result of the interplay between ${\bar g}$ and $v_F$ as if one cuts a fermionic line one
looses one factor $N$ since one closed fermionic loop disappears. This is
compensated by the fact that there is one running fermionic line less in the
diagram. 

\begin{figure}[tbp]
\begin{center}
\epsfxsize=\columnwidth 
\epsffile{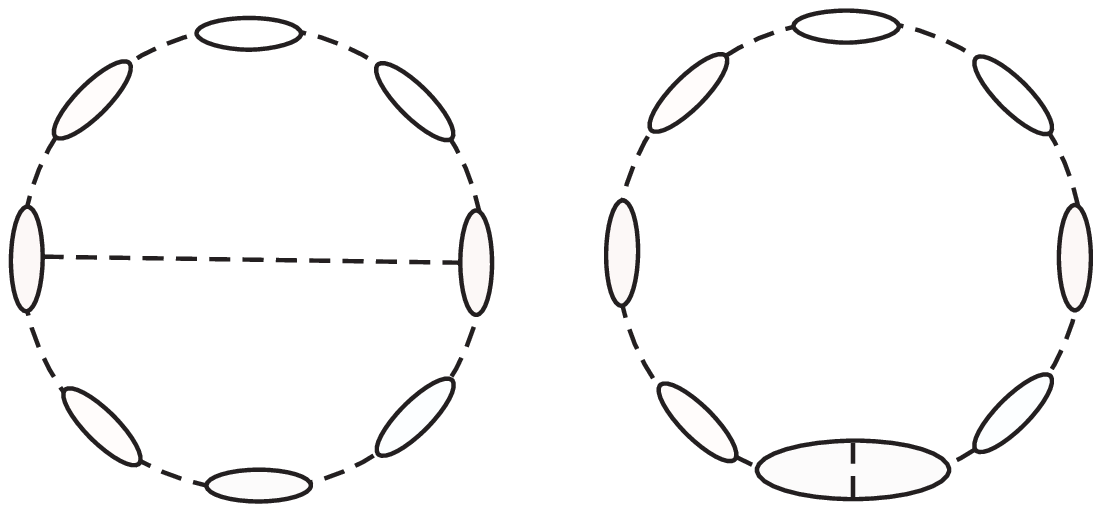} 
\end{center}
\caption{ The diagrams for the thermodynamic potential to order $1/N$.
}
\label{dia2}
\end{figure}
The diagrams which constitute the expansion in $\frac{1}{N}$ are then
 obtained by
reducing the number of internal fermionic loops at a given number of $%
\overline{g}$. The leading corrections are shown in Fig. \ref{dia2}.
 These diagrams have $m= n/2 -1$ and hence 
 are of
order $1/N$. Cutting fermionic or bosonic lines yields self energy and vertex 
corrections to order $1/N$. Higher order diagrams are obtained by further reducing the number of closed fermionic loops.

\section{Appendix C}
\label{append_vertex_ren}.

In this Appendix we  evaluate the polarization bubble with the
vertex correction. We first  compute this bubble explicitly. Then we
redo computations in a different way: we  first evaluate
 the effective four-fermion vertex
 and then contract two of its  opposite external legs (see 
Fig. \ref{four-vertex}) and convolute the vertex with the spin propagator.
 This last computation allows us to directly compute our theory with 
 the $\phi^4$ theory of critical behavior for the dynamical exponent $z=2$.
\subsection{direct computation}

The diagram which we need to compute  
is presented in Fig. \ref{four-vertex}c
In the analytical form we have for $\xi = \infty$
\begin{eqnarray}
\Pi_2 ({\bf q}, \Omega) &=& -\frac{2 {\bar g}^2 N}{(2\pi)^4} T^2 
\sum_{\omega, \omega^\prime} \int d^2 k d^2 k^\prime
~\frac{1}{\gamma|\omega - \omega^\prime| + ({\bf k}-{\bf k}^\prime)^2}\nonumber \\
&&\frac{1}{\epsilon_k - i \omega} 
\frac{1}{\epsilon_{k+q+Q} - i(\omega + \Omega)} \nonumber \\
&& 
\frac{1}{\epsilon_{k^\prime +q} - i (\omega^\prime + \Omega)} 
\frac{1}{\epsilon_{k^\prime +Q} - i \omega^\prime}
\label{C1}
\end{eqnarray}
where, as before $\epsilon_k = v_x k_x + v_y k_y,~\epsilon_{k+Q} = -v_x k_x + v_y k_y$, and $k$ measures a deviation from a hot spot. The combinatoric
 factor $2N$ comes from spin summation and summation over hot spots, and 
$\gamma = N {\bar g}/(4\pi v_x v_y)$.

Consider for simplicity the case $v_x =v_y$. Introducing first 
$k_x + k_y = k_+,~k_y -k_x = k_-$,  $k^\prime_x + k^\prime_y = k^\prime
_+,~k^\prime_y -k^\prime_x = k^\prime_-$, $q_+ = q_x + q_y,~q_- = q_y -q_x$,
and then $(k_+ + k^\prime_+) = 2 b_+$, $k_+ -  k^\prime_+ = b_-$, 
$(k_- + k^\prime_-) = 2 c_+$, $k_- -  k^\prime_- = c_-$, and substituting into (\ref{C1}) we obtain 
\begin{eqnarray}
&& \Pi_2 ({\bf q}, \Omega) = -\frac{2{\bar g}^2 N}{(2\pi)^4} \frac{1}{4 v^4_x}~
T^2 \sum_{\omega, \omega^\prime} \int d b_- d c_- \nonumber \\
&& ~\frac{1}{\gamma|\omega - \omega^\prime| + (b^2_- + c^2_-)/2}\nonumber \\
&&\times \int d b_+ 
\left[\frac{1}{b_+ + \frac{b_-}{2} - i \frac{\omega}{v_x}} \frac{1}{b_+ - 
\frac{b_-}{2} + q_- -
 i\frac{\omega^\prime + \Omega}{v_x}}\right] \nonumber\\
&&\times \int d c_+ 
\left[\frac{1}{c_+ - \frac{c_-}{2} - i \frac{\omega^\prime}{v_x}} 
\frac{1}{c_+ + \frac{c_-}{2} + q_+ -
 i\frac{\omega + \Omega}{v_x}}\right] 
\label{C2}
\end{eqnarray}
The integration over $ d b_+ $ and $ d c_+$ is straightforward and yields
\begin{eqnarray}
&& \Pi_2 ({\bf q}, \Omega)  = \frac{2{\bar g}^2 N}{(2\pi)^4} \frac{1}{4 v^4_x}~
T^2 \sum_{\omega, \omega^\prime} \int d b_- d c_-\nonumber \\
&& ~\frac{1}{\gamma|\omega - \omega^\prime| + (b^2_- + c^2_-)/2}\nonumber \\
&&\frac{[\mbox{sign}(\omega^\prime + \Omega) - \mbox{sign} \omega] [\mbox{sign} \omega^\prime -
\mbox{sign}(\omega + \Omega)]}{[b_- - q_- + i \frac{\omega^\prime - \omega 
+\Omega}{v_x}][c- + q_+ - i \frac{\omega - \omega^\prime 
+\Omega}{v_x}]}
\label{C3}
\end{eqnarray} 
Shifting the momenta $b_- - q_- = {\tilde b}_-,~
c_- + q_+ = {\tilde c}_-$ we rewrite (\ref{C3}) as
\begin{eqnarray}
&& \Pi_2 ({\bf q}, \Omega) = \frac{2{\bar g}^2 N}{(2\pi)^4} \frac{\pi^2}{4 v^4_x}~
T^2 \sum_{\omega, \omega^\prime} \int d {\tilde b}_- d {\tilde c}_-\nonumber \\
&&
 ~\frac{1}{\gamma|\omega - \omega^\prime| + (({\tilde b}_- + q_-)^2 + 
({\tilde c} - q_+)^2)/2}\nonumber \\
&&\frac{[\mbox{sign}(\omega^\prime + \Omega) - \mbox{sign} \omega] [\mbox{sign} \omega^\prime -
\mbox{sign}(\omega + \Omega)]}{[{\tilde b}_- + i \frac{\omega^\prime - \omega 
+\Omega}{v_x}][{\tilde c}_- - i \frac{\omega - \omega^\prime 
+\Omega}{v_x}]}
\label{C4}
\end{eqnarray} 
Next step is to observe that typical ${\tilde b}_-$ and ${\tilde c}_-$ are
of order of a typical frequency, and hence can be safely 
neglected in the bosonic propagator, in which typical momenta scale as a square  root of a typical frequency.
The remaining momentum integration then proceeds straightforwardly, and we 
obtain
\begin{eqnarray}
&& \Pi_2 ({\bf q}, \Omega) = \frac{2{\bar g}^2 N}{(2\pi)^4} \frac{\pi^4}{4 v^4_x}~
T\sum_{\tilde \omega} 
 ~\frac{\mbox{sign}({\tilde \omega} + \Omega)  \mbox{sign}(\Omega - {\tilde \omega})}{\gamma|{\tilde \omega}| + q^2} \nonumber \\
&& \times T\sum_{\omega^\prime}  (\mbox{sign}(\omega^\prime + \Omega) - 
\mbox{sign} (\omega^\prime + {\tilde \omega}))
\nonumber \\
&&\times  
(\mbox{sign} \omega^\prime - \mbox{sign}
(\omega^\prime + {\tilde \omega} + \Omega))]
\label{C7}
\end{eqnarray}
where $\tilde \omega = \omega - \omega^\prime$.
At $T\rightarrow 0$, the summation over $\omega^\prime$ is elementary and 
yields $4(|{\tilde \omega}| - \Omega)$. Substituting this result and the expression for $\gamma$ into
(\ref{C7}), evaluating the integral over ${\tilde \omega}$ up to an upper cutoff $A$, and re-analyzing the computational steps for $v_x \neq v_y$ 
(this adds a factor $Q(v) = (4/\pi) \tan^{-1}(v_x/v_y)$)  we finally obtain
\begin{eqnarray}
&& \Pi_2 ({\bf q}, \Omega) = \frac{Q(v)}{N}
({\bf q}^2 + \gamma|\Omega|)                   \nonumber \\
&&\times \left[\log{\frac{A \gamma}{{\bf q}^2 +              
\gamma|\Omega|}} + \log{\frac{q^2}{{\bf q}^2 + 
\gamma|\Omega|}}\right]                                
 + \frac{1}{N} (A - 2|\Omega|). 
\label{C8}
\end{eqnarray}
The constant contribution $(=A)$ 
 has to be neglected to avoid double counting as 
is comes from high energies and has  already been  
absorbed into a bare $\chi_0 ({\bf q}, \Omega)$.  
The logarithmical terms, on the other hand, come from low energies and have to be kept.
  
We see that to a logarithmical accuracy
 ${\bf q}^2$ and $\gamma |\omega|$ are renormalized in 
exactly the same way, i.e., the diffusion coefficient 
does not diverge as the system approaches
 a QCP. It does acquire a nonsingular $1/N$ correction 
which is beyond the scope of our analysis.  
We however checked that the logarithmical divergence at $q=0$ is 
fictitious and is eliminated when one included the self-energy of intermediate fermions.  

Let's also check whether $\omega \log \omega$ term at $T=0$ 
is collaborated by $T \log T$ term at a finite $T$, i.e., 
whether vertex corrections 
enforce $\omega/T$ scaling in the spin susceptibility. 
To address this issue we evaluate $\Pi_2 ({\bf k}, \Omega)$ 
at finite $T$ and $\Omega = q =0$. The summation over 
Matsubara frequencies in (\ref{C7}) is straightforward. Performing 
the summation over $\omega^\prime$ and substituting the 
explicit form of $\gamma$ we obtain
\begin{eqnarray}
\Pi_2 ({\bf q}, \Omega) &=& -\frac{2 Q(v) \gamma}{N}~\pi T 
\sum_{p=1}^{A/(2\pi T)}
 \frac{p-1}{p} \nonumber \\
&& = \frac{\gamma}{N} (- A + 2 \pi T )\log {\frac{A}{2 \pi T}}
\label{C9}
\end{eqnarray}
This result implies that finite $T$ accounts for the shift 
of the bosonic frequency by $2\pi T$, i.e., it just shifts 
a bosonic $\Omega = 2\pi T n$ from $n=0$ to $n=1$. Obviously 
then,  in real frequencies, $\Pi_2 ({\bf q}, \Omega)$ {\it does not} contain a 
singular $T \log T$ term, i.e., there is no $\Omega/T$ scaling 
in the spin susceptibility at the QCP.     

\subsection{four-boson vertex}
\label{spin-interaction}
 We next explicitly compute  the four-boson vertex in Fig.\ref{four-vertex}.
In analytical form it is given by
\begin{eqnarray}
b&=&\frac{Ng^{4}}{(2\pi )^{2}}\int 
\frac{d\omega^\prime d^{2}q^\prime}{i\omega^\prime -\epsilon _{q^\prime}}
\frac{1}{i(\Omega +\omega _{1}+\omega )-\epsilon _{Q+q^\prime+q_{1}+q}}\nonumber \\
&\times& 
\frac{1}{i(\omega^\prime +\omega _{1}+\omega _{2})-\epsilon _{q^\prime+q_{1}+q_{2}}} \nonumber \\
&&
\frac{1}{i(\omega^\prime +\omega +\omega _{2})-\epsilon _{Q+q^\prime+q+q_{2}}}\label{b-diagr}
\end{eqnarray}
Linearizing, as before, the fermionic dispersion near hot spots 
as  $\epsilon_q^\prime = v_x q^\prime_x + v_y q^\prime_y$,
$\epsilon_{q^\prime+Q} = -v_x q^\prime_x + v_y q^\prime_y$, and performing the momentum integration, we obtain
\begin{eqnarray}
b&=&\frac{Ng^{4}}{16\pi v_{x} v_{y}}\int d\omega^\prime 
\frac{\mbox{sign}(\omega^\prime )-\mbox{sign}(\omega^\prime +\omega _{1}+\omega _{2})}
{\epsilon _{q_{1}+q_{2}}-i(\omega _{1}+\omega _{2})} \nonumber \\
&\times &
\frac{\mbox{sign}(\omega^\prime+\omega _{1}+\omega )-
\mbox{sign}(\omega^\prime +\omega+\omega _{2})}
{\epsilon _{Q+q_{1}-q_{2}}-i(\omega _{1}-\omega _{2})}  \nonumber
\end{eqnarray}
The  integration  over $\omega^\prime $ is also straightforward and yields
\begin{equation}\label{b}
b=\frac{Ng^{4}}{8\pi v_{x}v_{y}}
\frac{|\omega _{2}+\omega|+|\omega _{2}-\omega |
-|\omega _{1}+\omega |-|\omega _{1}-\omega | }
{[i(\omega _{1}+\omega _{2})-\epsilon _{q_{1}+q_{2}}]
[i(\omega _{1}-\omega _{2})-\epsilon _{Q+q_{1}-q_{2}}]}
\end{equation}
This result is quoted in the main text.

\subsection{ First-order bosonic self-energy}
\label{2loop}

We next  explicitly compute the first oder bosonic self-energy by 
contracting the external legs in the four-boson vertex and convoluting 
it with the spin susceptibility.
The corresponding diagrams are presented in  Fig. \ref{four-vertex} b and c. 

The first diagram is obtained by contracting the adjacent external legs 
of the four-fermion vertex. In this situation
 $\omega_1 = \omega_2$ and ${\bf q}_1 = {\bf q}_2 $. 
Substituting this into Eq. (\ref{b}) and then into the expression 
for the bosonic self-energy, we find that 
 the self-energy  vanishes due to still presence of double poles.
This is consistent with our earlier result that the 
inclusion of the fermionic self-energy does not affect the form 
of the bosonic propagator. 

The second  diagram  
is obtained by contracting the external legs 
which are not adjacent to each other. In this situation, 
  $\omega = q =0$. Let's choose $\omega_1$ to be an external frequency 
(which we label  $\Omega$ for consistency of notations) 
 and at first 
set  the external $q_1 =0$ (this corresponds to external momentum 
 ${\bf q} = {\bf q}_1 + {\bf Q} = {\bf Q} =(\pi, \pi)$).
The bosonic self-energy is then given by
\begin{eqnarray}
&&\Pi_2 ({\bf Q}, \Omega)=-2\int b(\Omega , \omega _{2}, \omega =0)
\chi ({\bf q}_{2},\omega _{2})\frac{d\omega _{2}d^{2}q_{2}}{(2\pi )^{3}}
\nonumber \\
&=&  \frac{-Ng^{4}\chi_{0}}{(2\pi )^{4}v_{x}v_{y}}\int 
\frac{|\omega_2 |-|\Omega |}{[i(\Omega +\omega_2 )-\epsilon _{q_2}]
[i(\Omega -\omega_2 )-\epsilon _{Q-q_2}]} \nonumber \\
&\times& 
\frac{d\omega_2 d^{2}q_2}{\gamma |\omega_2 |+{\bf q}^{2}_2}   
  \label{2loop-polariz}
\end{eqnarray}
(the  combinatoric factor $2$ accounts for two possibilities to choose the
 external frequency, the
factor $-1$ comes from the summation over spin projections, which we reinstalled here).
One can easily make sure that to a logarithmical accuracy, 
we can neglect $\Omega$ in the denominator and set it as a lower cutoff in the integration over $\omega_2$. Subtracting the contribution from high energies
we then obtain 
\begin{eqnarray}
\Pi_2 ({\bf Q}, \Omega)&=& \frac{-Ng^{4}\chi_{0}}
{16\pi^{4}v_{x}v_{y}}|\Omega |
\int \frac{1}{(i\omega_2-\epsilon _{q_2})
(i\omega_2 -\epsilon _{Q+q_2})}                 \nonumber \\
&\times &\frac{d\omega_2 d^{2}q_2}{\gamma |\omega_2 |+{\bf q}^{2}} \nonumber
\end{eqnarray}
This integrand is exactly the same as for $\Delta g/g$ in the 
Appendix A. Using the results from Appendix A
 we  find
\begin{equation}\label{2loop-polariz-omega}
\Pi_2 ({\bf Q}, \Omega)=\gamma |\Omega |
\frac{Q(v_{x}/v_{y})}{2N\chi_{0}}|\log{\Omega }| 
\end{equation}

We next evaluate the correction to the polarization 
operator when both  $\Omega $ and ${\bf q}_1 ={\bf q}-{\bf Q}$ 
are nonzero. For simplicity we choose
 $v_{x}=v_{y}=v/\sqrt{2}$, and also set $q = q_x$. 
We will restore rotational symmetry at the end of computations.
Rotating the coordinates such that $\epsilon _{q_2}=v_F q_{2x}$ and 
$\epsilon _{Q+q_2}=v_F q_{2y}$ we can express 
$\Pi_2 ({\bf q}, \Omega)$ from (\ref{2loop-polariz}) as 
\begin{eqnarray}
&&\Pi_2 ({\bf q}, \Omega)=
  \frac{- 2 Ng^{4}\chi_{0}}{(2\pi )^{4}v_F^{2}}\int 
\frac{|\omega_2 |-|\Omega |}{i(\Omega +\omega_2 )-v_F q_{2x}}\nonumber \\
&\times&\frac{1}{i(\Omega -\omega_2 )+v_F q_{2 y}-v_F q_{y}} 
\frac{d\omega_2 d^{2}q_2}{\gamma |\omega_2 |+q_{2x}^{2}+q_{2y}^{2}} 
\label{2loop-polariz-full}
\end{eqnarray}
As before,  the dependence on $\Omega $ in the denominator 
can be eliminated by setting $\Omega$ as
 the lower limit on the integration over $\omega_2$.
Furthermore, as typical $\omega_2$ scale as $q_2^2$ and are 
obviously smaller than $q_2$, the $i \omega_2$ term in the 
fermionic propagators can be reduced to $i \delta \mbox{sign} (\omega_2)$.
With these simplifications, we obtain
\begin{eqnarray}
&& \Pi_2 ({\bf q}, \Omega)=
  \frac{ 4Ng^{4}\chi_{0}}{(2\pi )^{4}v_F^{2}} Re
\int\limits_{0}^{\infty }d\omega_2 \int  
\frac{|\omega_2 |-|\Omega |}{-i0_{+} +v_F q_{2x}}\nonumber \\
&\times&\frac{1}{-i0_{+}+v_F q_{2y}-v_F q_{y}} 
\frac{ d^{2}q_2}{\gamma |\omega_2 |+q_{2x}^{2}+q_{2y}^{2}} 
\end{eqnarray}
Now one can easily  evaluate first the integral over $q_{2x}$ 
and then over $q_{2y}$ and obtain
\begin{equation}
 \Pi_2 ({\bf q}, \Omega)= 
-\frac{Ng^{4}\chi_{0}}{4\pi^{2} v^{4}} \int \limits_{0}^{\infty }d\omega 
\frac{\omega_2 -|\Omega |}{\gamma \omega_2 +q_{y}^{2}} \label{2loop-polar-inf} 
\end{equation}
Evaluating the frequency integral,
 subtracting the contribution from high frequencies, restoring the rotational symmetry, and expressing $\gamma$ in terms of $g$ and $v_F$ we finally obtain 
\begin{equation}\label{2loop-polar-result}
\Pi_2 ({\bf q}, \Omega)= \frac{Q(v)}{N}({\bf q}^2 + \gamma|\Omega|) 
|\log(\gamma |\Omega | +{\bf q}^{2})| 
\end{equation}
To a logarithmical accuracy, this is indeed the same result as Eqn. (\ref{C8}).

\section{Appendix D}

In this Appendix, we evaluate the fermionic self-energy
  at finite
temperatures. We first consider the limit $N \rightarrow \infty$ at finite $\lambda$, and then discuss the modification of the results at finite $N$.

\subsection{$N \rightarrow \infty$ limit}

The sum we need to evaluate is given by  Eq. (\ref{seT}). It reads
\begin{equation}
\Sigma (k_{hs}, \Omega_m) = i \pi T \lambda 
~\sum_{n}~\frac{{\rm sign}\omega _{n}}
{\sqrt{1 + \frac{|\omega_{n-m}|}{\omega_{sf}}}}.            \label{B1}
\end{equation}
Using Poisson summation formula 
\begin{eqnarray}
T \sum_n F(n) &=& \frac{1}{2\pi} \int_{-\infty}^{\infty} F(x) dx  \nonumber \\
&+& \frac{1}{\pi} 
\sum_{p=1}^{\infty} \int_0^\infty\left(F(x) + F(-x)\right) 
\cos \frac{xp}{T} dx                                                \label{B2}
\end{eqnarray}
to separate $T=0$ and finite $T$ 
contributions to the self-energy, we obtain after integrating over $x$ in both
 terms in (\ref{B2})
\begin{equation}
\Sigma (k_{hs}, \Omega_m) \!=\! 2 i \lambda \!\left(\frac{\Omega_m}
{1 \!\!+ \!\sqrt {1 \!\!+ \!\frac{|\Omega_m|}{\omega_{sf}}}} 
\!\!+ \!P (\Omega_m, \!T) {\rm sign} \Omega_m\right)               \label{B3}
\end{equation}
The $T$ dependent term $P(\Omega_m, T)$ is expressed in terms of Fresnel 
integrals $S(x)$ and $C(x)$ as
\begin{eqnarray}
&&P (\Omega_m, T) = \sum_{p=1}^{\infty} 
\left(\frac{2 \pi T \omega_{sf}}{p}\right)^{1/2}                \nonumber \\
&\times& \Big[\cos{\frac{p \omega_{sf}}{T}}\left
[S\left(\sqrt{\frac{p(\Omega_m +\omega_{sf})}{T}}\right) - 
S\left(\sqrt{\frac{p\omega_{sf}}{T}}\right)\right]              \nonumber \\
&+&\sin{\frac{p \omega_{sf}}{T}}
\left[C\left(\sqrt{\frac{p(\Omega_m +\omega_{sf})}{T}}\right) - 
C\left(\sqrt{\frac{p\omega_{sf}}{T}}\right)\right] \Big]         \label{B4}
\end{eqnarray}
The Fresnel integrals are defined as
\begin{equation}
S(\sqrt{x}) = \frac{1}{\sqrt{2\pi}} 
\int^x_0 \frac{\sin {t}}{\sqrt{t}} dt~
C(\sqrt{x}) = \frac{1}{\sqrt{2\pi}} 
\int^x_0 \frac{\cos {t}}{\sqrt{t}} dt~                           \label{B5}
\end{equation}
At low $T \ll \omega_{sf}$, one can expand Fresnel integrals in powers of 
$1/\sqrt{x}$. In this limit, Eq (\ref{B4}) reduces to
\begin{eqnarray}
&&P (\Omega_m, T) = 
\frac{(\pi T)^2}{12 \omega_{sf}} \left(1 + 2 \left(
\frac{\omega_{sf}}{|\Omega_m| + \omega_{sf}}\right)^{1/2}\right)  \nonumber \\
&&- \frac{15}{8}~\frac{T^4}{\omega^3_{sf}} \left(2.1 + 0.9 f
\left(\frac{\omega_{sf}}{|\Omega_m|}\right)\right) + O(T^6)        \label{B6}
\end{eqnarray}
where $f(0) =0$, $f(\infty) =1$.
At $\Omega_m \ll \omega_{sf}$, one recovers a conventional Fermi liquid result
$P (\Omega_m, T) \approx (\pi T)^2/(4 \omega_{sf})$. At $\Omega_m \gg 
\omega_{sf}$, but still $T \ll \omega_{sf}$ (i.e., for large $m$), the leading 
functional dependence is still $T^2$, but the prefactor is reduced by $1/3$. 
Also notice that the subleading, $T^4$ corrections become comparable 
with $T^2$ terms starting from relatively small $T \sim 0.5~ \omega_{sf}$.

In the opposite limit $T \gg \omega_{sf}$, 
the contributions to the sum (\ref{B1}) 
from all $n \neq m$ can
be simplified by approximating 
\begin{equation}
\left(1 + \frac{|\omega_{n-m}|}
{\omega_{sf}}\right)^{-1/2} \rightarrow 
\left(\frac{\omega_{sf}}{|\omega_{n-m}|}\right)^{1/2}.            \label{B66}
\end{equation}
With this approximation, one can sum over $n$ explicitly:
\begin{eqnarray}
\Sigma (\Omega_m) &=& i \frac{\pi}{2} ~T {\bar \omega}^{1/2}~\sum_{n \neq m} 
\frac{{\rm sign} (2 n+1)}{|\omega_{n-m}|^{1/2}}                   \nonumber \\
&&= i \left(\frac{\pi T {\bar \omega}}{2}\right)^{1/2}
~\left(\zeta\left(\frac{1}{2}\right) - 
\zeta\left(\frac{1}{2}, 1+m\right)\right)                         \label{B7}
\end{eqnarray}
where $\zeta$ is a Zeta function.
The absence of the correlation length in this expression 
is a consequence of the fact that  the product $\lambda 
\sqrt{\omega_{sf}} = 0.5 \sqrt{\bar \omega}$ is independent of $\xi$,
One indeed would expect this independence in the quantum-critical regime. 

In real frequencies, the 
imaginary part of the self-energy at finite $T$ can be obtained using the
spectral representation . We have 
 at 
${\bf k} = {\bf k}_{hs}$ and $v_s \rightarrow \infty$, 
\begin{eqnarray}\label{Bb81}
\Sigma^{\prime \prime}(\Omega ) &=& 
\frac{\lambda v_{F} \xi}{2\pi ^{2}}\int_{-\infty }^{\infty} 
Im\frac{d\omega d^{2}k}
{1 + ({\bf k} \xi)^{2}- 
i\frac{\omega }{\omega _{sf}}}                           \nonumber \\ 
&\times&Im\frac{1}{v_{F}k_{x}-\Sigma ^{R}(\Omega -\omega )}
f(\Omega /2T,\omega /2T)
\end{eqnarray}
where 
\begin{equation}
f(x,y) = \tanh (x-y)+1/\tanh (y).
\label{B82}
\end{equation}

At $N \rightarrow \infty$ and finite $\xi$,  the momentum 
integration can be factorized and we obtain:

\begin{equation}\label{B81}
\Sigma^{\prime \prime}(\Omega ) = 
\frac{\lambda}{2}Im\int_{-\infty }^{\infty} 
\frac{d\omega }
{\sqrt{1 -
i\frac{\omega }{\omega _{sf}}}}~f(\Omega /2T,\omega /2T)
\end{equation}

Note that the presence of $\Sigma^R$ in the integrand is irrelevant as 
when momentum integration is factorized, the integral over $k_x$ just gives a
 constant density of states, independent on $\Sigma^R$.
It is convenient to separate the 
 contributions to $\Sigma^{\prime \prime}(\Omega)$ 
from static and dynamical spin fluctuations. 
This can be done in a standard way by replacing $f(x,y)$ by
\begin{equation}
f(x,y) = \frac{1}{y} + f_1 (x,y)
\label{Bbb1}
\end{equation}
 where 
\begin{equation}
f_{1}(x,y) = \tanh (x-y)+1/\tanh (y)-1/y,
\end{equation} 
Substituting the form of $f(x,y)$ into (\ref{B81}) we obtain
\begin{equation}\label{B83}
\Sigma^{\prime \prime}(\Omega )= \Sigma^{\prime \prime}_{st} (\Omega) + 
\Sigma_{dyn}(\Omega),
\end{equation}
where
\begin{equation}
\Sigma^{\prime \prime}_{st} (\Omega) = T\lambda\pi
\label{B84}
\end{equation}
and
\begin{equation} 
\Sigma_{dyn} (\Omega) \approx \frac{\lambda }{2}Im
 ~\int _{-\infty }^{\infty }
\frac{d\omega  }{\sqrt{1 -i\frac{\omega }{\omega _{sf}}}}
f_{1}(\Omega /2T,\omega /2T)
\label{B86}
\end{equation}
At $\Omega \gg \omega_{sf}$, typical $\omega \gg \omega_{sf}$, and
$\Sigma_{dyn} (\Omega)$ can be approximated by
\begin{equation}
\Sigma_{dyn} (\Omega) \approx \frac{\sqrt{\bar \omega}}{4\sqrt{2}}
 ~\int _{-\infty }^{\infty }
\frac{d\omega}{\sqrt{\omega}}
f_{1}(\Omega /2T,\omega /2T)
\label{B87}
\end{equation}
This result can be rewritten in the scaling form
\begin{equation}
\Sigma_{dyn} (\Omega) = 
\left(\frac{T\bar{\omega}}{2}\right)^{1/2} 
D\left(\frac{\Omega}{T}\right)  \label{B8}
\end{equation}
where  $D(x)$ is given by
\begin{eqnarray}
&D&(x) = \int_0^{\infty} \frac{dy}{\sqrt {y}}                   \nonumber \\
&\times&\left(\frac{1}{e^y -1} -\frac{1}{y} + 
\frac{1}{2}\left(\frac{1}{e^{y+x} +1} +
\frac{1}{e^{y-x}+1}\right)\right)                               \label{B9}
\end{eqnarray}
In the two limits, $D(x)$ behaves as
$D(x \gg 1) = \sqrt{x} -2.58 -0.822/x^{3/2} +...$,
 and $D(x \ll 1) = -1.516 + 0.105 x^2 + ...$.

The real part of $\Sigma_{dyn} (\Omega)$ is obtained from (\ref{B8})
by Kramers-Kronig transformation. It contains the same $\sqrt{T}$ dependence
as (\ref{B8}) but a different scaling function of $\Omega/T$.

\subsection{finite $N$}

We recall that the need to study a finite $N$  is related to the fact that 
at a nonzero temperature, some of $1/N$ terms also scale with $\xi$ and therefore become relevant for physical $N =8$. Here we single out and explicitly evaluate there terms. We, however, still will be neglecting regular $1/N$ contributions.

Our point of departure is Eqn (\ref{Bb81}).
 At finite $N$, the momentum integration in 
(\ref{Bb81}) cannot be factorized. 
Let's use (\ref{Bbb1}) and
consider the 
static and dynamical contributions to $\Sigma^{\prime \prime} (\Omega)$ separately. This separation requires care as 
 to single out the truly static contribution (the one from a boson with zero Matsubara frequency) 
we not only have to replace $f(x,y)$ by $1/y$ but also set $\omega =0$ in the fermionic propagator. 
The static contribution is given by 
\begin{eqnarray}\label{Bc81}
\Sigma^{\prime \prime}_{st}(\Omega ) &=&
\frac{\lambda v_{F} \xi^2}{2\pi ^{2}}\int_{-\infty }^{\infty}\!\!\!\!\! 
Im\frac{d\omega d^{2}k}
{1 + ({\bf k} \xi)^{2}- 
i\frac{\omega }{\omega _{sf}}}\frac{2T}{\omega} \nonumber \\
&\times &Im\frac{1}{v_{F}k_{x}-\Sigma ^{R}(\Omega)}
\end{eqnarray}
The integration over $\omega$ is straightforward. Also, since 
$Im\Sigma ^{R}>0$, we can  integrate over $k_{x}$ by close the integration contour  
over the lower half plane. Introducing then
$\tilde{\Sigma }=\Sigma ^{R}(\Omega )/v_{F}\xi^{-1}$, we obtain
\begin{eqnarray}\label{Bd81}
\Sigma^{\prime \prime}_{st}(\Omega ) &=& T\lambda Im\int _{\infty }^{\infty } 
\frac{dx}{\sqrt{x^{2}+1}+\tilde{\Sigma }}
\frac{1}{\sqrt{x^{2}+1}}                          \nonumber\\
&=& 2T\lambda Im\frac{\log \left[\tilde{\Sigma }+ 
\sqrt{1 +\tilde{\Sigma }^{2}}\right]}{\sqrt{1+\tilde{\Sigma }^{2}}}
\end{eqnarray}
As we discuss in the main text, at $T \ll N \omega_{sf}$, the 
 fermionic self-energy in the integrand can 
be approximated by its $T=0$ value, Eqn (\ref{setot}). 
The latter, however, can be neglected only if ${\tilde \Omega} \ll 1$ or 
$\Omega \ll {\bar \Omega} = N^2 \omega_{sf}/9$. For infinite $N$, this scale is above ${\bar \omega}$ and is irrelevant for our study. For finite $N$, however,
${\bar \Omega}$ scales with $\omega_{sf}$ and definitely 
become smaller that ${\bar \omega}$ if $\xi \rightarrow \infty$ 
at a finite $N$. 
Obviously, in this situation, ${\tilde \Sigma}$ cannot be neglected, and 
the $N = \infty$ result for $\Sigma^{\prime \prime}_{st}$,
 Eq. (\ref{B84}) is invalid.

In the  quantum critical regime, $\Sigma^2 = i \Omega/{\bar \Omega}$.
 Substituting this into (\ref{Bd81}) we 
obtain the result cited in (\ref{seT101}).

The dynamical contribution to $\Sigma^{\prime \prime} (\Omega)$
 consists of two parts $\Sigma_{dyn}(\Omega) = \Sigma^{1}_{dyn}(\Omega) + 
\Sigma^{2}_{dyn}(\Omega)$. The first contribution accounts for the difference between $\Sigma^R (\Omega - \omega)$ and $\Sigma^R (\Omega)$:
 \begin{eqnarray}\label{Be81}
&&\Sigma^{1}_{dyn}(\Omega ) = 
\frac{\lambda v_{F} \xi}{2\pi ^{2}}\int_{-\infty }^{\infty}\!\!\!\!\! 
Im\frac{d\omega d^{2}k}
{1 + ({\bf k} \xi)^{2}- 
i\frac{\omega }{\omega _{sf}}}~\frac{2T}{\omega +i0} \nonumber \\
&\times& Im\left[\frac{1}{v_{F}k_{x}-\Sigma ^{R}(\Omega -\omega )}
-\frac{1}{v_{F}k_{x}-\Sigma ^{R}(\Omega )}
\right]
\end{eqnarray}
This integral is  
 nonsingular  when $\xi \rightarrow \infty $. We can then  
use a conventional $N \rightarrow \infty$ approach and 
factorize  the momentum integration.  Performing it,
 we immediately find that $\Sigma^{1}_{dyn}(\Omega )$ vanishes to  
leading order in $1/N$. 

The second contribution to the dynamical 
part is given by (\ref{Bb81})
with $f_{1}$ instead of $f$.
As $f_1 (x,y)$ is non-singular 
at $y \rightarrow 0$ (i.e., at $\omega \rightarrow 0$), the leading (in $1/N$)
 piece in $\Sigma ^{2}_{dyn}$ can again be obtained by 
 factorizing the momentum integration. This piece is then the 
same as in (\ref{B8}).

Finally, we found that 
the momentum integration in  (\ref{Bb81}) 
can be performed explicitly at arbitrary $N$ 
by, e.g., transforming to polar coordinates. Furthermore, by 
explicitly evaluating $Im G (k, \Omega)$ and pulling the 
$Im$ symbol for bosonic propagator out of  frequency integral, we can obtain 
the full expression for $\Sigma (\Omega, T)$, not only its imaginary part. 
  Performing the momentum integration, we found  
\begin{eqnarray}\label{Bbbbb81}
\Sigma (\Omega,  T ) &=& 
\frac{\lambda}{4\pi i}\int_{-\infty }^{\infty}\!\!\!\!\! 
 d\omega~~  f(\Omega /2T,\omega /2T)~ \nonumber \\
&& (S_R (\Omega, \omega) - S_A (\Omega, \omega))~ \mbox{sign}(\Omega - \omega)
\end{eqnarray}
where
\begin{eqnarray}
S_{R,A} (\Omega, \omega)&=& 
\frac{1}{\sqrt{{\tilde \Sigma}^2_{R,A} + \Pi_R}}      \nonumber \\
&\times &\log{\frac{\sqrt{{\tilde \Sigma}^2_{R,A}} - 
\sqrt{{\tilde \Sigma}^2_{R,A} + \Pi_R}}{\sqrt{{\tilde \Sigma}^2_{R,A}} 
+ \sqrt{{\tilde \Sigma}^2_{R,A} + \Pi_R}}}
\label{Bcccc81}
\end{eqnarray}
and $\Pi_R = \Pi_R (\omega) = 1 - i \omega/\omega_{sf}$ and 
${\tilde \Sigma}^2_{R,A} = \Sigma_{R,A} (\Omega - \omega)/(v_F \xi^{-1})$.

This expression is of limited use as it  contains 
regular $1/N$ terms which are beyond our accuracy as we neglected 
$1/N$ vertex corrections. Nevertheless, it is useful for the 
understanding how small are 
 regular $1/N$ terms.  
The $\Sigma^{\prime \prime}$ from Eq. (\ref{Bbbbb81}) 
is plotted in Fig.~\ref{sigmaT_exact}.


\begin{references}

\bibitem{BM} 
J.G. Bednortz and K.A. Muller, \rmp {\bf 60}, 585 (1988).

\bibitem{AFM}
S. Chakravarty, B. I. Halperin, and D.R. Nelson, \prb {\bf 39}, 2311 (1989); 
S.M. Hayden {\it et al}, \prl {\bf 66}, 821 (1991); 
T. Imai {\it et al}, \prl {\bf 70}, 10002 (1993); 
D. C. Johnson, \prl {\bf 62}, 957 (1989); 
B. Keimer {\it et al}, \prl {\bf 67}, 1930 (1991).

\bibitem{photo_over} 
 K. Gofron {\it et al}, \prl {\bf 73}, 3302 (1994);
D.S. Dessau {\it et al}  \prl {\bf 71}, 2781 (1993);
Jian Ma {\it et al}, \prb {\bf 51}, 3832 (1995); 
J.~C. Campuzano {\em {\it et al}.}, \prb {\bf 53}, R 14 737 (1996).

\bibitem{photo_under} 
M.~R. Norman {\em {\it et al}.}, \prl {\bf 79}, 3506 (1997); 
J. C. Campuzano {\em {\it et al}.,} \prl {\bf 83}, 3709 (1999); 
Z.~X. Shen {\em {\it et al}.}, Science {\bf 280}, 259 (1998); 


\bibitem{tunn_under}  
Ch. Renner {\it et al}, \prl {\bf 80}, 149 (1998); 
 Y. DeWilde {\it et al}, {\em ibid} {\bf 80}, 153 (1998); 
V.M. Krasnov {\it et al},  \prl {\bf 84}, 5860 (2000);
J. F. Zasadzinski {\it et al},  \prl  to appear.

\bibitem{girsh_science} 
G. Blumberg {\it et al}, Science, {\bf 278}, 1427 (1997).

\bibitem{tsue_RMP}  
D. A. Wollmann, D. J. Van Harlingen, W. C. Lee, 
D. M. Ginsberg, and A. J. Leggett, \prl {\bf 71}, 2134 (1993); 
C.C. Tsuei and J.R. Kirtley, \rmp {\bf 72}, 969 (2000).

\bibitem{valla}   
T. Valla {\it et al}, Science {\bf 285}, 2110 (1999).

\bibitem{kaminski}  
A. Kaminski {\it et al}, \prl {\bf 84}, 1788 (2000).

\bibitem{timusk} 
e.g. A. Puchkov, D. Basov, and T. Timusk, J. Phys.: 
Cond. Matter {\bf 8}, 10049 (1996).

\bibitem{transport} 
R. C. Yu {\it et al}, \prl {\bf 69}, 1431 (192); 
S.J. Hagen, Z.Z. Wang, and N.P. Ong, \prb {\bf 40}, 9389 (1989).

\bibitem{extra_over} see e.g. Z. Yusof {\it et al}, cond-mat/0104367

\bibitem{anderson} 
P.W. Anderson, {\it The Theory of Superconductivity 
in the High $T_{c}$ Cuprates} 
(Princeton University Press, Princeton, New Jersey, 1997).

\bibitem{spin_charge_sep} 
T. Senthil and M.P.A. Fisher, \prb {\bf 63}, 4521 (2001); 
N. Nagaosa and P.A. Lee, \prb {\bf 61}, 9166 (2000); 
P.A. Lee and X-G Wen, cond-mat/0008419,  
D.A. Ivanov, P.A. Lee, and X-G Wen, cond-mat/990931.

\bibitem{varma1}  
C. Varma {\it {\it et al}.}, \prl {\bf 63}, 1996 (1989), 
{\em ibid} {\bf 64}, 497 (1990);   
P. Littlewood and C. Varma, \prb {\bf 46}, 405 (1992); 
G. Kotliar {\it et al}, Europhys. Lett. {\bf 15}, 655 (1991).

\bibitem{grilli}
C. Castellani, C. DiCastro, and M. Grilli, \prl {\bf 75}, 4650 (1995);  
J. Phys. Chem. Solids {\bf 59}, 1694 (1998).

\bibitem{david_review} 
D. Pines, Z. Phys. B {\bf 103}, 129 (1997) and references therein.

\bibitem{subir21} 
S. Sachdev, Science, {\bf 288}, {475} (2000); 
 M. Vojta and S. Sachdev, \prl {\bf 83}, 3916 (1999); 
see also N. Read and S. Sachdev, \prl {\bf 66}, 1773 (1991).

\bibitem{scs}
S.  Sachdev, A.V.  Chubukov, and A. Sokol, \prb {\bf 51}, 14874 (1995).

\bibitem{chub2}  
A. Chubukov, Europhys. Lett. {\bf 44}, 655 (1997).
 
\bibitem{NFL_at_crit_general} 
S. Sachdev, {\em Quantum Phase Transitions}, Cambridge University Press 
(Cambridge, 1999).

\bibitem{subir} 
A. Chubukov, S. Sachdev, and J. Ye, \prb {\bf 49}, 11919 (1994).

\bibitem{varma} 
C.M. Varma, \prb {\bf 55}, 14554 (1997). 

\bibitem{ddw} 
S. Chakravarty, R. Laughlin, D. Morr, and C. Nayak, cond-mat/0005443. 

\bibitem{scal_pr}  
D.J. Scalapino, Phys. Rep. {\bf 250}, 329 (1995).

\bibitem{aeppli} 
G. Aeppli {\em {\it et al}.}, Science {\bf 278}, 1432 (1997) 
and references therein.

\bibitem{nmr} 
A. Millis. H. Monien, and D. Pines, \prb {\bf 42}, 197 (1990),  

\bibitem{keimer1} 
P. Bourges in 
{\it The gap Symmetry and Fluctuations in High Temperature Superconductors}, 
J. Bok, G. Deutscher, D. Pavuna, and S.A. Wolf eds. 
(Plenum Press, 1998), pp. 349-371 and references therein.
  
\bibitem{mont_pines}  
P. Monthoux and D. Pines, \prb {\bf 47}, 6069 (1993); 
{\em ibid} {\bf 50}, 16015 (1994).

\bibitem{landau}  
E.M. Lifshitz and L.P. Pitaevskii, {\it Statistical Physics, p.2} 
(Butterworth-Heinemann, 1995, Oxford)

\bibitem{agr}  
see, e.g., J. Annett, N. Goldenfeld, and S. Renn in {\it 
Physical properties of high temperature superconductors}, ed. by D. Ginsberg
(World Scientific, NJ 1990).

\bibitem{shraiman_siggia} 
B. I. Shraiman and E. D. Siggia, \prl {\bf 62}, 1564 (1989). 

\bibitem{dagotto}  
E. Dagotto, A. Nazarenko, and M. Boninsegni, \prl {\bf 73}, 728 (1994); 
E. Dagotto, A. Nazarenko, and A. Moreo, \prl {\bf 74}, 310 (1995); 
S. Haas, E. Dagotto, and A. Moreo, \prl,to appear; 
A. Nazarenko, K.J.E. Vos, S. Haas, E. Dagotto, and
R.J. Gooding, \prb, {\bf 51}, 8676 (1995). 

\bibitem{swz} 
J.R.Schrieffer, X.G.Wen, and S.C.Zhang, \prb {\bf 39}, 11663 (1989); 
A.V.Chubukov and D.M.Frenkel, {\em ibid} {\bf 46}, 11884 (1992).

\bibitem{castro_neto} 
A. H. Castro-Neto, cond-mat/0102281.

\bibitem{hanke} 
S-C Zhang, Science, {\bf 275}, 1089 (1997); 
E. Arigoni, M.G. Zacher, and W. Hanke, 
cond-mat/01051125 and references therein.

\bibitem{sushkov} 
O.P. Sushkov, cond-mat/0002421 and references therein.

\bibitem{shadow_FS}  
P. Aebi {\it et al}, \prl {\bf 72}, 2757 (1994); 
{\em ibid} {\bf 74}, 1886 (1995).

\bibitem{acs}  
Ar Abanov, A. V. Chubukov, and J. Schmalian, cond-mat/0005163.

\bibitem{acf}  
Ar. Abanov, A. Chubukov, and A.M. Finkel'stein,
cond-mat/9911445,
Europhys. Lett., {\bf 54}, 488 (2001) 

\bibitem{sw_dsc} 
C. Stemmann, C. P\'{e}pin, and M. Lavagna, \prb {\bf 50}, 4075 (1994);
D.~Z. Liu, Y. Zha, and K. Levin, \prl {\bf 75}, 4130 (1995); 
I. Mazin and V. Yakovenko, {\em ibid.} {\bf 75}, 4134 (1995);
 A. Millis and H. Monien, \prb {\bf 54}, 16172 (1996); 
N. Bulut and D. Scalapino, \prb {\bf 53}, 5149 (1996); 
D.~K. Morr and D. Pines, \prl {\bf 81}, 1086 (1998); 
S. Sachdev and M. Vojta, Physica B {\bf 280}, 333 (2000);
Y.-J. Kao {\em et al.}, \prb  {\bf 61}, R11898 (2000);
F. Onufrieva and P. Pfeuty, {\tt cond-mat/9903097};
O. Tchernyshyov {\em et al.}, {\tt cond-mat/0009072};
J. Brinckmann and P.A. Lee, \prl {\bf 82}, 2915 (1999);
M. R. Norman and H. Ding, \prb {\bf 57}, R11089 (1998);
M.~R. Norman, \prb {\bf 61}, 14751 (2000);
Ar. Abanov and A. Chubukov, \prl {\bf 83}, 1652 (1999);

\bibitem{eli_ph}  
D. J. Scalapino, {\em The electron-phonon interaction
and strong coupling superconductors, }in {\em Superconductivity}, in {\em %
Superconductivity}, Vol. 1, p. 449, Ed. R. D. Parks, Dekker Inc. N.Y. 1969; 
 J. R. Schrieffer, {\em Theory of Superconductivity}
(Benjamin, Reading, Mass., 1966).

\bibitem{hf_exp}  
H. von Lohneysen {\it et al}, \prl {\bf 72}, 3262 (1994); 
D. Bitko, T. Rosenbaum and G. Aeppli, \prl {\bf 77}, 940 (1966); 
S. Raymond {\it et al}, J. Low Temp. Phys. {\bf 107}, 295 (1997);
 F. Steglich {\it et al}, Z. Phys. B {\bf 103}, 235 (1997);
Y. Aoki {\it et al}, J. Phys. Soc. Jpn, {\bf 66}, 2993 (1997).

\bibitem{hf_dome} 
N. Mathur {\it et al}, Nature {\bf 394}, 39 (1998).

\bibitem{nfl_hf} 
D. Vollhardt, \rmp, {\bf 56}, 99 (1984); 
T Morya and J. Kawabata, J. Phys. Soc. Jpn., {\bf 34}, 639 (1973); 
K.B. Blagoev {\it et al}, \prl {\bf 82}, 133 (1999).

\bibitem{hf_th} 
see e.g., P. Coleman, Nature {\bf 406}, 508 (2000); 
P. Coleman, C. P\'{e}pin, Q. Si, and R. Ramazashvili, cond-mat/0105006; 
A. Rosh, \prl {\bf 82}, 4280 (1999); 
A. Rosh, cond-mat/0103446 and references therein; 
Q. Si {\it et al}, cond-mat/0011477.

\bibitem{frac_power}  
O. Stockert {\it et al}, \prl {\bf 80}, 5627 (1998); 
A. Schr\"{o}der {\it et al}, {\em ibid}, {\bf 80}, 5623 (1988); 
A. Schr\"{o}der {\it et al}, Nature {\bf 407}, 351 (2000).

\bibitem{st_pines}  
B.P. Stojkovic and D. Pines, \prb {\bf 55}, 8576 (1997); 
{\em ibid} {\bf 56}, 11931 (1997).

\bibitem{ami}  
L.B. Ioffe, A.I. Larkin, A.J. Millis, and B.L. Altshuler, JETP
Lett., {\bf 59}, 65 (1994);  
B. L. Altshuler, L.B. Ioffe, and A. J. Millis, \prb {\bf 52}, 5563 (1995).

\bibitem{im} 
L.B. Ioffe and A. J. Millis, \prb {\bf 51}, 16151 (1995); 
see also  A.M. Tsvelik and M. Reizer, \prb {\bf 48}, 9887 (1993).

\bibitem{karl} 
D. Manske, I. Eremin, and K.H. Bennemann, \prb {\bf 63}, 054517 (2001) 
and references therein.

\bibitem{dahm} 
T. Dahm and L. Tewordt, \prl {\bf 74}, 793 (1995);
T. Dahm, D. Manske and L. Tewordt, \prb {\bf 58}, 12454 (1998). 

\bibitem{wheatley} M.L. Lercher and J.M. Wheatley, JMMM {\bf 185}, 384 (1998),
 \prb {\bf 63}, 012403 (2000).

\bibitem{infD} 
A. Georges {\it et al}, \rmp, {\bf 68}, 13 (1996).

\bibitem{hertz}  
J.A. Hertz, \prb {\bf 14}, 1165 (1976).

\bibitem{millis} 
A.J. Millis, \prb {\bf 48}, 7183 (1993). 

\bibitem{chub1}   
A. Chubukov, \prb {\bf 52}, R3840 (1995). The expression
for the vertex correction in this paper should contain $0.5 
\log \sin \phi/2$ instead of $\log \cos \phi/2$.

\bibitem{cs}  
A. Chubukov and J. Schmalian, \prb {\bf 57}, R11085 (1998).

\bibitem{ch_m} 
A.Chubukov and D. Morr, \prl {\bf 81}, 4716 (1998).
 Ar. Abanov and A. Chubukov, \prl, {\bf 83}, 1652 (1999).

\bibitem{cmm} 
A. Chubukov, D. Morr, and  P. Monthoux, \prb  {\bf 56}, 7789 (1997).

\bibitem{ac_qcp}  
Ar. Abanov and A. Chubukov, \prl {\bf 84}, 5608 (2000).


\bibitem{ach}  
R. Haslinger, Ar. Abanov, and A. Chubukov, \prb {\bf 63}, 020503(R) (2001).

\bibitem{hubbard} 
C. H. Pao and N. E. Bickers, \prl {\bf 72}, 1870 (1994); 
Ph. Monthoux and D.J. Scalapino, \prl {\bf 72}, 1874 (1994); 
S. Grabowski, M. Langer, J. Schmalian, and K. H. Bennemann,
Europhys. Lett. {\bf 34}, 219 (1996); 
J. Schmalian, S. Grabowski, and K. H. Bennemann, \prb {\bf 56}, R509 (1997);  
N. Bulut and D. Scalapino, \prb {\bf 53}, 5149 (1996);
D. J. Scalapino and S.R. White, \prb {\bf 58}, 1347 (1998); 
S. Trugman, \prb {\bf 37}, 1597 (1988).

\bibitem{heinz} 
H.J. Schulz, \prl {\bf 64}, 1445 (1990); 
I. Dzyaloshinskii and V. Yakovenko, JETP {\bf 94}, 344 (1988) 

\bibitem{exp_orb} 
A. Chubukov and K.A. Musaelian, \prb {\bf 50}, 6238 (1994); 
A. Singh and Z. Tesanovic, \prb {\bf 41}, 614 (1990). 

\bibitem{ioffe_larkin} 
L. Ioffe and A. I. Larkin, \prb {\bf 39}, 8988 (1989).

\bibitem{dombre} 
Th. Dombre, J. Physique, {\bf 51}, 847 (1990).

\bibitem{cm2} 
A. Chubukov and K.A. Musaelian, \prb {\bf 5}, 12605  (1995).

\bibitem{stripes} 
S. A. Kivelson, E. Fradkin, and V. J. Emery, Nature {\bf 393}, 550 (1998); 
J. Zaanen, Nature {\bf 404}, 714 (2000) and references therein.

\bibitem{hr} R. Hlubina and T.M. Rice, \prb, {\bf 52}, 13043 (1995).

\bibitem{ARPES_FS}   
M. R. Norman {\it et al.}, \prl {\bf 79}, 3506 (1997); 
Z-X. Shen {\it et al}, Science {\bf 280}, 259 (1998); 
T. Sato {\it et al}, \prb to appear.  

\bibitem{lutt} 
J.M. Luttinger and J.C. Ward, Phys. Rev. {\bf 118}, 1417 (1960); 
J.M. Luttinger, Phys. Rev. {bf 119}, 1153 (1960).

\bibitem{johnson1} 
P.D. Johnson {\it et al}, cond-mat/0102260

\bibitem{bogdanov1} 
P.V. Bogdanov {\it et al}, cond-mat/0004349 

\bibitem{neutron_extract} 
P. Bourges {\it et al}, Science {\bf 288}, 1234 (2000) and references therein.

\bibitem{nmr_extract}  
V. Barzykin {\it et al}, {\em ibid} {\bf 49}, 1544 (1994)

\bibitem{chiral} 
S.B. Treiman, R. Jackiw, and D.J. Gross, 
in {\it Lectures on Current Algebra and its Applications} 
(Princeton University Press, Princeton, 1972).

\bibitem{eliashberg} 
G.M. Eliashberg, Sov. Phys. JETP {\bf 11}, 696 (1960).
On recent discussion on the applicability of the Eliashberg approach 
to phonons see A.S. Alexandrov, cond-mat/0102189 and references therein.

\bibitem{migdal}  
A.B. Migdal, Sov. Phys. JETP, {\bf 7}, 996 (1958).

\bibitem{mahan}  G.D. Mahan, Many-Particle Physics, Plenum Press, 1990.

\bibitem{andy92} 
A.J. Millis, \prb {\bf 45}, 13047 (1992).

\bibitem{georges} 
S. Sachdev and A. Georges,  preprint, cond-mat/9503158.

\bibitem{kadanoff} 
L. Kadanoff, Phys. Rev. {\bf 132}, 2073 (1963).

\bibitem{ch-morr} 
A. Chubukov and D. Morr, Phys. Rep., {\bf 288}, 355 (1997).

\bibitem{ks}  
A.P. Kampf and J.R. Schrieffer, Phys. Rev B {\bf 42}, 7967 (1990).
 J.R. Schrieffer, J. Low Temp.  Phys. {\bf 99}, 397 (1995).  

\bibitem{schr}  
J.R. Schrieffer, J. Low Temp.  Phys. {\bf 99}, 397 (1995). 

\bibitem{trembl}  
I. Vilk and A. M. S. Tremblay, J. Phys. I {\bf 7}, 1309 (1997).

\bibitem{ssp} 
J. Schmalian, D. Pines, and B. Stojkovic, \prl {\bf 80}, 3839 (1998).

\bibitem{sadovskii} M.V. Sadovskii, cond-mat/0102111.

\bibitem{cms} 
A. Chubukov, D. Morr, and K. Shakhnovich, Philos. Mag. B {\bf 74}, 563 (1996). 

\bibitem{lp} 
M. Lavagna, cond-mat/0102119; 
M. Lavagna and C. P\'{e}pin, \prb {\bf 62}, 6450 (2000).

\bibitem{im2} 
L. B. Ioffe and A. J. Millis, private communication

\bibitem{im3}  
L. B. Ioffe and A. J. Millis, Phys. Rev B {\bf 58}, 11631 (1998), 
see also  A. T. Zheleznyak, V. Yakovenko, H. D. Drew, and I. I. Mazin, 
\prb {\bf 57}, 3089 (1998).

\bibitem{shulga} 
see e.g, S.V. Shulga, in {\it Proceedings of the Albena Worksop, 1998},
Kluwer Publishers, Dortrecht, p.323 (2001) and references therein. 

\bibitem{sum_rule} 
D. B. Tanner and T. Timusk, {\it Optical Properties of 
High-Temperature Superconductors} in {\it Physical Properties of 
High-Temperature Superconductors} 
D.M. Ginsberg, ed (World Scientific, Singapure, 1992).

\bibitem{homes} 
C.C. Homes {\it et al}, \prl {\bf 71}, 1645 (1993).

\bibitem{basov} 
D. N. Basov {\it et al} \prl {\bf 77}, 4090 (1996).

\bibitem{tu} 
J.J. Tu {\it et al}, cond-mat/01004208.

\bibitem{qui} 
M.A. Quijada {\it et al}, \prb {\bf 60}, 14917 (1999).

\bibitem{a_ge} 
A.A. Abrikosov and V.M. Genkin, Sov. Phys. JETP {\bf 38}, 417 (1974).

\bibitem{klein} 
M.V. Klein and S.B. Dierker,  \prb {\bf 29}, 4976 (1984).

\bibitem{dev}  
T.P. Devereaux {\em et al.}, \prl {\bf 72}, 396 (1994); 
\prb {\bf 54}, 12523 (1996); 
T.P. Devereaux and A.P. Kampf, Int. J. Mod. Phys. B  {\bf 11}, 2093 (1997).

\bibitem{dev2} 
H. Monien and A. Zavadowskii, \prb {\bf 41}, 8798 (1990);
W.C. Wu and A. Griffin, \prb {\bf 51}, 1190 (1995); 
T. Devereauz and D. Einzel, {\em ibid} {\bf 51}, 15357 (1995). 

\bibitem{bl} 
A. Chubukov and D. Frenkel, \prb {\bf 52}, 9760 (1995) and references therein.

\bibitem{mcmillan}  
W.L. McMillan, Phys. Rev. {\bf 167}, 331 (1968).

\bibitem{ml}  
Ph. Monthoux and G.G. Lonzarich, \prb {\bf 59}, 14598
(1999).

\bibitem{ferro} 
R. Roussev and A. Millis, cond-mat/0005449Z; 
Wang {\it et al}, cond-mat/0104097 


\bibitem{onufr} 
F. Bouis {\it et al}, cond-mat/9906369.

\bibitem{katanin}
V. Yu Irkhin, A.A. Katanin and M.I Katsnelson,
 cond-mat/0102381 and references 
therein.

\bibitem{va} 
E. Abrahams and C. Varma, Proc. Natl. Acad. of Sciences {\bf 97}, 5714 (2000).

\bibitem{orenst} 
S. Spielman {\it et al}, \prl {\bf 73}, 1537 (1994).


\end{references}
\end{document}